\documentclass[fleqn,usenatbib]{mnras}

\usepackage{newtxtext,newtxmath}

\usepackage[T1]{fontenc}

\usepackage{setspace}

\DeclareRobustCommand{\VAN}[3]{#2}
\let\VANthebibliography\thebibliography
\def\thebibliography{\DeclareRobustCommand{\VAN}[3]{##3}\VANthebibliography}

\usepackage{graphicx}	
\usepackage{amsmath}	

\usepackage{amssymb}	
\usepackage{longtable}
\usepackage{pdflscape}

\newcommand{\edited}[1]{{#1}}

\title[Catalogue of IPHAS PNe]{Grantecan spectroscopic observations and confirmations of Planetary Nebulae candidates in the Northern Galactic Plane}

\author[A. Ritter et al.]{
A. Ritter,$^{1,2}$
Q. A. Parker$^{1,2}$\thanks{E-mail: quentinp@hku.hk}, L. Sabin,$^{3}$,
P. Le D\^u$^{4,5}$, 
L. Mulato$^{4,6}$,
D. Patchick$^{7}$
\\
$^{1}$ The Department of Physics, CYM Physics Laboratory, The University of Hong Kong, Hong Kong (S.A.R.)\\
$^{2}$Laboratory for Space Research, The University of Hong Kong, Room 405b, Cyberport 4, 100 Cyberport Rd, Hong Kong (S.A.R.)\\
$^{3}$Instituto de Astronom\'{\i}a, Universidad Nacional Aut\'onoma de M\'exico, Apdo. Postal 106, 22800 Ensenada, B.C., Mexico\\
$^{4}$2SPOT, 38690 Chabons, France\\
$^{5}$Kermerrien Observatory, 29840 Porspoder, France\\
$^{6}$Cornillon Observatory, 30630 Cornillon, France\\
$^{7}$Deep Sky Hunters Consortium, 90025 Los Angeles, USA
}

\date{Accepted XXX. Received YYY; in original form ZZZ}

\pubyear{2022}

\begin{document}
\label{firstpage}
\pagerange{\pageref{firstpage}--\pageref{lastpage}}
\maketitle

\begin{abstract}
We present Grantecan 10~m telescope (GTC) spectroscopic confirmations of 55 faint Planetary Nebulae (PNe) candidates discovered largely in the INT Photometric H$\alpha$ Survey of the Northern Galactic Plane (IPHAS) by our pro-am collaboration. We confirm 46 of them as `True' (T), 4 as `Likely' (L) and 5 as `Possible' (P) PNe and including \edited{5} new PNe central star (CSPN) discoveries. This was from observations of 62 new candidates yielding a maximum PN discovery success rate of 89\%.  The sensitivity and longer wavelength coverage of IPHAS allows PNe to be found in regions of greater extinction and at these lower Galactic latitudes, including PNe in a more advanced evolutionary state and at larger distances compared to previously known Galactic PNe. We use an holistic set of observed characteristics and optical emission-line diagnostics to confirm candidates. Plasma properties have been determined in a self-consistent way using PyNeb. This work is facilitated by the functionality of our powerful, multi-wavelength database `HASH' (Hong Kong, Australian Astronomical Observatory, Strasbourg Observatory H-alpha Planetary Nebula catalogue) that federates known imaging, spectroscopy and other pertinent data for all Galactic T, L, P PNe and the significant numbers of mimics. Reddenings, corrected radial velocities and PNe electron density and temperature estimates are provided for these new PNe where possible. 
\end{abstract}




\section{Introduction}
Planetary Nebulae (PNe) are the expanding shells of ionized gas ejected from low- to intermediate-mass stars ($\sim1$ to 8~M\textsubscript{\(\odot\)}) towards the end of their lives. They are powerful astrophysical tools for studying late stage stellar evolution \citep{Kwitter2022} and plasma physics \citep{Hajduk2021}, as well as investigating the chemical evolution of the whole Galaxy \citep{Dopita1997,Maciel2003}. Due to their very bright emission lines they can be observed to great distances across our own galaxy and detected and traced in other nearby galaxies in the local group and beyond.

Since Charles Messier observed the first PN (Dumbell Nebula in Vulpecula) in 1764, over three thousand eight hundred of these beautiful objects have now been discovered in our Galaxy \citep{Parker2016}. Before the advent of the SHS \citep{2005MNRAS.362..689P} and IPHAS \citep{2005MNRAS.362..753D} narrow-band H$\alpha$ surveys of the Southern and Northern Galactic Planes respectively, the vast majority of the previously known PNe ($\sim$1,500) were compiled from over 200 years of observations from a wide variety of telescopes and spectrographs into the Strasbourg-ESO Catalogue and its supplement \citep{1992secg.book.....A,1996fsse.book.....A}, and the largely overlapping but independent compendium of Kohoutek \citet{2001A&A...378..843K}. However, these new H$\alpha$ surveys led to a more than doubling of known Galactic PNe as reported in \citet{2006MNRAS.373...79P,2008MNRAS.384..525M} for the Southern Galactic plane and \citet{2014MNRAS.443.3388S} for the Northern Galactic plane. These `new' PNe were not simply more of the same but are generally fainter, more evolved/extended, more obscured and also, in many cases, more compact compared to the previous catalogues. Other notable numbers of new PNe discoveries have also come from the Deep Sky Hunters (DSH) `amateur’ consortium and more recently from a dedicated group of French amateurs, e.g. \citet{2012RMxAA..48..223A}. This is via painstaking, fresh analysis of the now on-line broad and narrow-band Digital Sky Survey plates, e.g. \citet{2010PASA...27..156J}, \citet{2006A&A...447..921K,2012IAUS..283..414K,2014apn6.confE..48K}. This includes our own recent professional-amateur (pro-am) collaboration (Le D\^u et al., A\&A, in press) that reports the discovery of 210 PNe. Many of these new discoveries are from outside the narrow Galactic latitude confines of the Milky-Way H$\alpha$ surveys (ie $\sim \pm10$~degrees for the SHS and $\pm5$~degrees for IPHAS). We are indeed currently in a new golden age of PNe discovery in our own Galaxy where pro-am collaborations are playing an increasingly important role with their access to bespoke, dedicated facilities and time.

\section{Observations}
\begin{figure}
    \centering
    \includegraphics[width=0.45\textwidth]{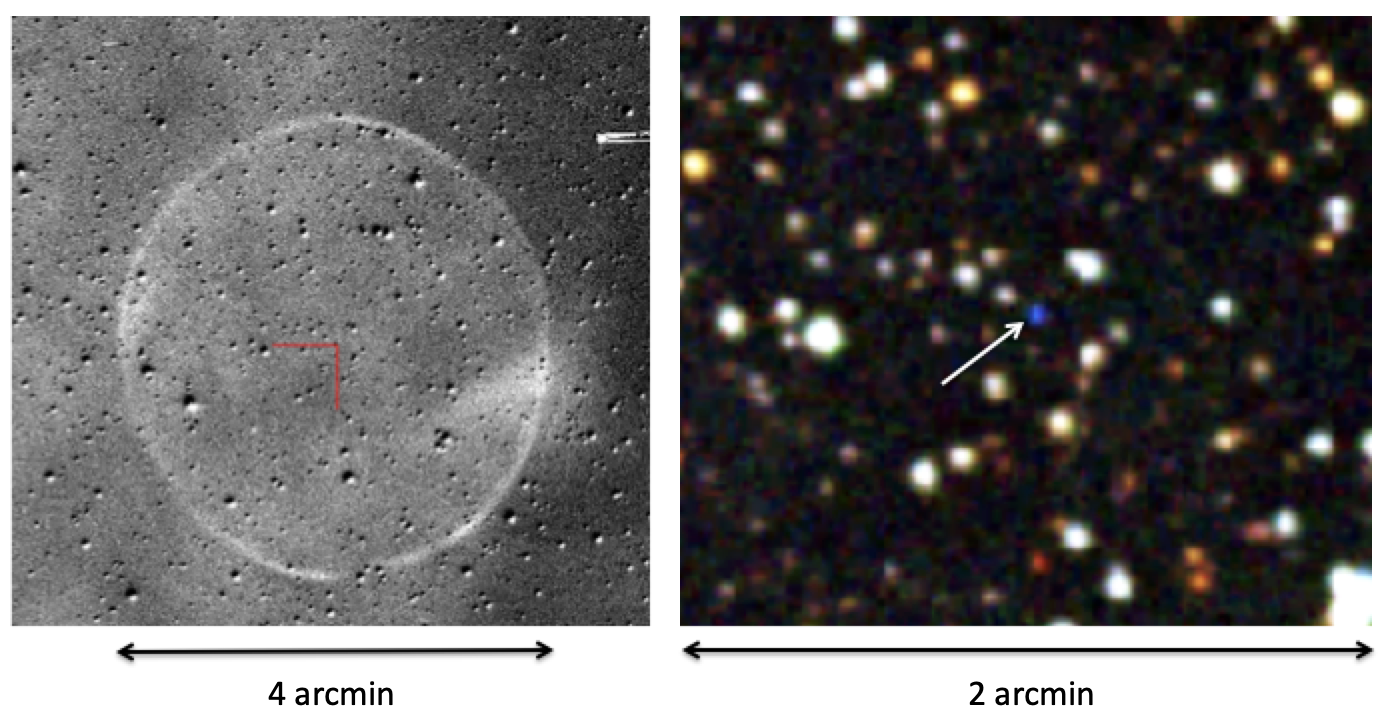}
    \caption{Left panel: Quotient image of a newly confirmed PNe `Ju 1' (HASH ID~4408) that has a diameter of 240\arcsec and with the position of the \edited{independently} discovered CSPN indicated by a half-cross in red. The quotient is obtained by dividing the IPHAS H$\alpha$ image by the equivalent broad-band R (red) image. Right panel: A $2\times2$ arcminute RGB broad band image created from the `SSS' i-band, r-band and B$_j$ band deep photographic data centred on the faint, blue CSPN (arrowed).}
    \label{fig:Fig-CSPN}
\end{figure}

A total of 62 new PN candidates were found by careful scrutiny of the IPHAS survey data and other data for more ``out of plane" candidates by our combined pro-am team. A total of 55 subsequently newly confirmed PNe were carefully examined across the available HASH imagery compiled as part of the process of ingesting any new PN. These data also provided all the \edited{independently} discovered PN central stars (CSPN) by looking for faint blue stars at/near the geometric centres of these new PNe. Even for large angular size PNe the CSPN is nearly always located at its exact centre (if visible in the available surveys) and is usually the only really blue star in the vicinity, giving confidence in its veracity. In Fig.~\ref{fig:Fig-CSPN} we give example images of one of the newly confirmed PNe `Ju 1' (HASH ID~4408) that is 4\arcmin~in diameter and its \edited{independently} found CSPN. The left panel is a H$\alpha$ quotient image clearly showing the slightly oval PNe with \edited{the} CSPN indicated by a red half-cross. The right panel gives a $2\times2$ arcminute RGB broad band image created from the `SSS' \citep{Hambly2001} i-band, r-band and B$_j$ band deep photographic data where the faint blue CSPN is clearly evident. 

All candidates were observed as part of a long-term ``filler" program with the OSIRIS long slit spectrograph on the Spanish 10.4~m Gran Telescopio Canarias (GTC) on La Palma in the Canary Islands. Such projects make decent use of time not suitable for the most challenging observing programs. This list is comprised of mainly large, very low surface brightness candidates not previously observed spectroscopically. It was supplemented by a further 16 PNe identified previously but that needed deeper spectroscopy for firmer confirmation. This provides a total of 78 GTC spectroscopic observations reported here. Of the 16 previously observed PNe candidates, one has been demoted as a result of these new observations from a `T'(True) to a `P' (probable), IPHASX J023538.6+633823 has been upgraded from `P' to `T'; IPHASX J055242.8+262116 from `L' (Likely) to T and one, `Pa30', is now known as a supernova remnant (SNR) of the historical Chinese guest star SN~1181~AD \citep{2021ApJ...918L..33R}. The other 12 remain as `T' PNe. 

The observations were performed between March 1st 2016 and May 17th 2018, under the programs GTC4-16AMEX, GTC12-17AMEX, GTC8-17BMEX and GTC11-18AMEX. Due to the constraints inherent to the filler mode, the observations were conducted under different seeing (up to 2\arcsec), moon phase (dark to bright) and sky transparency (photometric to non-photometric) conditions. Such conditions are no serious impediment to the successful spectroscopic follow-up of even very low surface brightness PNe that emit most of their light in narrow emission lines.

The spectroscopic data were obtained using two 2048$\times$4096 Marconi CCD44-82 detectors with a pixel size of 15~$\mu$m/pix. The plate scale was 0.254\arcsec~per pixel with 2$\times$2 binning adopted. We used the R1000B grating that provides long spectral coverage over the entire optical band from 3630\AA\ to 7500\AA, a dispersion of 2.12\AA/pix and a spectral resolution of 2.15\AA. Such a configuration allows detection of all significant optical emission lines from ionised plasmas, indispensable for the diagnostic identification and analysis of different kinds of nebulae, including PNe. The concurrent observation of 
spectrophotometric standard stars allowed flux calibration, necessary to determine the logarithmic extinction (cH$\beta$) from the corrected Balmer decrement. The chosen spectrograph setup also provides a resolving power sufficient to separate nearby diagnostic emission lines such as \,[S\,{\sc ii}] 6716\AA\, and \,[S\,{\sc ii}] 6731\AA\ used here for electronic density N$_{e}$ estimates from the [SII]~6716~/~6731\AA\ line ratio (see later). The slit width was typically set to 0.8\arcsec (despite the variable seeing) to retain decent spectral resolution and the total exposure time ranged from  2000 to 3600~seconds depending on the surface brightness of the candidate nebulae. \\
\edited{In Table~\ref{tab:1} we present summary data for the 55 discovered PN presented here and the 16 re-observations. The table lists, in order, the \edited{IAU PNG designation, PN usual name}, HASH ID number, positional information (RA/DEC J2000 and Galactic latitude and longitude), PN status as T, L, P, whether there is a CSPN detected, the angular diameter in H$\alpha$ in arcseconds, the PN morphological classification following the 'ERBIAS sparm' scheme outlined in \citep{2006MNRAS.373...79P}, the heliocentric radial velocity in kms$^{-1}$\edited{, slit position angle 0 for East to West, positive East towards South,
negative East towards North), and exposure time.}}







\section{Data Reduction}
\begin{figure}
    \hspace*{-4mm}
    \includegraphics[width=65mm, angle=270]{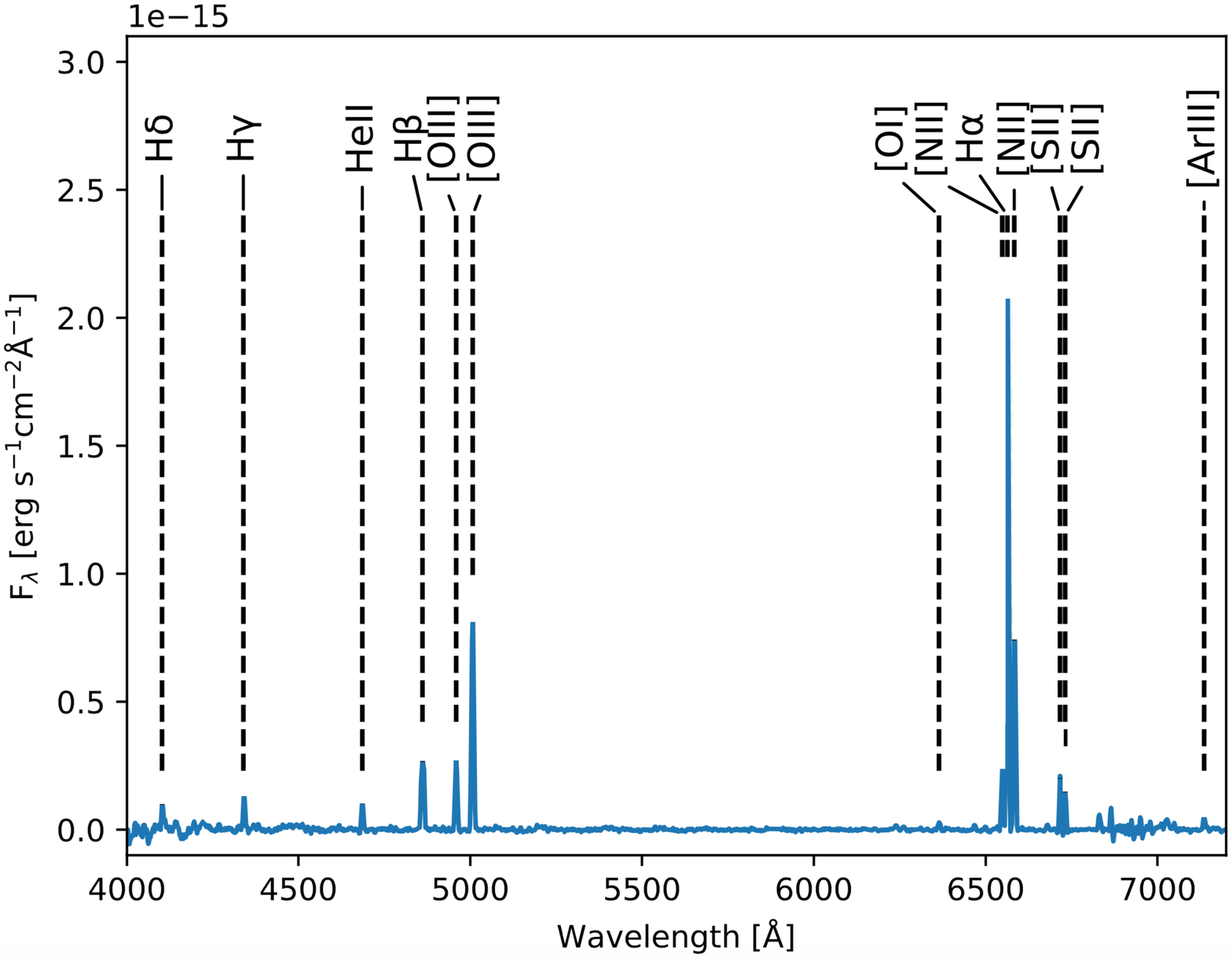}
    \caption{GTC spectrum of newly confirmed PN Ju-1 revealing a high excitation spectrum with prominent [OIII]~5007\&4959\AA\, H$\beta$,  HeII~4686\AA\ and H$\gamma$ emission in the blue and H$\alpha$, [NII], [SII] and [ArIII] lines in the red.}
    \label{fig:Ju-1-GTC-spectrum}
\end{figure}

Standard CCD spectrograph reduction steps were employed. These included combination of individual frames with cosmic ray rejection, bias subtraction, illumination correction, flat-fielding and wavelength calibration via standard calibration arc lamps. We used the IRAF based \href{https://www.inaoep.mx/~ydm/gtcmos/gtcmos.html}{GTCMOS pipeline} \citep{2016MNRAS.460.1555G} built for the purpose. The resulting 2-D images were then visually inspected to identify suitable sky and nebula extraction areas. This is given the extended nature of most of our PN candidates across the slit and in order to avoid contamination by stars. Any clearly identified PN central stars  (CSPN), if falling on the slit, were extracted separately. The sky-subtracted, 1-D spectra of the PNe candidates were then co-added if necessary and subsequently flux calibrated to produce the final spectrum used for evaluation. An example flux-calibrated PN spectrum of Ju-1, first confirmed from these GTC data, is shown in Fig.~\ref{fig:Ju-1-GTC-spectrum} and displays the most common PN emission lines as marked. 

\section{Data analysis}
The first task was to use the observed spectral signatures and emission line ratios in an holistic combination with all other available imagery (including for object morphology and environment), object measurements and separate observations available conveniently in HASH, to make a decision on the true nature of the observed nebulae.  This is following the robust precepts we have previously established, e.g.  \citeauthor{Frew2010}, (\citeyear{Frew2010}) and \citeauthor{Parker2022}, (\citeyear{Parker2022}).

Of the 62 new PN candidates observed for this program we confirm 55, with 46 as True, 4 as Likely and 5 as Possible PNe. This gives a maximum confirmation rate of 89\% with the remaining 7 rejected candidates being identified as a mixture of the various kinds of the usual PNe mimics (such as H~II regions, parts of Supernova remnants and objects of unknown nature, but not PNe). These 7 rejected PNe and their GTC spectra can be examined, if required, via their unique HASH ID's as follows: 8241, 8248, 8448, 11583, 31955, 31956 and 31957.

\begin{figure}
    \centering
    \includegraphics[width=0.45\textwidth]{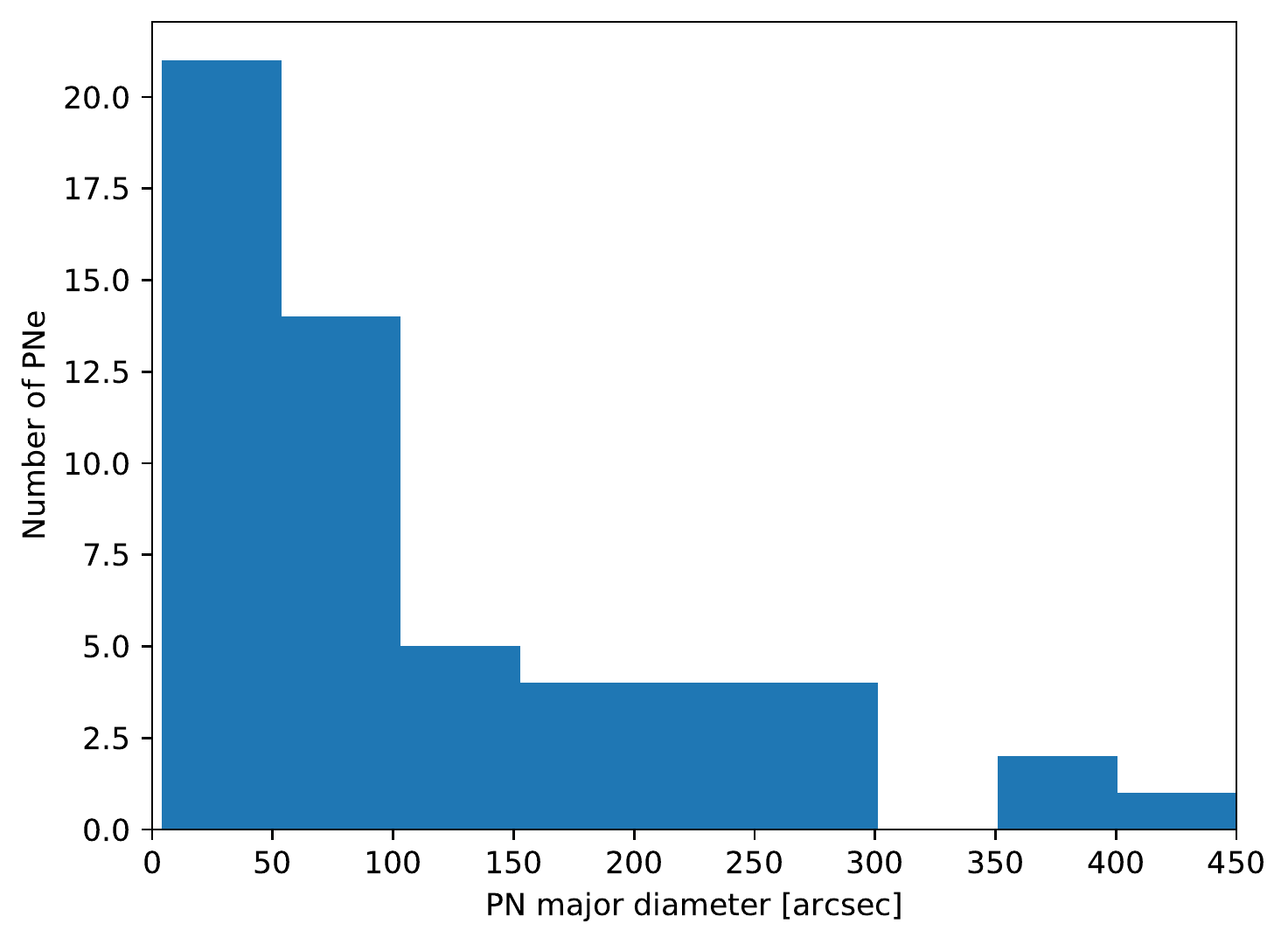}
\caption{\edited{Histogram of major axis diameters in arcseconds for the 55 new PNe confirmed here. About 38\% of the sample are less than 50\arcsec~in size but only 3 are less than 10\arcsec~across while 34 (62\%) have diameters $>$50\arcsec.}}
 \label{fig:diameters}
\end{figure}

\edited{\subsection{Major axis diameters}
A histogram of the major axis diameters in arcseconds (\arcsec) measured from the IPHAS H$\alpha$ imagery for the 55 new PNe confirmed here are presented in Fig.~\ref{fig:diameters}. Only 38\% of the sample are less than 50\arcsec~in size with only 3 less than 10\arcsec~across while 34 (62\%) have diameters $>$50\arcsec. The average diameter for the sample is 111\arcsec~with $\sigma = 107$\arcsec~confirming the wide angular size range but generally more extended nature of this sample. For comparison, the average angular diameter for all 2695 True PN in HASH, as at September 2022, is $\sim$56\arcsec.
}

\edited{\subsection{Spectral analysis}}
For the 55 new, confirmed PNe and for the 16 re-observed, spectral analysis was undertaken after first applying the standard heliocentric velocity correction in kms$^{-1}$. The spectral resolution employed is insufficient to determine useful kinematic information from the PNe, such as the PN shell's expansion velocity, but systemic velocities are reasonably well determined with typical velocity error of $\pm$15~kms$^{-1}$. We then fit all available emission lines above the background noise level in each spectrum  with Gaussian profiles (recall there is little continuum in PNe spectra). Uncertainties were estimated by adding noise (the standard deviation of the spectrum around the emission lines) to the fitted profiles and re-fit 100 times. Using the Python package \href{https://pypi.org/project/PyNeb/}{PyNeb} developed by V.~Luridiana, C.~Morisset, and R.A.~Shaw \citep{2015A&A...573A..42L} we first calculated the reddening and corrected for it. We also simultaneously calculated nebula electron densities and, where possible, electron temperatures (given the principle diagnostic emission lines for the latter can be very weak). The emission line ratios are all used to self consistently determine such electron densities and temperatures as these variables are not independent but weakly correlated. This has been typically ignored in the past when determining electron densities for PNe by assuming a $T_e$ of 10,000K. 

To estimate the uncertainties of these physical characteristics we added 500 Monte-Carlo mock observations with the uncertainties from the line intensities to each spectrum. From this work, because of the limited S/N in many cases due to very low surface brightness and despite use of a 10~m-class telescope, only 27 objects ($\sim39\%$\edited{, measured line intensities shown in Tab~\ref{tab:line_intensities}}) out of combined total of 70 T,L,P PNe yielded both electron temperatures and densities from the diagnostic emission line ratios (this is from 55 new confirmations and 16 re-observations with 15 of these being confirmed). These PyNeb results are based on the usual [SII]~6717/6731~\AA\ and [OII]~3737/3729~\AA\ ratios traditionally used for plasma electron density estimates but accounting for electron temperature and then the usual 
[OIII]~4363/(5007+4959)~\AA\ or [NII]~5755/(6548+6583)~\AA\ ratios for electron temperature, while simultaneously accounting for density variations. Out of these 27 objects, 7 yielded the electron temperatures and densities from both sets of line ratios used to estimate these characteristics. The comparison of those 7 independently estimated values shows good agreement as can be seen for all the values given in Tab.~\ref{tab:electrons}.

\begin{figure}
    \centering
    \includegraphics[width=0.45\textwidth]{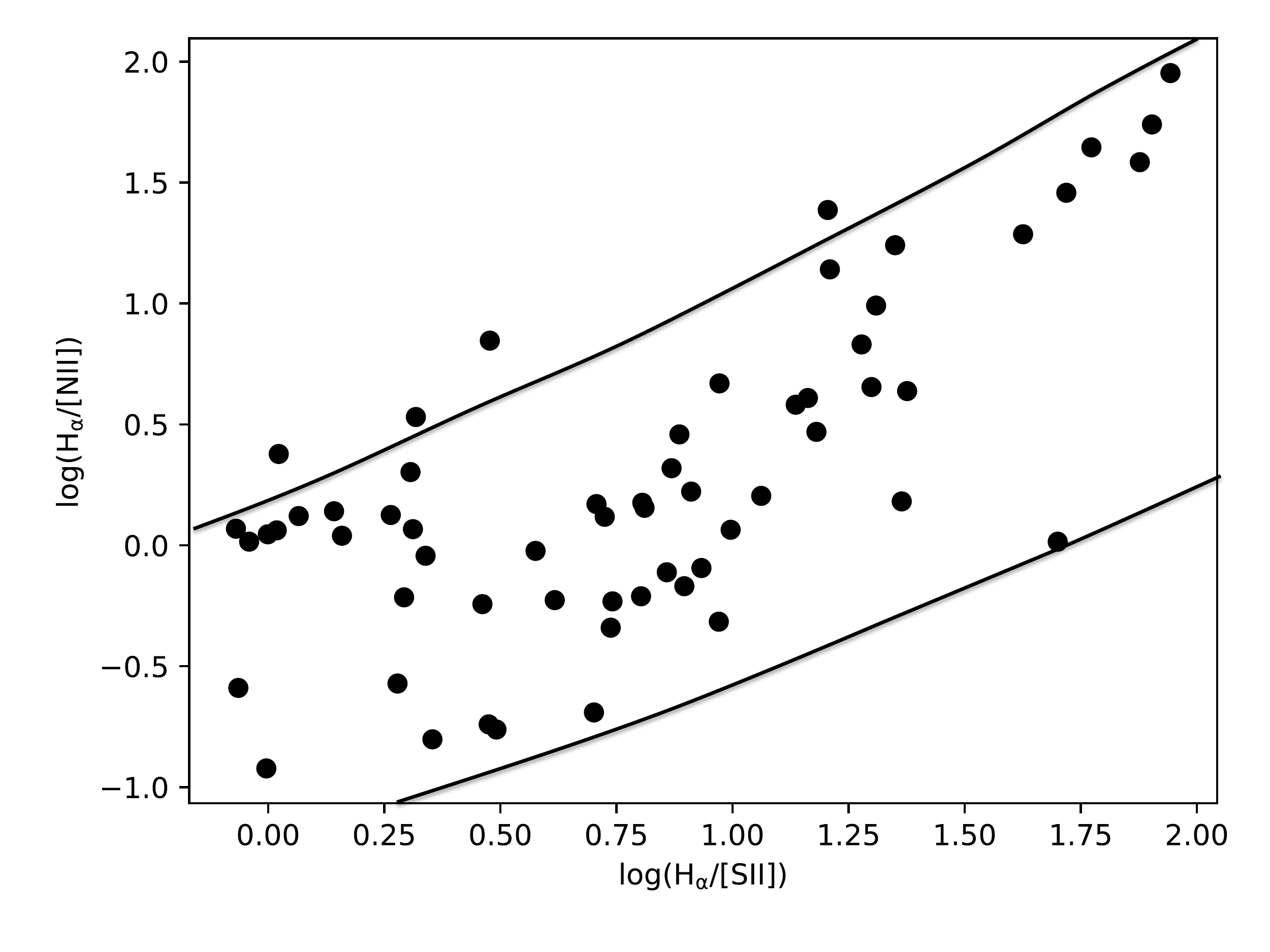}
    \caption{Standard log $\mathrm{(H_\alpha)/[NII]}$ versus log $\mathrm{(H_\alpha)/[SII]}$ `BPT' diagnostic diagram for the confirmed PNe, where [NII] refers to the sum of the flux of the two red nitrogen lines at 6548 and 6584 \AA, and [SII] refers to the equivalent sum of the flux of the two red sulphur lines at 6717 and 6731 \AA. Nearly all points lie within the normal PNe range as indicated by the two black lines following the tracks in Fig.4 of \citep{Frew2010}. A few points below log $\mathrm{(H_\alpha)/[SII]}$ of 0.5 and above log $\mathrm{(H_\alpha)/[NII]}$ of 0.0 overlap slightly with the typical SNR and Galactic HII locii.}
    \label{fig:GTC-fig3.pdf}
\end{figure}

In Fig.~\ref{fig:GTC-fig3.pdf} we present the traditional `BPT' diagnostic diagram (often used to separate different kinds of nebula source) \citep{1981PASP...93....5B} of all T,L,P PNe in this study following \citet{Frew2010}. In this particular example we plot emission line flux ratios of $\mathrm{(H_\alpha)/[NII]}$ versus log $\mathrm{(H_\alpha)/[SII]}$, where [NII] refers to the sum of the fluxes of the two [NII] at 6548 and 6584~\AA, and [SII] refers to the equivalent sum of the flux of the two [SII] lines at 6717 and 6731~\AA. Nearly all points lie within the normal PNe range as indicated by the two black line tracks - see Fig.4 in \citep{Frew2010}.

In Table~\ref{tab:electrons} we present, where possible, estimates of the interstellar reddening E(B-V) and PN electron temperature and density determined from the measured flux calibrated emission lines as produced by PyNeb. Of the total PNe sample only 27 spectra had sufficiently high S/N to enable these measurements from detected lines.

\subsection{\edited{Central stars and \textit{Gaia} eDR3 distances}}
\begin{figure}
    \centering
    \includegraphics[width=0.45\textwidth]{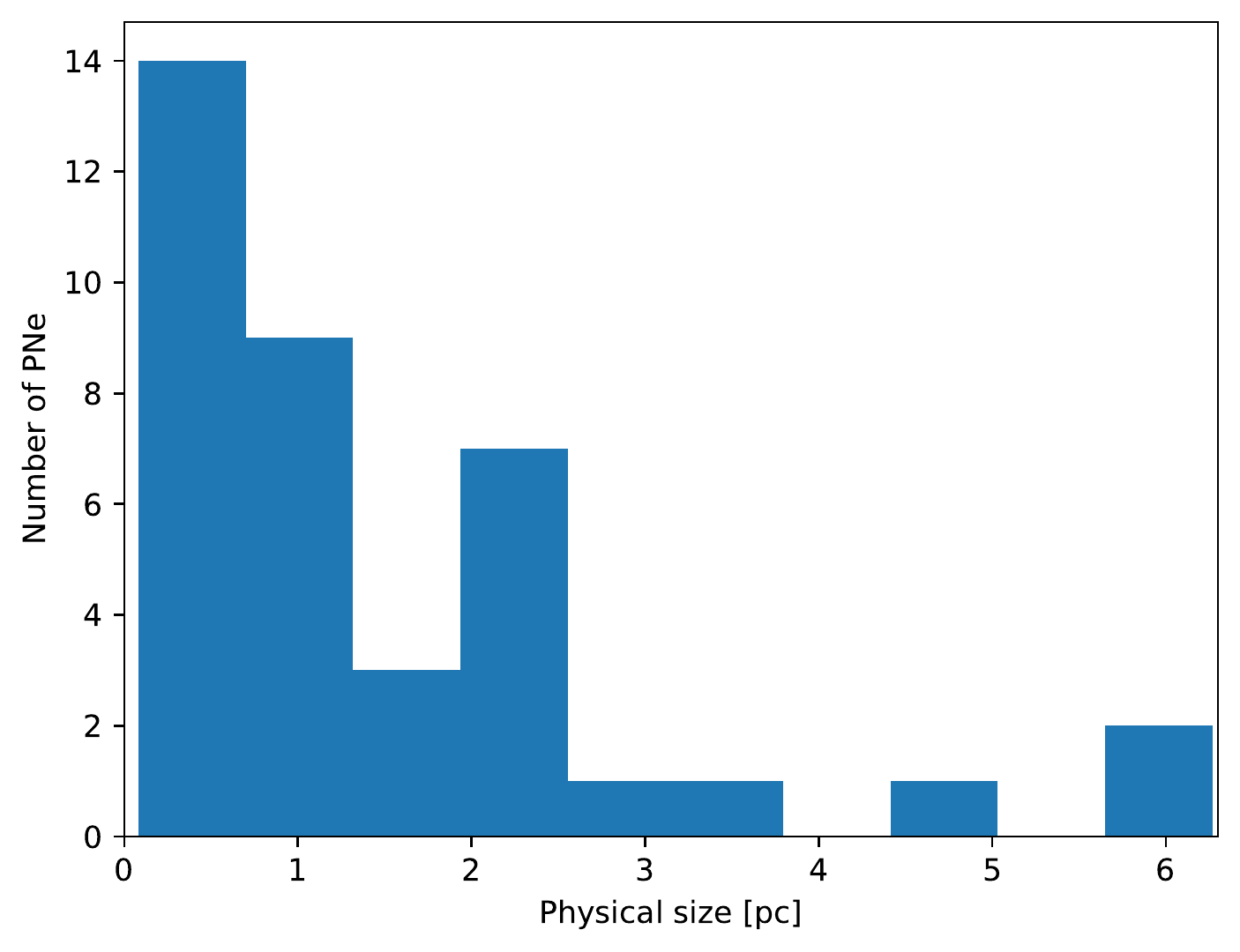}
    \caption{\edited{Physical sizes of the 38 PNe with \citet{Bailer-Jones2021} geometric distances, calculated from the HASH angular diameters.}}
    \label{fig:sizes}
\end{figure}

\edited{The identification of the Central Stars of PNe (CSPNe) is often difficult. They are generally faint and can be in a crowded field. While automated searches for CSPNe in \textit{GAIA} have been developed \citep[e.g.][]{Gonzalez-Santamaria2021, Chornay2021}, \citet{ParkerRitter2022} have shown that purely automated procedures can lead to miss-identifications. In fact 4 CSPNe reported in \citet{Gonzalez-Santamaria2021} for PNGs 037.6-04.7, 059.2+01.0, 105.7+02.2, 149.1+08.7 are likely wrong, as well as 1 CSPN reported in \citet{Chornay2021} (PNG 037.6-04.7). We therefore carefully inspected each PN for possible CSPN making use of the imagery conveniently provided by the HASH Database. For 36 new and 10 already known PN candidates (including Pa 30 which turned out to be a SNR and has subsequently been left out of the CSPNe analysis) a likely CSPN could be identified. 
Out of those 45 CSPNe, 41 could be identified in the GAIA DR3, with 38 (33 new, 5 already known PNe) having distances determined in \citet{Bailer-Jones2021}. Only 1 CSPN has a radial velocity determined in GAIA DR3 \citep{GAIADR32022, TheGaiaMission2016}. The Bailer-Jones et al. distances, GAIA DR3 radial velocities (where available), and the physical sizes (calculated from the geometric distances and HASH angular diameters) are shown in Tab.~\ref{tab:distances}. A histogram of the physical sizes of the 38 PNe with distances is shown in Fig.~\ref{fig:sizes}.}

\section{Conclusions}
We have spectroscopically confirmed 46 true, 4 likely, and 5 possible usually low surface brightness, generally large angular size PNe in the Northern Galactic plane for a total of 55 new PNe. These have now been incorporated into HASH together with their usually determined characteristics such as accurate position(s), morphology and angular size and presence of any CSPN. This PNe sample was effectively too difficult to confirm spectroscopically on smaller aperture telescopes. We have further refined the confidence and spectral characteristics for 16 previously observed PNe candidates, where the existing spectroscopy was too poor, and updated their HASH entries accordingly (with one being rejected as a PN). Seven new candidates observed were rejected as PNe based on our GTC spectroscopy. They remain in HASH as part of our large catalogue of various mimics. Where the PNe spectral S/N permits we have also calculated electron temperatures and densities from either the [NII] / [SII] or the [OIII] / [SII] line ratios for 27 objects. We have also included estimates of interstellar reddening E(B-V) towards these PNe. We have found 37 of the 55 newly confirmed PNe to have credible blue CSPN identified from deep multi-wavelength images where PanSTARRS \citep{Chambers2016} and SDSS \citep{1998AJ....116.3040G} imagery often play the key role. These CSPN co-ordinates have been added to HASH and cross checked against Gaia EDR3. This work has added about $\sim2\%$ to the current total of confirmed Galactic PNe in HASH of over 3848 T,L,P entries.

\section*{Acknowledgements}
QAP thanks the Hong Kong Research Grants Council for GRF research support under grants 17326116 and 17300417. AR thanks HKU for the provision of a postdoctoral fellowship. LS acknowledges support by UNAM PAPIIT project IN110122 (Mexico). This work has made use of data from the European Space Agency (ESA) mission
{\it Gaia} (\url{https://www.cosmos.esa.int/gaia}), processed by the {\it Gaia}
Data Processing and Analysis Consortium (DPAC,
\url{https://www.cosmos.esa.int/web/gaia/dpac/consortium}). Funding for the DPAC
has been provided by national institutions, in particular the institutions
participating in the {\it Gaia} Multilateral Agreement.

\edited{\section{Data Availability}
The data underlying this article are available in the article itself and in its associated online material freely accessible from the HASH database found here: \url{http://hashpn.space} by simply entering the unique HASH ID number for each source as provided.}

\onecolumn
\begin{landscape}
\fontsize{8}{10}\selectfont
 \begin{longtable}{ | *{14}{l|} }
\caption{Summary of key data for the 55 newly confirmed PNe including ID, HASH ID, position (RA/DEC J2000 and Galactic l/b), T,L,P status, whether there is a CSPN detected, angular size, Morphology, heliocentric spectroscopic radial velocity\edited{, slit position angle (0 for East to West, positive East towards South, negative East towards North), and the exposure time (if observed in more than one night then the exposure times are added together)}. The last 16 entries in the table are for previously observed candidates that required better spectra for final confirmation. 'IX' stands for 'IPHASX'.}	
	\\	\hline
		\edited{IAU PNG} & \edited{Target Name} & HASH ID & RA & DEC & l & b & Status & CS & Ang. Diam. ["] & Morphology & $\mathrm{v_{rad} [km s^{-1}]}$&PA [$^\circ$]&$\mathrm{t_{exp}}$ [s]\\
		\endhead  
	\hline
037.6-04.7&BMP J1917+0200&2502&19:17:07.30&02:00:10.15&37.6558&-4.7916&T&y&76x64&Ea&33.8&-90&2*1200\\
037.9-03.4&Abell 56&390&19:13:06.10&02:52:47.89&37.9727&-3.4965&T&y&206x182&Rmr&-3.2&73&2*1000 + 2*1000\\
038.7-02.4&IX J191058.9+040350&8193&19:10:58.90&04:03:50.40&38.7811&-2.4816&T&n&11&Ba&19.6&0&2*1200\\
040.5-00.0&IX J190543.8+064413&8190&19:05:43.80&06:44:13.34&40.5534&-0.0911&T&n&14&Ea&77.6&-90&2*1000\\
040.6-01.5&IRAS 19086+0603&8528&19:11:04.80&06:08:45.06&40.6406&-1.5428&T&n&32x22&Bam&89.5&-43&2*1000\\
043.8+02.1&IXJ190333&8506&19:03:38.50&10:42:27.47&43.8464&2.1878&P&n&450&A&41.1&54&\\
051.3+01.8&PM 1-295&452&19:19:18.80&17:11:48.08&51.3613&1.8143&L&y&20&Rm&-19.2&-90&2*1000\\
057.6+01.8&PM 1-305&4819&19:31:41.30&22:43:38.75&57.6121&1.8763&T&y&15&Eas&11.8&19&30\\
058.1-00.8&IX J194301.3+215424&8566&19:43:01.30&21:54:24.91&58.1786&-0.8113&T&n&14&B&17.8&-45&2*1250 + 2*1250\\
058.9+09.0&Si 1-2&486&19:06:07.20&27:12:58.50&58.9398&9.0837&T&y&60&Ra&33.8&11&2*1300\\
059.2+01.0&Ou 3&10957&19:38:17.54&23:45:48.70&59.255&1.0531&T&y&90&Rar&-12.8&17&2*1250\\
060.0-04.3&Abell 68&495&20:00:10.60&21:42:56.02&60.0457&-4.3343&T&y&38&Bas&-20.1&-25&2*1000\\
060.5+05.6&Pa 21&15561&19:23:15.00&27:07:34.36&60.5899&5.6419&T&n&4x2&S&-23.8&-20&2*1250\\
062.1+03.1&IX J193617.5+272051&8206&19:36:17.61&27:20:51.90&62.1658&3.1954&T&y&9&Ra&-82.3&14&2*1250\\
062.4+00.6&IX J194645.3+262211&8214&19:46:45.30&26:22:11.50&62.4691&0.6838&P&n&290&A&14.6&14&2*1200\\
062.5-01.8&IX J195627.3+250648&8219&19:56:27.30&25:06:48.56&62.5056&-1.8359&L&n&24&E&20.5&-19&2*1300\\
063.1+00.8&IX J194745.5+270150&8217&19:47:45.50&27:01:50.70&63.1529&0.8232&P&n&207&Ea&-35.2&12&2*1300\\
064.9-09.1a&Pa 15&15551&20:29:07.63&23:11:09.30&64.9665&-9.1329&T&y&12&Ers&-1.9&-26&2*500\\
066.1+04.7&IX J193849.6+313744&8210&19:38:49.70&31:37:44.83&66.1861&4.7925&T&y&160&Ear&-9.2&-3&2*1300\\
066.5-14.8&Kn 45&4359&20:53:03.94&21:00:10.90&66.5074&-14.8979&T&y&145x138&Ears&-53.9&26&2*1200\\
066.9-07.8&Kn 19&4356&20:29:20.60&25:32:39.84&66.9484&-7.8206&T&y&74x73&Rars&15.5&14&2*1200\\
067.3-02.6&IX J201058.0+284455&8232&20:10:58.00&28:44:55.00&67.3074&-2.6396&T&n&48x36&Bap&29.2&-6&2*1300\\
069.6-03.7&Kn 20&10881&20:21:26.47&30:05:38.01&69.6977&-3.7869&T&n&16&Rr&26.4&-10&2*1250\\
070.5+11.0&Kn 61&10899&19:21:38.94&38:18:57.20&70.524&11.0068&T&y&100x92&Rrs&-21.5&-6&2*1300\\
073.4+01.5&IRAS 20084+3604&8230&20:10:17.90&36:13:09.73&73.4897&1.5595&P&n&13.9&Eam&-88.2&31&2*1250\\
075.0-07.2&Pa 27&15564&20:48:58.37&32:18:14.80&75&-7.2043&T&y&72x60&Eam&-24.7&15&3*800\\
075.3+05.5&Pa 22&15565&19:58:13.08&39:54:40.80&75.3581&5.5347&T&y&49x45&Eas&-33.4&46&2*1250\\
075.5+01.7&Ju 1&4408&20:15:21.45&38:02:43.80&75.5701&1.7233&T&y&240&Rr&31.5&67&2*1200\\
076.8-08.1&Pa 28&15566&20:58:10.94&33:08:33.10&76.8932&-8.1711&T&y&133x123&Ears&10.5&-3&2*1300\\
078.4-07.2&Pa 29&15567&20:59:43.50&34:54:23.00&78.4594&-7.2741&T&y&4&E&-194.6&-18&30\\
079.8-10.2&Alves 1&10960&21:15:06.60&33:58:18.01&79.8888&-10.2639&T&y&270&Es&21.4&-2&2*1300\\
082.1-07.8&Kn 24&560&21:13:37.70&37:15:37.44&82.1173&-7.8014&T&y&190&Bams&-6.0&34&2*1300 + 2*1000\\
082.5-06.2&Kn 25&4362&21:09:20.20&38:36:06.12&82.5312&-6.2696&T&y&79x57&Bmp&-17.9&-80&2*1200\\
093.8-00.2&LDu 18&17066&21:29:52.30&50:54:21.64&93.8654&-0.2238&T&n&21&Rar&-51.7&-151&2*1250\\
094.5-00.8a&LDu 1&10959&21:36:05.80&50:54:09.29&94.5844&-0.8915&T&y&132x120&Rars&-36.6&-45&2*1300 + 2*1000\\
098.3-04.9&Pa 41&15568&22:10:13.64&50:04:33.40&98.3084&-4.928&T&y&102x82&Ea&-48.0&5&2*1300\\
099.1+05.7&KTC 1&4367&21:28:11.00&58:52:34.68&99.1885&5.7209&T&y&22x16&E&-50.8&62&300\\
099.7-08.8&HaWe 15&602&22:30:33.43&47:31:23.30&99.7159&-8.8978&T&y&295x180&Em&-3.7&14&2*1300\\
103.7+07.2&Kn 30&4368&21:47:24.50&63:05:09.60&103.766&7.2839&T&y&13x12&Eamrs&-130.2&-44&2*1250\\
107.0+21.3&K 1-6&617&20:04:14.28&74:25:36.00&107.0357&21.3839&T&n&198x160&Ea&-6.0&-78&2*1200\\
111.2-03.0&We 2-260&8266&23:22:23.70&57:46:27.98&111.2606&-3.0654&T&n&162x132&Bams&-15.6&26&1600 + 2*1300\\
120.4-01.3&Ou 2&10956&00:30:56.74&61:24:34.30&120.4821&-1.3647&T&y&76&Emrs&-98.7&78&2*1200 + 2*1300\\
129.2-02.0&We 2-5&655&01:42:37.88&60:09:47.20&129.2637&-2.0778&T&y&210x165&Bamrs&-33.4&45&2*1000 + 2*1250\\
129.6-05.6&KLSS 2-8&658&01:40:05.84&56:34:54.60&129.6292&-5.6559&T&y&90x75&Er&-97.3&80&2*1300\\
136.8-13.2&Kn 58&10896&02:12:27.84&47:27:10.10&136.8472&-13.2179&T&y&75x52&B&-133.4&-90&2*1250\\
138.1+04.1&Sh 2-200&670&03:10:58.86&62:47:54.90&138.128&4.1193&T&n&360x345&Eamrs&-48.9&52&2*1000\\
139.3+04.8&KK 26&4393&03:23:04.90&62:47:11.76&139.311&4.8418&T&n&106x54&Bams&-3.7&-18&2*1200\\
147.1-09.0&HaWe 3&4495&03:16:34.00&46:53:37.39&147.1045&-9.0531&T&y&38x36&Ears&-5.5&2&2*30\\
147.2+08.3&Kn 33&4330&04:32:38.10&60:20:12.12&147.2155&8.365&T&y&17x16&Ras&41.5&24&30\\
151.0-00.4&Ou 1&8458&04:07:21.58&51:24:22.40&151.009&-0.457&L&y&95x75&I&-1.4&0&2*1000\\
154.8+05.9&Kn 36&4333&04:55:24.50&52:59:15.00&154.8942&5.9869&T&n&59x50&Ear&15.5&21&1800\\
164.8-09.8&Kn 51&10890&04:25:26.86&35:06:07.80&164.806&-9.8386&T&y&84x60&Iams&32.8&-8&2*1300\\
174.6-05.2&IX J051152.2+302751&8313&05:11:52.20&30:27:51.19&174.6695&-5.2616&P&y&375x245&Er&1.8&0&2*1200\\
175.6+11.4&Kn 62&15571&06:23:55.42&38:15:14.50&175.6316&11.4563&T&y&126&R&88.6&3&2*1300\\
182.3-03.7&IX J053650.8+245616&8331&05:36:50.80&24:56:16.69&182.364&-3.7717&L&n&300x225&B&-8.3&-43&2*1500\\
\hline
040.7+03.4&IX J185322.1+083018&4424&18:53:22.10&08:30:18.00&40.7292&3.4413&T&n&110&Rar&-45.3&-40&2*1500\\
059.1-01.4&Ra 17&8215&19:47:28.90&22:28:23.81&59.1857&-1.4214&T&y&27&Ras&-13.3&18&2*1300\\
060.2+00.8&Kn 11&10878&19:41:19.10&24:30:52.56&60.2489&0.8219&T&n&9&B&-11.0&-17&1000\\
086.1+05.4&We 1-10&571&20:31:52.36&48:52:49.70&86.1906&5.4601&T&y&195x185&Rar&-3.2&11&2*1000\\
086.9-03.4&Ou 5&15806&21:14:20.03&43:41:36.00&86.9108&-3.4821&T&y&33x20&Bmp&-38.0&-14&2*1000\\
095.1+00.9&KKR 62&4431&21:30:44.90&52:41:48.84&95.1902&0.9889&T&n&56x45&B&-3.7&73&2*1250\\
095.2+25.4&Kn 59&10897&18:41:41.90&65:11:57.98&95.2776&25.4542&T&y&4&Ramrs&4.5&-88&2*900\\
097.4+12.3&KnFe 1&10889&20:38:09.19&61:55:02.90&97.425&12.3683&T&y&42x36&Ers&-18.8&65&2*1200\\
098.9+03.0&IX J214032.5+564751&10285&21:40:32.60&56:47:51.61&98.9933&3.0779&T&n&9.4&Rr&-64.9&-59&2*540\\
100.3+02.8&Cr 1&4386&21:49:11.69&57:27:19.70&100.3154&2.8183&P&n&120x106&Es&-43.9&-16&2*1250\\
105.7+02.2&FsMv 1&5240&22:25:56.90&60:11:48.12&105.7755&2.2642&T&y&88&Eas&-82.3&54&2*1300\\
109.4+07.7&Kn 31&4369&22:27:39.19&66:44:09.50&109.4006&7.72&T&y&80&Rars&-31.6&56&2*1300\\
123.0+04.6&Pa 30&15569&00:53:11.20&67:30:02.40&123.0998&4.6295&SNR&y&171x156&&9.1&-60&2*1300\\
134.1+03.0&IX J023538.6+633823&4425&02:35:39.40&63:38:23.93&134.1943&3.0706&T&n&135&Ea&-7.4&-64&2*1300\\
149.1+08.7&Kn 34&4332&04:45:18.65&59:09:24.60&149.1755&8.7933&T&y&60x57&Rars&51.1&20&2*1200\\
183.0+00.0&IX J055242.8+262116&9824&05:52:42.80&26:21:16.09&183.02190&0.01762&T&y&16.1&Ramr&76.7&-43&4*750\\
	\hline
\label{tab:1}
\end{longtable}

\clearpage
\fontsize{8}{8}\selectfont
\begin{longtable}{ | *{10}{l|} }
\caption{\edited{Line intensities for each PN from the 27 PNe spectra with sufficient S/N of which 23 are for the new PNe discovered. "n.d." stands for not detected.}}
\\	\hline
IAU PNG & $\mathrm{H}_\alpha$ & $\mathrm{H}_\beta$ & [NII] 5755\AA & [NII] 6548\AA & [NII] 6584\AA & [OIII] 4363\AA & [OIII] 5007\AA & [SII] 6716\AA & [SII] 6731\AA \\
\endhead  
\hline
037.6-04.7 & $1.3\text{e-}15\pm8.3\text{e-}18$ & $3.0\text{e-}16\pm7.7\text{e-}18$ & $2.8\text{e-}17\pm9.6\text{e-}18$ & $5.6\text{e-}16\pm6.8\text{e-}18$ & $1.8\text{e-}15\pm7.2\text{e-}18$ & n.d. & $4.1\text{e-}16\pm6.7\text{e-}18$ & $2.6\text{e-}16\pm7.6\text{e-}18$ & $2.0\text{e-}16\pm8.0\text{e-}18$ \\
037.9-03.4 & $7.2\text{e-}15\pm1.1\text{e-}16$ & $1.6\text{e-}15\pm7.5\text{e-}17$ & $2.5\text{e-}16\pm8.3\text{e-}17$ & $8.9\text{e-}15\pm1.2\text{e-}16$ & $2.6\text{e-}14\pm1.2\text{e-}16$ & n.d. & $2.7\text{e-}15\pm6.8\text{e-}17$ & $9.5\text{e-}16\pm1.0\text{e-}16$ & $4.7\text{e-}16\pm1.1\text{e-}16$ \\
040.6-01.5 & $2.9\text{e-}15\pm5.6\text{e-}18$ & $8.5\text{e-}17\pm4.7\text{e-}18$ & $1.8\text{e-}17\pm3.3\text{e-}18$ & $2.6\text{e-}15\pm5.3\text{e-}18$ & $8.3\text{e-}15\pm5.6\text{e-}18$ & n.d. & $5.8\text{e-}16\pm4.0\text{e-}18$ & $8.2\text{e-}16\pm5.2\text{e-}18$ & $7.1\text{e-}16\pm5.3\text{e-}18$ \\
057.6+01.8 & $9.8\text{e-}14\pm4.2\text{e-}17$ & $1.0\text{e-}14\pm3.8\text{e-}17$ & n.d. & $8.5\text{e-}16\pm4.6\text{e-}17$ & $1.7\text{e-}15\pm3.9\text{e-}17$ & $6.9\text{e-}16\pm6.0\text{e-}17$ & $8.6\text{e-}14\pm3.8\text{e-}17$ & $6.2\text{e-}16\pm4.3\text{e-}17$ & $6.8\text{e-}16\pm3.7\text{e-}17$ \\
058.9+09.0 & $1.6\text{e-}14\pm3.3\text{e-}17$ & $4.7\text{e-}15\pm1.0\text{e-}16$ & $7.1\text{e-}17\pm3.9\text{e-}17$ & $2.8\text{e-}15\pm3.1\text{e-}17$ & $9.0\text{e-}15\pm3.2\text{e-}17$ & n.d. & $3.5\text{e-}14\pm9.1\text{e-}17$ & $1.6\text{e-}15\pm2.8\text{e-}17$ & $1.3\text{e-}15\pm2.9\text{e-}17$ \\
059.1-01.4 & $2.2\text{e-}15\pm1.5\text{e-}17$ & $3.6\text{e-}16\pm4.8\text{e-}17$ & $8.2\text{e-}17\pm1.5\text{e-}17$ & $5.1\text{e-}16\pm1.7\text{e-}17$ & $1.4\text{e-}15\pm1.5\text{e-}17$ & n.d. & $1.6\text{e-}15\pm3.7\text{e-}17$ & $1.3\text{e-}15\pm1.3\text{e-}17$ & $8.6\text{e-}16\pm1.5\text{e-}17$ \\
060.0-04.3 & $5.4\text{e-}14\pm3.1\text{e-}17$ & $1.2\text{e-}14\pm3.3\text{e-}17$ & n.d. & $3.2\text{e-}15\pm2.8\text{e-}17$ & $8.8\text{e-}15\pm2.8\text{e-}17$ & $3.5\text{e-}16\pm5.3\text{e-}17$ & $5.8\text{e-}14\pm3.7\text{e-}17$ & $1.6\text{e-}15\pm3.0\text{e-}17$ & $1.1\text{e-}15\pm2.5\text{e-}17$ \\
060.5+05.6 & $9.3\text{e-}16\pm4.5\text{e-}18$ & $1.3\text{e-}16\pm7.6\text{e-}18$ & $1.7\text{e-}17\pm6.4\text{e-}18$ & $2.7\text{e-}16\pm4.5\text{e-}18$ & $8.8\text{e-}16\pm4.4\text{e-}18$ & n.d. & $2.7\text{e-}15\pm7.6\text{e-}18$ & $4.0\text{e-}17\pm3.9\text{e-}18$ & $6.8\text{e-}17\pm4.3\text{e-}18$ \\
069.6-03.7 & $4.0\text{e-}15\pm9.7\text{e-}18$ & $5.1\text{e-}16\pm2.9\text{e-}17$ & $5.5\text{e-}17\pm1.5\text{e-}17$ & $6.0\text{e-}16\pm1.1\text{e-}17$ & $1.8\text{e-}15\pm1.1\text{e-}17$ & n.d. & $4.2\text{e-}15\pm2.3\text{e-}17$ & $2.7\text{e-}16\pm8.5\text{e-}18$ & $2.1\text{e-}16\pm7.8\text{e-}18$ \\
073.4+01.5 & $1.1\text{e-}15\pm3.5\text{e-}18$ & $1.1\text{e-}18\pm1.2\text{e-}18$ & $7.0\text{e-}18\pm3.1\text{e-}18$ & $1.3\text{e-}16\pm3.3\text{e-}18$ & $4.1\text{e-}16\pm3.5\text{e-}18$ & n.d. & n.d. & $2.9\text{e-}16\pm3.6\text{e-}18$ & $2.4\text{e-}16\pm3.2\text{e-}18$ \\
075.3+05.5 & $1.5\text{e-}15\pm5.6\text{e-}18$ & $5.1\text{e-}16\pm9.4\text{e-}18$ & n.d. & $3.0\text{e-}17\pm6.7\text{e-}18$ & $5.6\text{e-}17\pm5.8\text{e-}18$ & $4.9\text{e-}17\pm1.5\text{e-}17$ & $3.2\text{e-}15\pm7.8\text{e-}18$ & $3.0\text{e-}17\pm1.1\text{e-}17$ & $3.6\text{e-}17\pm1.1\text{e-}17$ \\
078.4-07.2 & $3.8\text{e-}15\pm4.6\text{e-}18$ & $1.1\text{e-}15\pm9.4\text{e-}18$ & n.d. & $1.4\text{e-}17\pm4.1\text{e-}18$ & $2.8\text{e-}17\pm3.7\text{e-}18$ & $7.8\text{e-}17\pm1.7\text{e-}17$ & $3.8\text{e-}15\pm9.9\text{e-}18$ & $1.8\text{e-}17\pm6.4\text{e-}18$ & $2.6\text{e-}17\pm6.4\text{e-}18$ \\
082.5-06.2 & $2.1\text{e-}15\pm1.1\text{e-}17$ & $5.8\text{e-}16\pm1.9\text{e-}17$ & n.d. & $1.9\text{e-}16\pm1.1\text{e-}17$ & $5.3\text{e-}16\pm1.0\text{e-}17$ & $5.4\text{e-}17\pm2.9\text{e-}17$ & $5.0\text{e-}15\pm1.8\text{e-}17$ & $9.5\text{e-}17\pm1.1\text{e-}17$ & $4.6\text{e-}17\pm9.1\text{e-}18$ \\
086.9-03.4 & $4.6\text{e-}14\pm1.5\text{e-}17$ & $6.9\text{e-}15\pm3.2\text{e-}17$ & $3.1\text{e-}17\pm1.5\text{e-}17$ & $5.9\text{e-}16\pm1.5\text{e-}17$ & $1.8\text{e-}15\pm1.6\text{e-}17$ & $2.5\text{e-}16\pm6.3\text{e-}17$ & $3.9\text{e-}14\pm2.7\text{e-}17$ & $6.2\text{e-}16\pm1.8\text{e-}17$ & $4.6\text{e-}16\pm1.7\text{e-}17$ \\
095.1+00.9 & $7.0\text{e-}15\pm2.5\text{e-}17$ & $6.2\text{e-}16\pm4.0\text{e-}17$ & $6.4\text{e-}17\pm2.6\text{e-}17$ & $2.8\text{e-}15\pm2.3\text{e-}17$ & $8.5\text{e-}15\pm2.3\text{e-}17$ & n.d. & $5.7\text{e-}15\pm4.0\text{e-}17$ & $6.3\text{e-}16\pm2.1\text{e-}17$ & $4.7\text{e-}16\pm2.3\text{e-}17$ \\
095.2+25.4 & $1.0\text{e-}14\pm1.3\text{e-}17$ & $3.5\text{e-}15\pm1.2\text{e-}17$ & $8.8\text{e-}17\pm1.1\text{e-}17$ & $2.8\text{e-}15\pm1.5\text{e-}17$ & $8.3\text{e-}15\pm1.4\text{e-}17$ & $7.5\text{e-}17\pm2.7\text{e-}17$ & $2.5\text{e-}15\pm1.1\text{e-}17$ & $2.6\text{e-}15\pm1.2\text{e-}17$ & $2.0\text{e-}15\pm1.4\text{e-}17$ \\
097.4+12.3 & $4.4\text{e-}15\pm9.5\text{e-}18$ & $1.3\text{e-}15\pm8.2\text{e-}18$ & $1.7\text{e-}17\pm1.0\text{e-}17$ & $2.9\text{e-}16\pm1.0\text{e-}17$ & $8.6\text{e-}16\pm9.1\text{e-}18$ & $1.7\text{e-}16\pm1.8\text{e-}17$ & $1.2\text{e-}14\pm8.9\text{e-}18$ & $1.9\text{e-}16\pm8.8\text{e-}18$ & $1.3\text{e-}16\pm9.0\text{e-}18$ \\
098.9+03.0 & $3.4\text{e-}15\pm1.4\text{e-}17$ & $4.5\text{e-}16\pm3.6\text{e-}17$ & $5.7\text{e-}17\pm1.2\text{e-}17$ & $7.6\text{e-}16\pm1.5\text{e-}17$ & $2.1\text{e-}15\pm1.7\text{e-}17$ & $1.5\text{e-}16\pm5.0\text{e-}17$ & $3.6\text{e-}15\pm3.2\text{e-}17$ & $1.8\text{e-}16\pm1.6\text{e-}17$ & $1.6\text{e-}16\pm1.5\text{e-}17$ \\
099.7-08.8 & $1.3\text{e-}14\pm5.8\text{e-}17$ & $4.0\text{e-}15\pm4.0\text{e-}17$ & $1.4\text{e-}16\pm5.7\text{e-}17$ & $4.2\text{e-}15\pm5.4\text{e-}17$ & $1.3\text{e-}14\pm5.1\text{e-}17$ & $2.9\text{e-}16\pm8.2\text{e-}17$ & $2.2\text{e-}14\pm3.8\text{e-}17$ & $1.1\text{e-}15\pm5.2\text{e-}17$ & $6.6\text{e-}16\pm4.8\text{e-}17$ \\
111.2-03.0 & $2.2\text{e-}15\pm8.4\text{e-}17$ & $2.2\text{e-}18\pm1.5\text{e-}92$ & $2.4\text{e-}16\pm8.2\text{e-}17$ & $4.8\text{e-}15\pm1.1\text{e-}16$ & $1.4\text{e-}14\pm8.3\text{e-}17$ & n.d. & $1.8\text{e-}15\pm7.9\text{e-}17$ & $1.3\text{e-}15\pm7.5\text{e-}17$ & $1.0\text{e-}15\pm7.7\text{e-}17$ \\
120.4-01.3 & $4.4\text{e-}15\pm1.2\text{e-}17$ & $8.4\text{e-}16\pm1.5\text{e-}17$ & n.d. & $1.3\text{e-}16\pm1.1\text{e-}17$ & $3.3\text{e-}16\pm1.0\text{e-}17$ & $1.2\text{e-}16\pm2.2\text{e-}17$ & $1.0\text{e-}14\pm1.4\text{e-}17$ & $1.2\text{e-}16\pm1.1\text{e-}17$ & $9.9\text{e-}17\pm1.1\text{e-}17$ \\
129.2-02.0 & $7.1\text{e-}15\pm6.0\text{e-}17$ & $1.5\text{e-}15\pm3.0\text{e-}18$ & $2.2\text{e-}16\pm3.6\text{e-}17$ & $1.0\text{e-}14\pm4.7\text{e-}17$ & $3.1\text{e-}14\pm5.1\text{e-}17$ & n.d. & $2.2\text{e-}15\pm9.2\text{e-}18$ & $1.3\text{e-}15\pm4.6\text{e-}17$ & $9.6\text{e-}16\pm5.4\text{e-}17$ \\
136.8-13.2 & $2.0\text{e-}14\pm3.6\text{e-}17$ & $6.3\text{e-}15\pm8.3\text{e-}17$ & n.d. & n.d. & $6.9\text{e-}16\pm3.9\text{e-}17$ & $7.4\text{e-}16\pm1.6\text{e-}16$ & $5.5\text{e-}14\pm7.2\text{e-}17$ & $2.2\text{e-}16\pm3.3\text{e-}17$ & $1.6\text{e-}16\pm3.2\text{e-}17$ \\
139.3+04.8 & $3.1\text{e-}14\pm3.7\text{e-}17$ & $4.3\text{e-}15\pm4.1\text{e-}17$ & $1.2\text{e-}16\pm2.7\text{e-}17$ & $5.1\text{e-}15\pm3.8\text{e-}17$ & $1.6\text{e-}14\pm3.6\text{e-}17$ & $3.3\text{e-}16\pm7.2\text{e-}17$ & $5.1\text{e-}14\pm3.3\text{e-}17$ & $2.8\text{e-}15\pm3.9\text{e-}17$ & $2.1\text{e-}15\pm4.5\text{e-}17$ \\
147.2+08.3 & $2.5\text{e-}14\pm1.7\text{e-}17$ & $5.0\text{e-}15\pm3.8\text{e-}18$ & $1.6\text{e-}16\pm1.8\text{e-}17$ & $4.4\text{e-}15\pm1.6\text{e-}17$ & $1.3\text{e-}14\pm1.8\text{e-}17$ & $1.7\text{e-}16\pm3.3\text{e-}17$ & $1.3\text{e-}14\pm3.9\text{e-}18$ & $9.7\text{e-}15\pm1.6\text{e-}17$ & $8.0\text{e-}15\pm1.7\text{e-}17$ \\
151.0-00.4 & $6.9\text{e-}15\pm1.9\text{e-}17$ & $1.8\text{e-}15\pm2.0\text{e-}17$ & $6.1\text{e-}17\pm1.6\text{e-}17$ & $1.5\text{e-}15\pm1.9\text{e-}17$ & $4.5\text{e-}15\pm2.0\text{e-}17$ & n.d. & $7.9\text{e-}15\pm1.9\text{e-}17$ & $2.0\text{e-}15\pm1.5\text{e-}17$ & $1.4\text{e-}15\pm1.6\text{e-}17$ \\
182.3-03.7 & $1.5\text{e-}15\pm3.6\text{e-}17$ & $1.9\text{e-}16\pm6.4\text{e-}17$ & $1.0\text{e-}16\pm3.6\text{e-}17$ & $1.1\text{e-}16\pm3.8\text{e-}17$ & $5.1\text{e-}16\pm3.4\text{e-}17$ & n.d. & $1.3\text{e-}15\pm6.3\text{e-}17$ & $8.5\text{e-}16\pm3.6\text{e-}17$ & $5.8\text{e-}16\pm3.5\text{e-}17$ \\
\hline
\label{tab:line_intensities}
\end{longtable}

\end{landscape}
\begin{longtable}{ | *{7}{l|} }
\caption{Estimates of the Interstellar reddening E(B-V) and nebula electron temperature and density from the 27 PNe spectra with sufficient S/N of which 23 are for the new PNe discovered. Note the large errors on many measures.}
\\ \hline
IAU PNG& E(B-V) & $\mathrm{T_{e^-}([NII],[SII]) [K]}$ & $\mathrm{T_{e^-}([OIII],[SII]) [K]}$ & $\mathrm{\rho_{e^-}([NII],[SII]) [\mathrm{cm^{-3}}]}$ & $\mathrm{\rho_{e^-}([OIII],[SII]) [\mathrm{cm^{-3}}]}$\\
\endhead  
\hline
037.6-04.7&$0.391 \pm 0.018$&$11093 \pm 2246$& &$159 \pm 70$&\\
037.9-03.4& $0.373 \pm 0.030$ & $8603 \pm 1346$ & & $101 \pm 84$&\\
040.6-01.5&$2.145 \pm 0.047$&$7220 \pm 475$& &$306 \pm 16$&\\
057.6+01.8&$1.049 \pm 0.003$& &$13180 \pm 355$& &$1124 \pm 169$\\
058.9+09.0&$0.129 \pm 0.015$&$8580 \pm 1927$& &$151 \pm 34$&\\
059.1-01.4&$0.676 \pm 0.097$&$32334 \pm 9084$& &$7 \pm 5$&\\
060.0-04.3&$0.419 \pm 0.002$& &$10487 \pm 508$& &$18 \pm 12$\\
060.5+05.6&$0.793 \pm 0.045$&$12796 \pm 2644$& &$5921 \pm 4999$&\\
069.6-03.7&$0.860 \pm 0.037$&$18620 \pm 4348$& &$166 \pm 85$&\\
073.4+01.5&$5.067 \pm 0.791$&$49961 \pm 23354$& &$344 \pm 86$&\\
075.3+05.5&$0.022 \pm 0.014$& &$13365 \pm 1927$& &$3136 \pm 5360$\\
078.4-07.2&$0.187 \pm 0.010$& &$15836 \pm 1652$& &$5141 \pm 6731$\\
082.5-06.2&$0.226 \pm 0.027$& &$13228 \pm 1089$& &$108 \pm 120$\\
086.9-03.4&$0.723 \pm 0.003$&$12817 \pm 3837$&$11351 \pm 855$&$90 \pm 11$&$92 \pm 10$\\
095.1+00.9&$1.182 \pm 0.048$&$9370 \pm 1647$& &$110 \pm 57$&\\
095.2+25.4&$0.002 \pm 0.002$&$8770 \pm 1128$&$17976 \pm 3975$&$204 \pm 12$&$240 \pm 17$\\
097.4+12.3&$0.161 \pm 0.006$&$11809 \pm 3480$&$13689 \pm 668$&$73 \pm 57$&$80 \pm 63$\\
098.9+03.0&$0.820 \pm 0.064$&$16044 \pm 2598$&$21227 \pm 2873$&$431 \pm 255$&$491 \pm 293$\\
099.7-08.8&$0.121 \pm 0.008$&$7526 \pm 1043$&$13089 \pm 460$&$17 \pm 14$&$16 \pm 0$\\
111.2-03.0&$5.053 \pm 0.033$&$51714 \pm 23756$& &$363 \pm 293$&\\
120.4-01.3&$0.524 \pm 0.014$& &$13600 \pm 1066$& &$446 \pm 306$\\
129.2-02.0&$0.436 \pm 0.008$&$8213 \pm 491$& &$110 \pm 67$&\\
136.8-13.2&$0.090 \pm 0.012$& &$12945 \pm 1117$& &$365 \pm 381$\\
139.3+04.8&$0.805 \pm 0.005$&$8859 \pm 1028$&$11503 \pm 850$&$95 \pm 32$&$97 \pm 35$\\
147.2+08.3&$0.465 \pm 0.000$&$10112 \pm 649$&$14359 \pm 1198$&$262 \pm 5$&$287 \pm 6$\\
151.0-00.4&$0.276 \pm 0.004$&$10206 \pm 2104$& &$62 \pm 17$&\\
182.3-03.7&$0.950 \pm 0.304$&$44981 \pm 24900$& &$99 \pm 50$&\\
\hline
\label{tab:electrons}
\end{longtable}

\twocolumn

\onecolumn
\begin{spacing}{1.6}
\begin{longtable}{ | *{7}{l|} }
	\caption{\edited{Central stars with \citet{Bailer-Jones2021} distances, physical diameters from HASH angular diameters and $r_{geo}$, and GAIA DR3  radial velocities (where available).}}\label{tab:distances}\\
		\hline
IAU PNG & HASH ID & GAIA ID & $r_{geo}$ [pc] & $r_{phot}$ [pc] & physical diameter [pc] & $v_{rad}$~[$\mathrm{km~s^{-1}}$]\\
\endhead  
\hline
037.9-03.4&390&4268179207028750592&$4836^{+3985}_{-2300}$&$1081^{+316}_{-228}$&$4.830^{+3.980}_{-2.297}$\ $\mathrm{x}4.268^{+3.516}_{-2.029}$&-\\
051.3+01.8&452&4514868732516293760&$2230^{+102}_{-87}$&$2254^{+152}_{-112}$&$0.216^{+0.010}_{-0.008}$\ $\mathrm{x}0.216^{+0.010}_{-0.008}$&-\\
057.6+01.8&4819&2019351991760108672&$3326^{+1606}_{-1281}$&$6727^{+2418}_{-1671}$&$0.242^{+0.117}_{-0.093}$\ $\mathrm{x}0.242^{+0.117}_{-0.093}$&-\\
058.9+09.0&486&2036937138334923648&$2633^{+1258}_{-857}$&$2201^{+1375}_{-809}$&$0.766^{+0.366}_{-0.249}$\ $\mathrm{x}0.766^{+0.366}_{-0.249}$&-\\
059.2+01.0&10957&2020279846433107200&$2948^{+2911}_{-1302}$&$11210^{+2509}_{-1690}$&$1.287^{+1.270}_{-0.568}$\ $\mathrm{x}1.287^{+1.270}_{-0.568}$&-\\
060.0-04.3&495&1826936121144576896&$2844^{+1514}_{-931}$&$12300^{+1296}_{-1178}$&$0.524^{+0.279}_{-0.172}$\ $\mathrm{x}0.524^{+0.279}_{-0.172}$&-\\
062.1+03.1&8206&2025166763313254400&$7091^{+4254}_{-3595}$&$26909^{+6721}_{-3839}$&$0.309^{+0.186}_{-0.157}$\ $\mathrm{x}0.309^{+0.186}_{-0.157}$&-\\
064.9-09.1a&15551&1830887113098897920&$3961^{+1098}_{-804}$&$3949^{+1598}_{-982}$&$0.230^{+0.064}_{-0.047}$\ $\mathrm{x}0.230^{+0.064}_{-0.047}$&-\\
066.1+04.7&8210&2033203609193896064&$7477^{+3235}_{-2852}$&$21607^{+3845}_{-13178}$&$5.800^{+2.510}_{-2.212}$\ $\mathrm{x}5.800^{+2.510}_{-2.212}$&-\\
066.5-14.8&4359&1814597642173959168&$2819^{+1722}_{-1023}$&$1060^{+81}_{-77}$&$1.982^{+1.211}_{-0.719}$\ $\mathrm{x}1.886^{+1.152}_{-0.685}$&-\\
066.9-07.8&4356&1832550223218972800&$5567^{+2741}_{-2137}$&$4957^{+2748}_{-2150}$&$1.997^{+0.984}_{-0.767}$\ $\mathrm{x}1.970^{+0.970}_{-0.756}$&-\\
070.5+11.0&10899&2052811676760671872&$5882^{+2441}_{-1775}$&$4195^{+1936}_{-1393}$&$2.852^{+1.183}_{-0.861}$\ $\mathrm{x}2.624^{+1.089}_{-0.792}$&-\\
075.0-07.2&15564&1859955657931121536&$1436^{+23}_{-24}$&$1430^{+27}_{-26}$&$0.501^{+0.008}_{-0.008}$\ $\mathrm{x}0.418^{+0.007}_{-0.007}$&$-47.4 \pm 2.3$\\
075.3+05.5&15565&2072773550888053888&$7103^{+2837}_{-2086}$&$22112^{+3220}_{-3585}$&$1.687^{+0.674}_{-0.496}$\ $\mathrm{x}1.550^{+0.619}_{-0.455}$&-\\
075.5+01.7&4408&2060926897208561280&$1426^{+741}_{-386}$&$1767^{+1416}_{-884}$&$1.659^{+0.862}_{-0.449}$\ $\mathrm{x}1.659^{+0.862}_{-0.449}$&-\\
076.8-08.1&15566&1865874672618774656&$3881^{+1459}_{-1161}$&$2355^{+512}_{-397}$&$2.503^{+0.941}_{-0.749}$\ $\mathrm{x}2.315^{+0.870}_{-0.692}$&-\\
078.4-07.2&15567&1866682878077985920&$4373^{+2464}_{-1880}$&$1979^{+3729}_{-1410}$&$0.085^{+0.048}_{-0.036}$\ $\mathrm{x}0.085^{+0.048}_{-0.036}$&-\\
079.8-10.2&10960&1866922365452368768&$1903^{+600}_{-371}$&$1099^{+919}_{-105}$&$2.492^{+0.785}_{-0.486}$\ $\mathrm{x}2.492^{+0.785}_{-0.486}$&-\\
082.1-07.8&560&1868658082001121664&$2567^{+996}_{-861}$&$1742^{+233}_{-238}$&$2.365^{+0.917}_{-0.793}$\ $\mathrm{x}2.365^{+0.917}_{-0.793}$&-\\
098.3-04.9&15568&1976783887984709120&$2199^{+320}_{-265}$&$2284^{+423}_{-292}$&$1.088^{+0.158}_{-0.131}$\ $\mathrm{x}0.874^{+0.127}_{-0.105}$&-\\
099.1+05.7&4367&2179544655458448512&$4854^{+1228}_{-1072}$&$4037^{+767}_{-699}$&$0.518^{+0.131}_{-0.114}$\ $\mathrm{x}0.377^{+0.095}_{-0.083}$&-\\
099.7-08.8&602&1986574557983855104&$1302^{+224}_{-143}$&$1093^{+113}_{-111}$&$1.863^{+0.320}_{-0.205}$\ $\mathrm{x}1.137^{+0.195}_{-0.125}$&-\\
103.7+07.2&4368&2216463232261024896&$5845^{+6396}_{-2553}$&$6513^{+1207}_{-2129}$&$0.368^{+0.403}_{-0.161}$\ $\mathrm{x}0.340^{+0.372}_{-0.149}$&-\\
120.4-01.3&10956&430204780732841600&$1586^{+823}_{-378}$&$1901^{+1675}_{-757}$&$0.584^{+0.303}_{-0.139}$\ $\mathrm{x}0.584^{+0.303}_{-0.139}$&-\\
129.2-02.0&655&509636112062310016&$3429^{+1599}_{-1398}$&$2837^{+992}_{-845}$&$3.492^{+1.628}_{-1.423}$\ $\mathrm{x}2.743^{+1.279}_{-1.118}$&-\\
129.6-05.6&658&409721566300550656&$1187^{+789}_{-340}$&$871^{+601}_{-410}$&$0.518^{+0.344}_{-0.148}$\ $\mathrm{x}0.432^{+0.287}_{-0.124}$&-\\
136.8-13.2&10896&354941216942517504&$2843^{+2130}_{-1084}$&$1386^{+45}_{-338}$&$1.034^{+0.775}_{-0.394}$\ $\mathrm{x}0.717^{+0.537}_{-0.273}$&-\\
147.1-09.0&4495&434853485833190528&$4089^{+1470}_{-955}$&$3633^{+843}_{-556}$&$0.753^{+0.271}_{-0.176}$\ $\mathrm{x}0.714^{+0.257}_{-0.167}$&-\\
147.2+08.3&4330&471551198232719744&$2393^{+2979}_{-1243}$&$1326^{+1629}_{-454}$&$0.197^{+0.246}_{-0.102}$\ $\mathrm{x}0.186^{+0.231}_{-0.096}$&-\\
151.0-00.4&8458&250719647917343232&$1998^{+903}_{-643}$&$830^{+227}_{-136}$&$0.920^{+0.416}_{-0.296}$\ $\mathrm{x}0.727^{+0.328}_{-0.234}$&-\\
164.8-09.8&10890&176269718435866112&$953^{+197}_{-109}$&$1250^{+137}_{-146}$&$0.388^{+0.080}_{-0.044}$\ $\mathrm{x}0.277^{+0.057}_{-0.032}$&-\\
174.6-05.2&8313&156214321404677376&$3448^{+1086}_{-707}$&$3734^{+876}_{-578}$&$6.270^{+1.974}_{-1.286}$\ $\mathrm{x}4.096^{+1.290}_{-0.840}$&-\\
175.6+11.4&15571&955255358615209856&$3960^{+1683}_{-1557}$&$1767^{+470}_{-318}$&$2.419^{+1.028}_{-0.951}$\ $\mathrm{x}2.419^{+1.028}_{-0.951}$&-\\
\hline
086.1+05.4&571&2179832585761932032&$2389^{+543}_{-411}$&$1843^{+271}_{-179}$&$2.259^{+0.514}_{-0.388}$\ $\mathrm{x}2.143^{+0.487}_{-0.368}$&-\\
086.9-03.4&15806&1970016153397634048&$6192^{+1846}_{-1344}$&$8937^{+1603}_{-1598}$&$0.991^{+0.295}_{-0.215}$\ $\mathrm{x}0.600^{+0.179}_{-0.130}$&-\\
097.4+12.3&10889&2195278765626033024&$3605^{+1502}_{-1353}$&$5237^{+3492}_{-1653}$&$0.734^{+0.306}_{-0.275}$\ $\mathrm{x}0.629^{+0.262}_{-0.236}$&-\\
109.4+07.7&4369&2218796151111558016&$3115^{+2702}_{-1230}$&$10382^{+2133}_{-1538}$&$1.208^{+1.048}_{-0.477}$\ $\mathrm{x}1.208^{+1.048}_{-0.477}$&-\\
183.0+00.0&9824&3430688759283627776&$2803^{+777}_{-495}$&$2605^{+627}_{-443}$&$0.219^{+0.061}_{-0.039}$\ $\mathrm{x}0.219^{+0.061}_{-0.039}$&-\\
\hline
\end{longtable}
\end{spacing}
\twocolumn



\bibliographystyle{mnras}
\bibliography{references} 




\appendix

\section{PNe images and spectra}

In this appendix we conveniently provide a table of graphical images and spectra for all GTC PNe observations divided into 2 sub indices. \edited{Table A1 presents 1-D GTC spectra for the 55 new confirmed PNe; Table A2 gives Spectra of the previously observed candidates that required better spectra for final confirmation. Table A3 provides images of the 55 newly confirmed PNe.} The table provides the \edited{IAU PNG ID;} unique HASH ID number; an RGB optical image from the IPHAS H$\alpha$, broad-band r and i data when available and then the quotient image from dividing the IPHAS H$\alpha$ by the broad-band r. If there is no IPHAS H$\alpha$ imagery available as the object falls outside of rather narrow Galactic latitude range of $\pm5$~degrees then the best available on-line optical image is presented. After this there is then the mid-infrared WISE \citep{2010AJ....140.1868W} 321 (3.4\micron, 4.6\micron~and 12\micron~band combination) as an RGB image; the NVSS radio image (when available), \citep{Condon1998} and finally the GALEX image when available, \citep{2005ApJ...619L...1M}. \edited{Table A4 gives Images of the previously observed candidates that required better spectra for final confirmation in the same way as Table A3.}

\clearpage
\onecolumn
\begin{longtable}{ *{2}{l} }
    \caption{\edited{Spectra of the 55 newly confirmed PNe.}}\label{tab:spectra1}\\
    \endhead  
\includegraphics[width=0.48\textwidth]{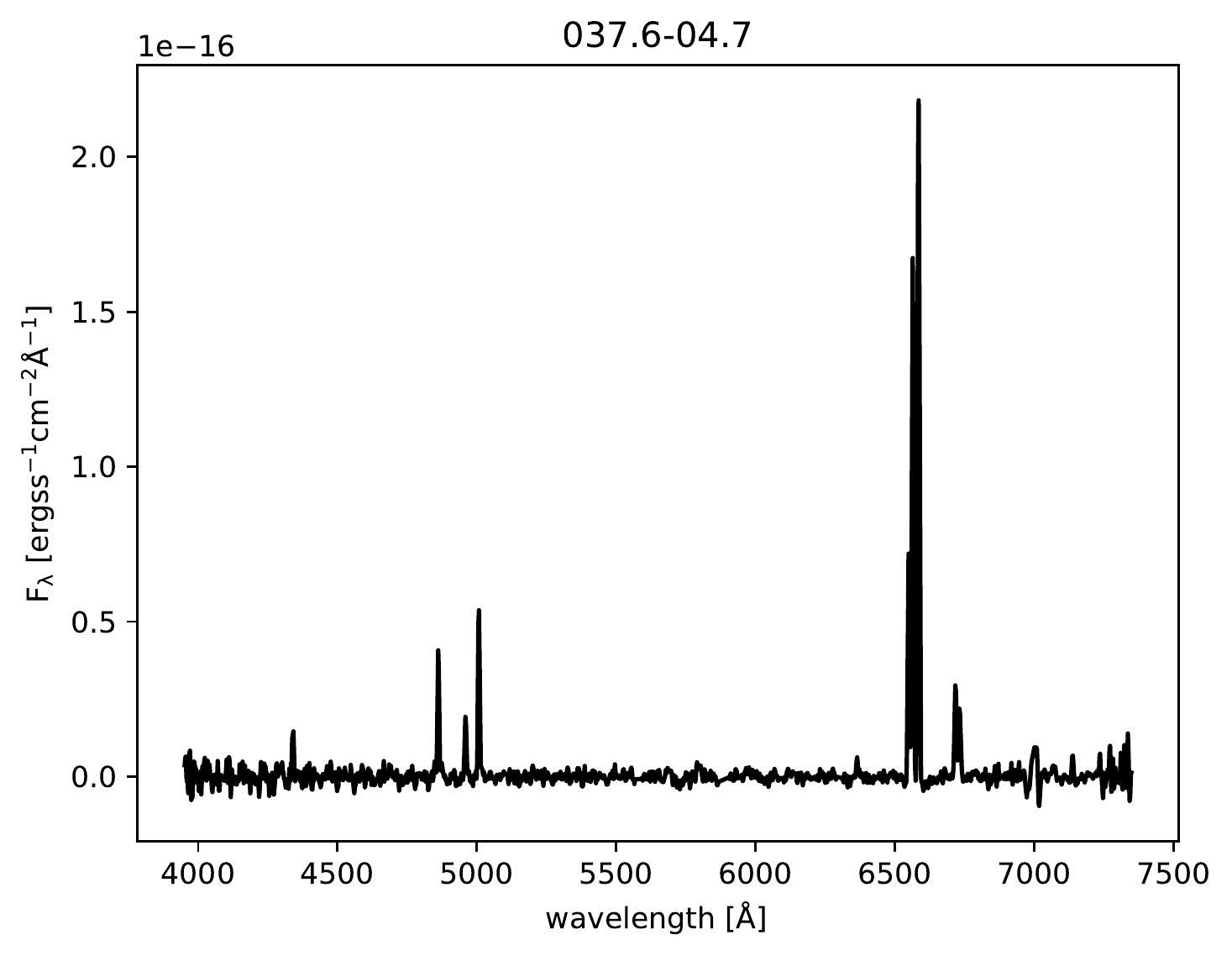}& \includegraphics[width=0.48\textwidth]{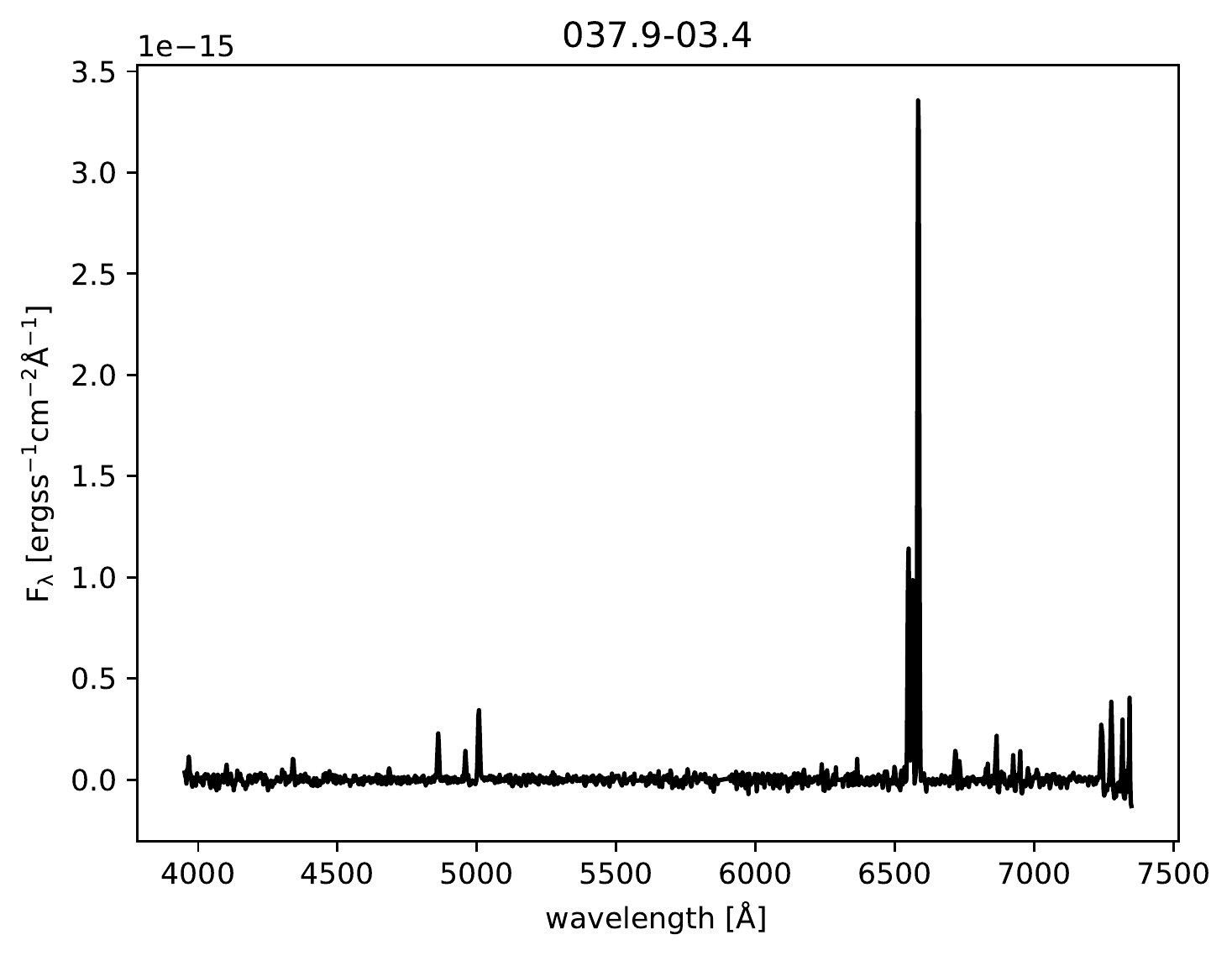}\\
\includegraphics[width=0.48\textwidth]{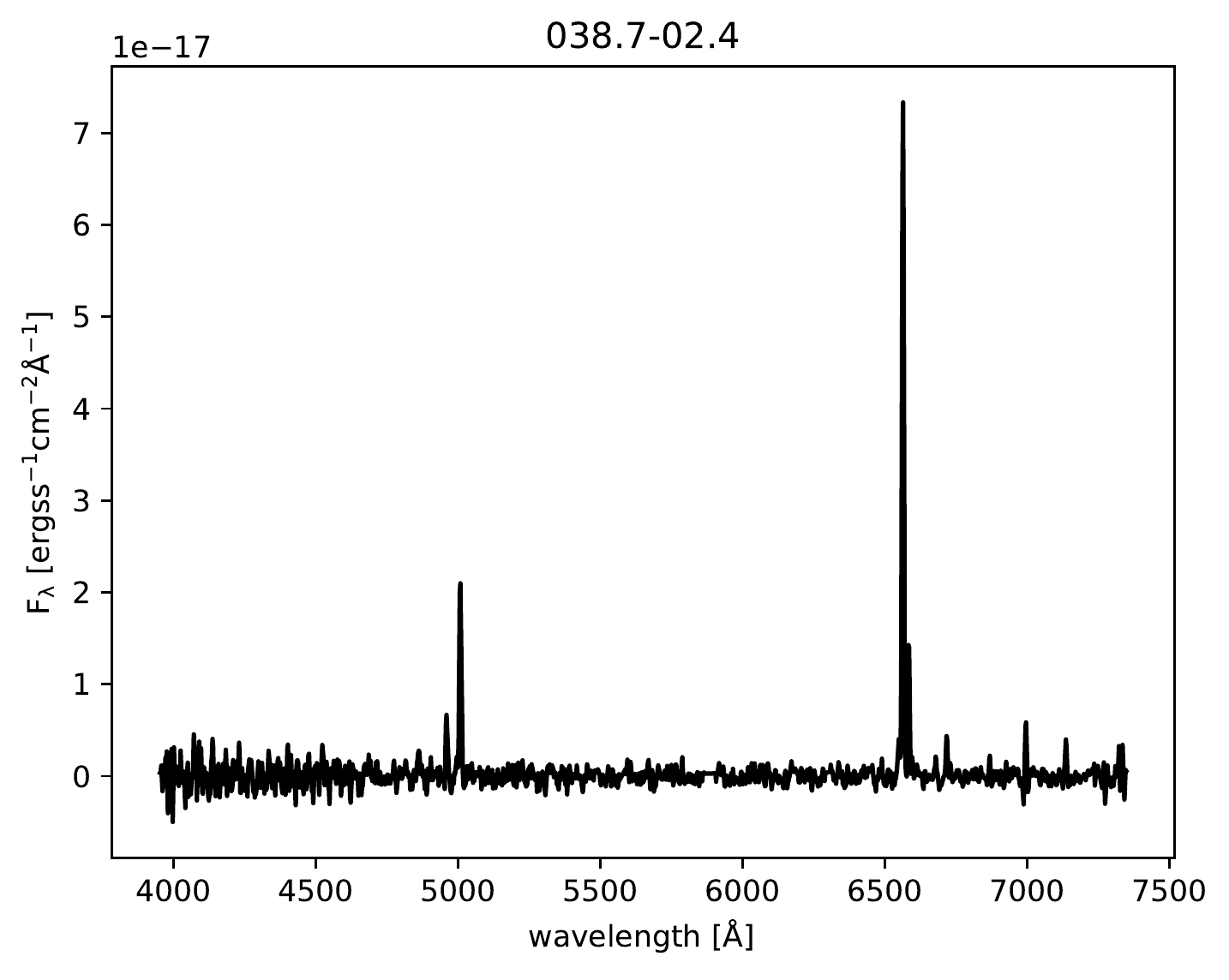}&\includegraphics[width=0.48\textwidth]{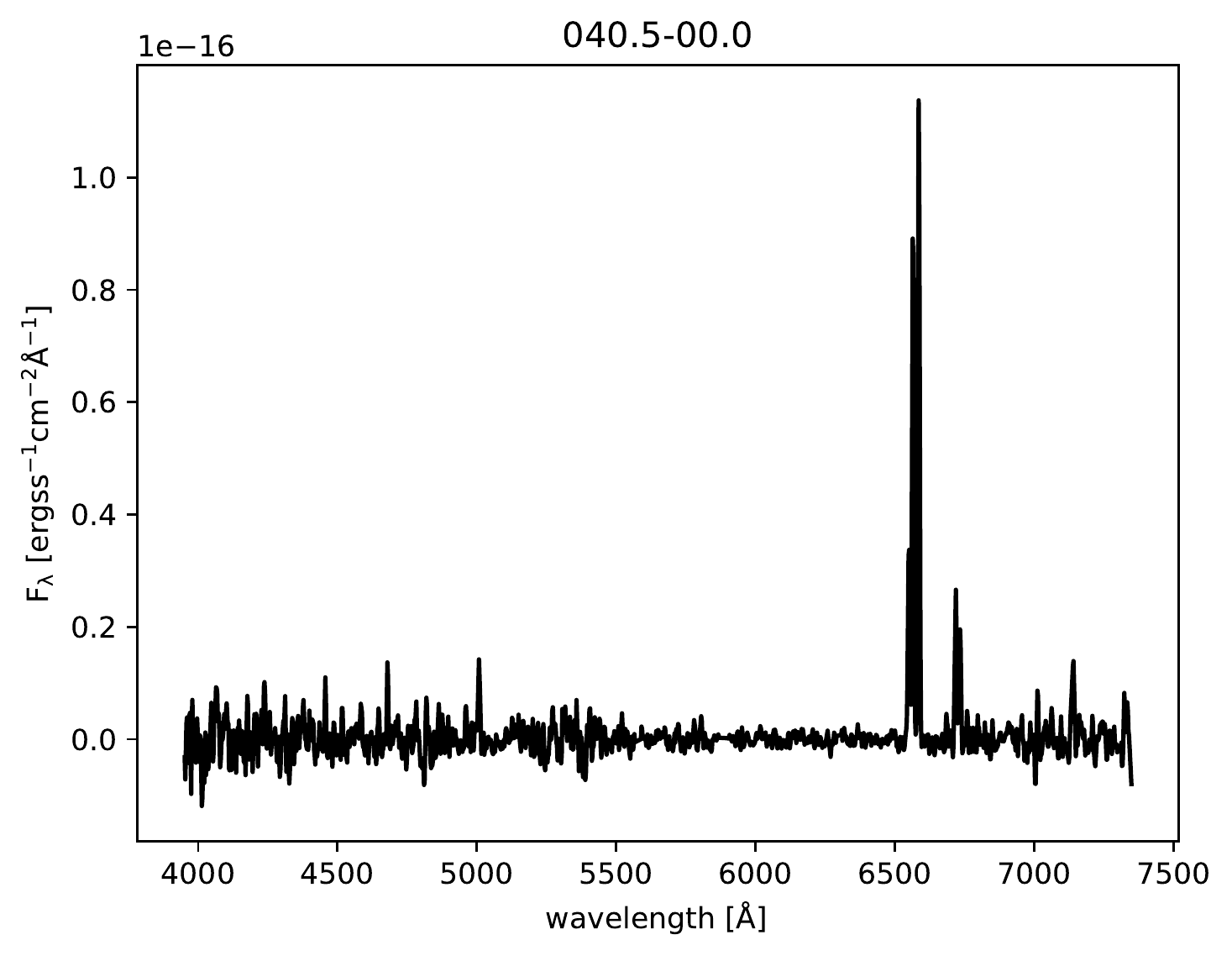}\\
\includegraphics[width=0.48\textwidth]{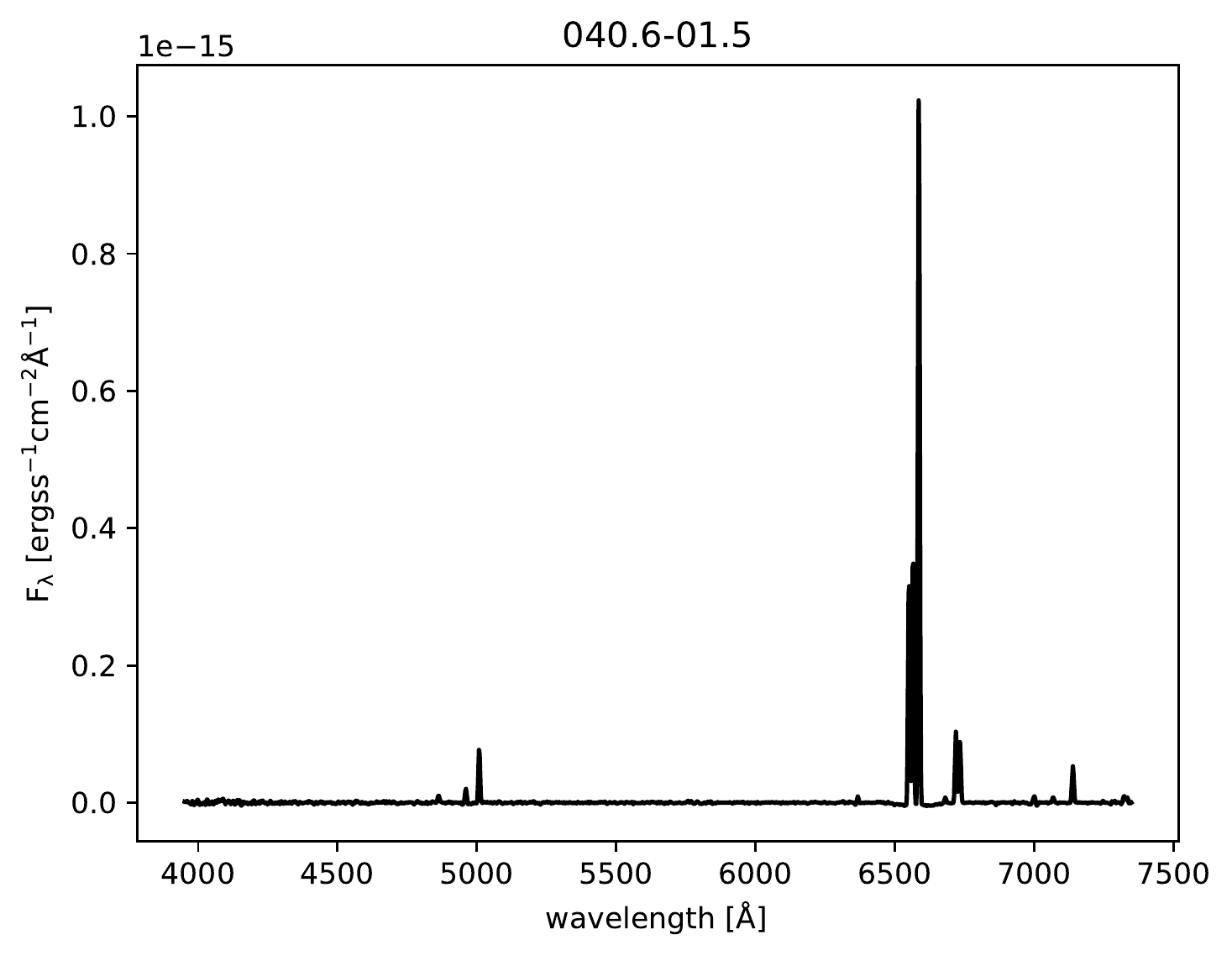}& \includegraphics[width=0.48\textwidth]{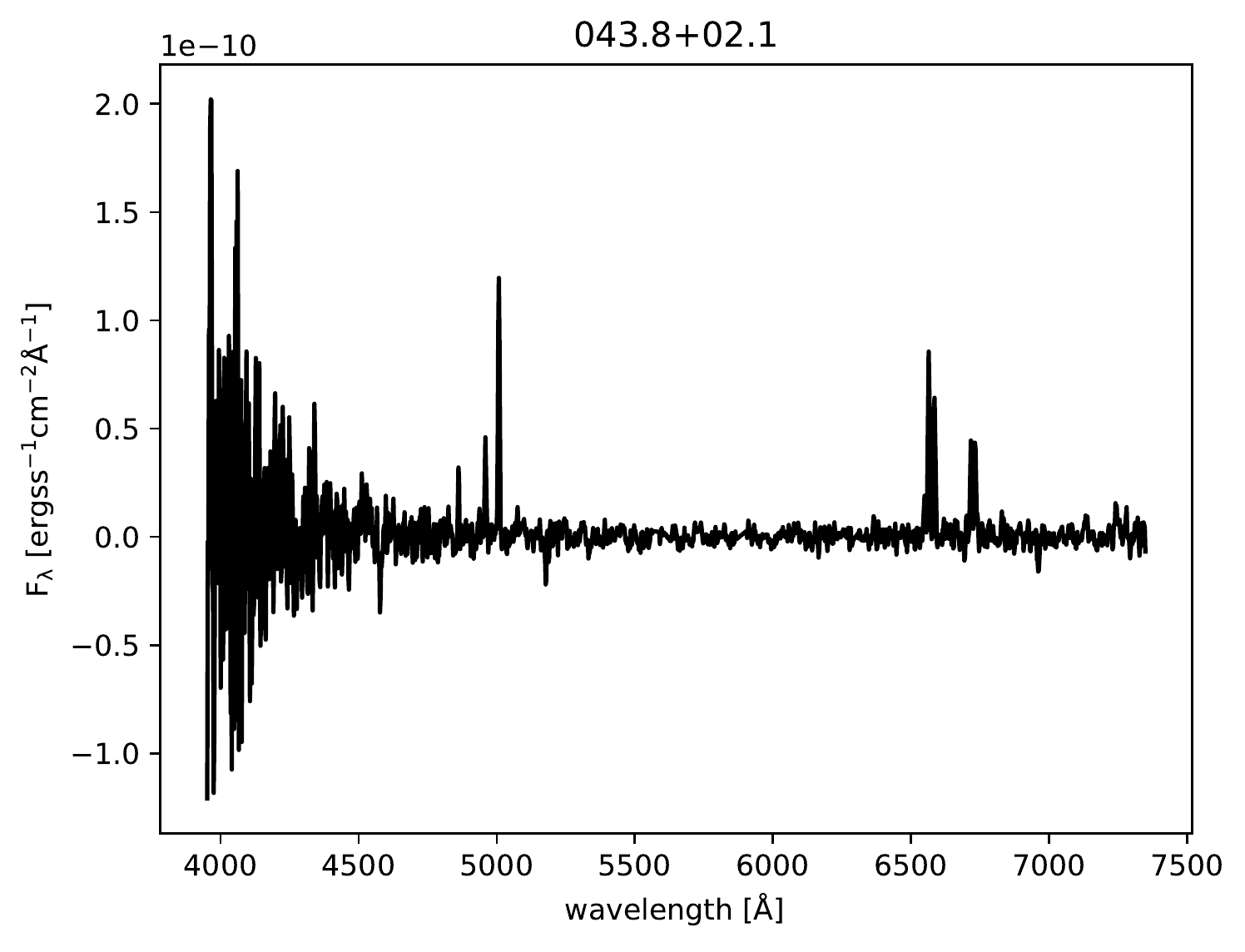}\\
 \includegraphics[width=0.48\textwidth]{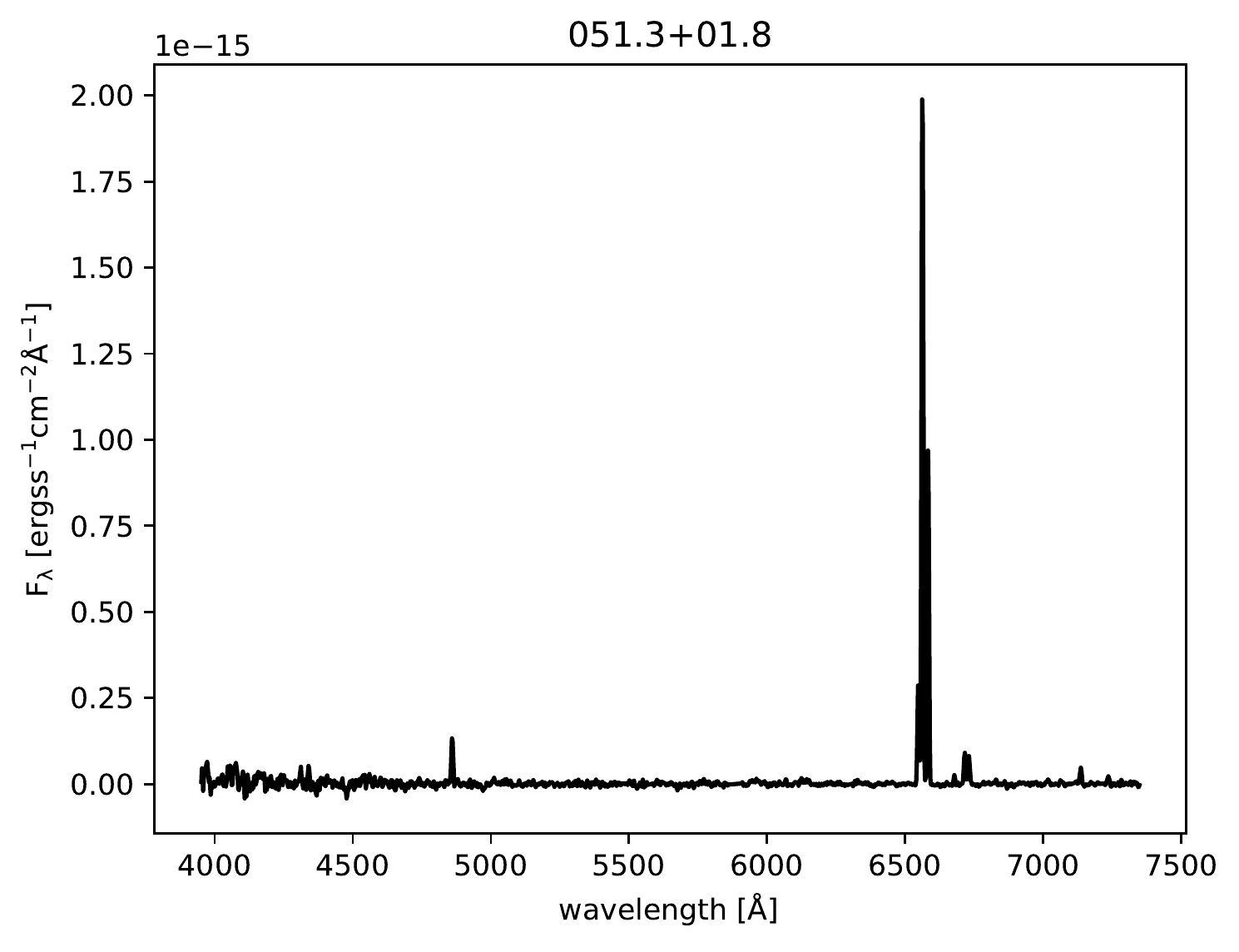}& \includegraphics[width=0.48\textwidth]{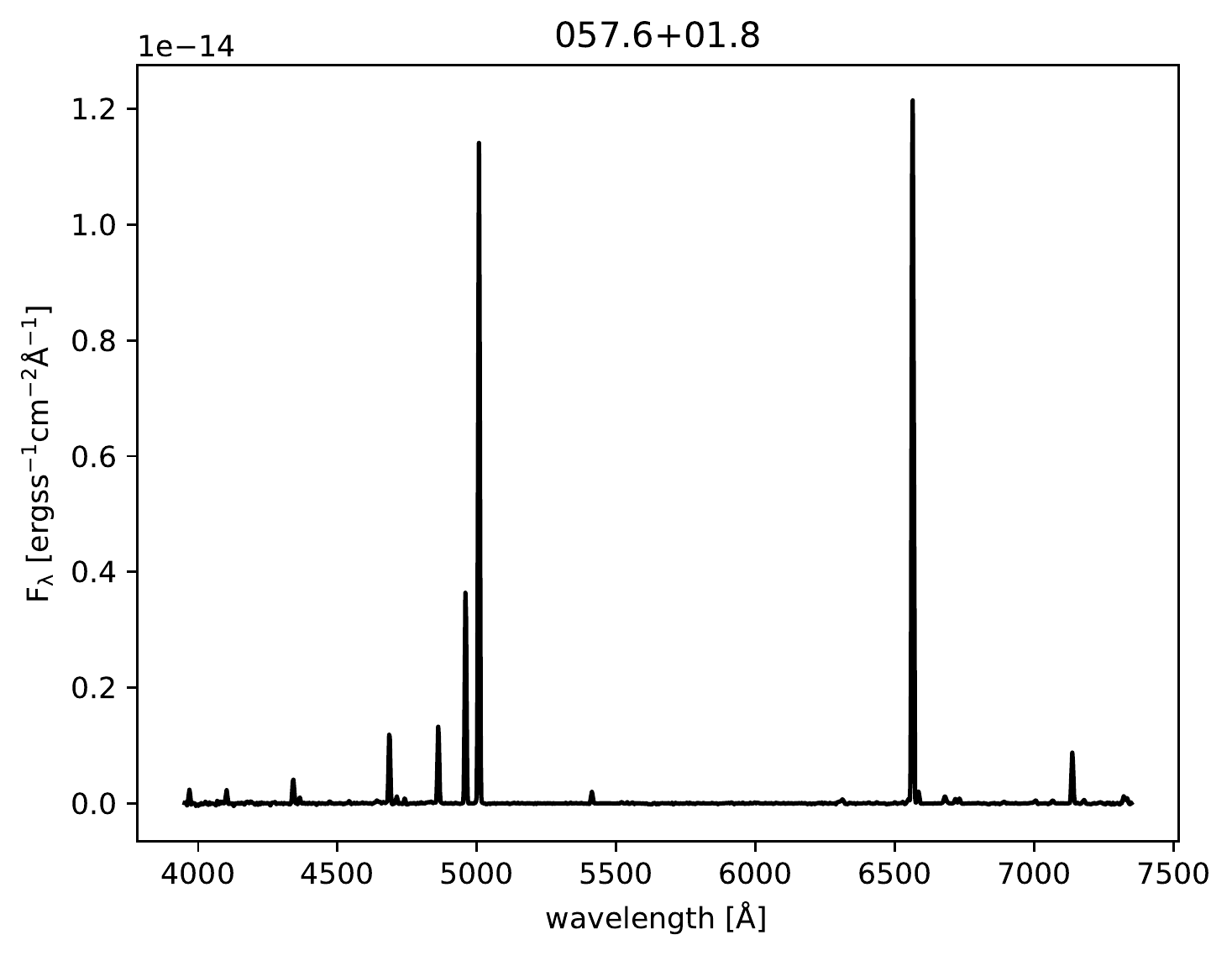}\\
 \includegraphics[width=0.48\textwidth]{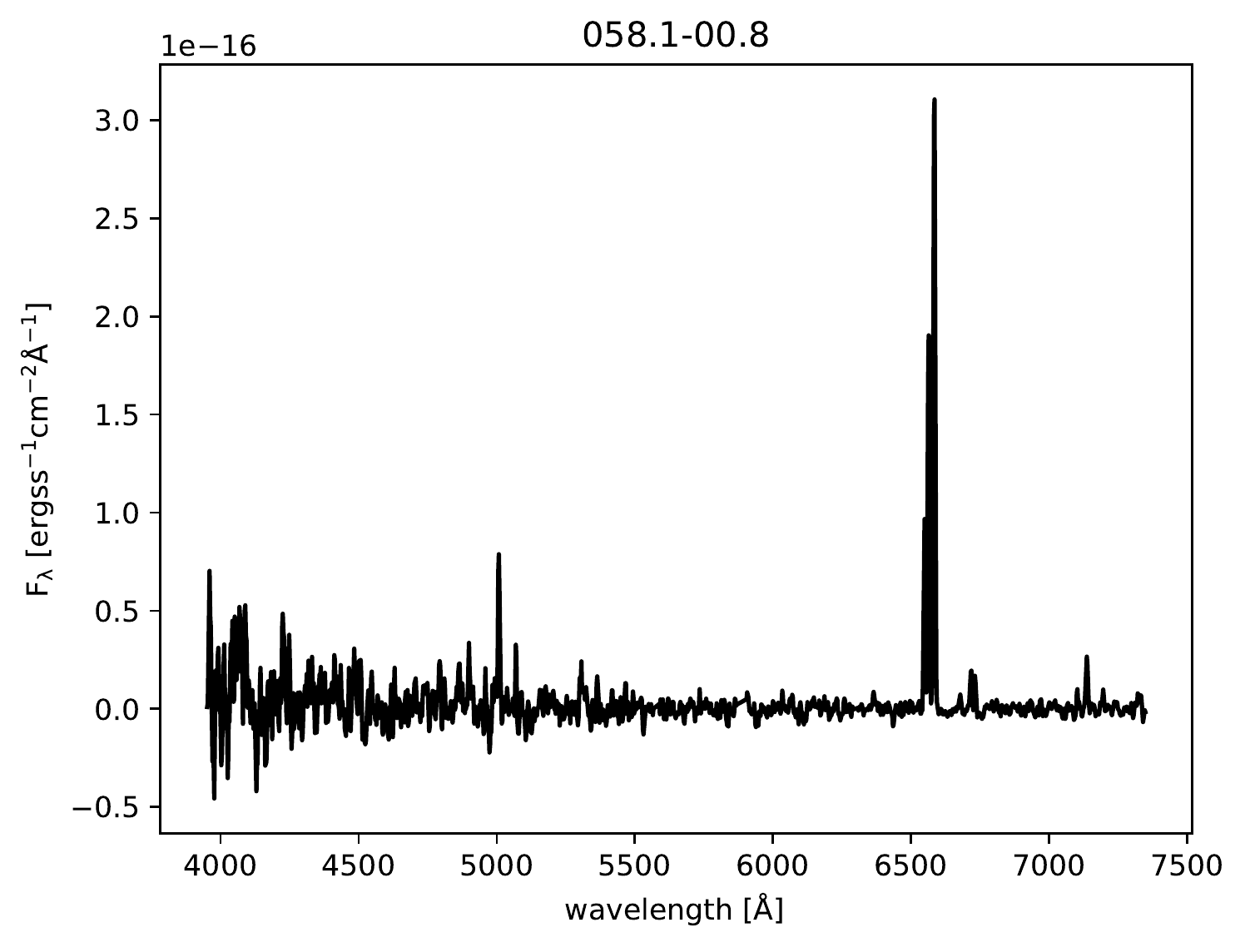}&\includegraphics[width=0.48\textwidth]{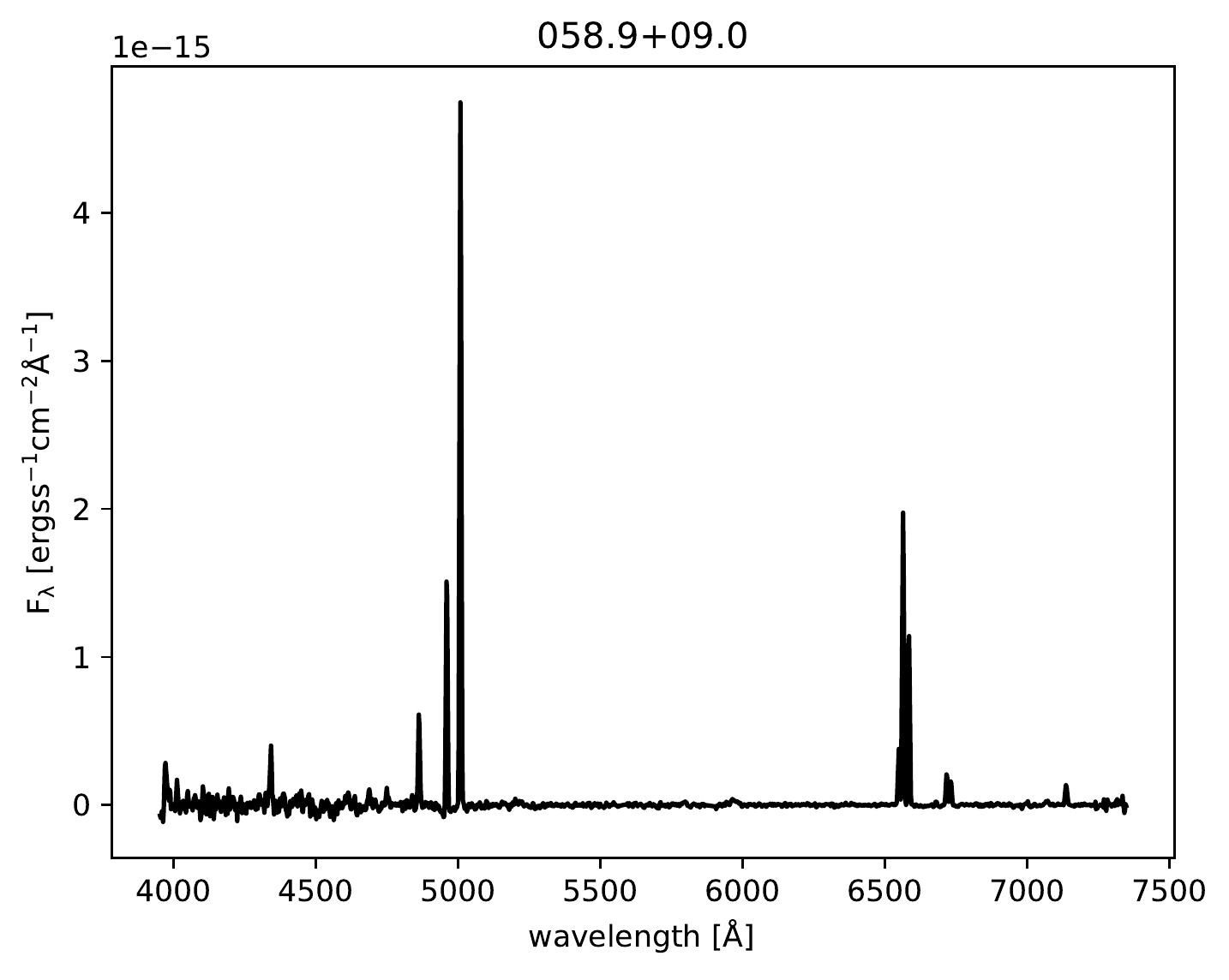}\\
 \includegraphics[width=0.48\textwidth]{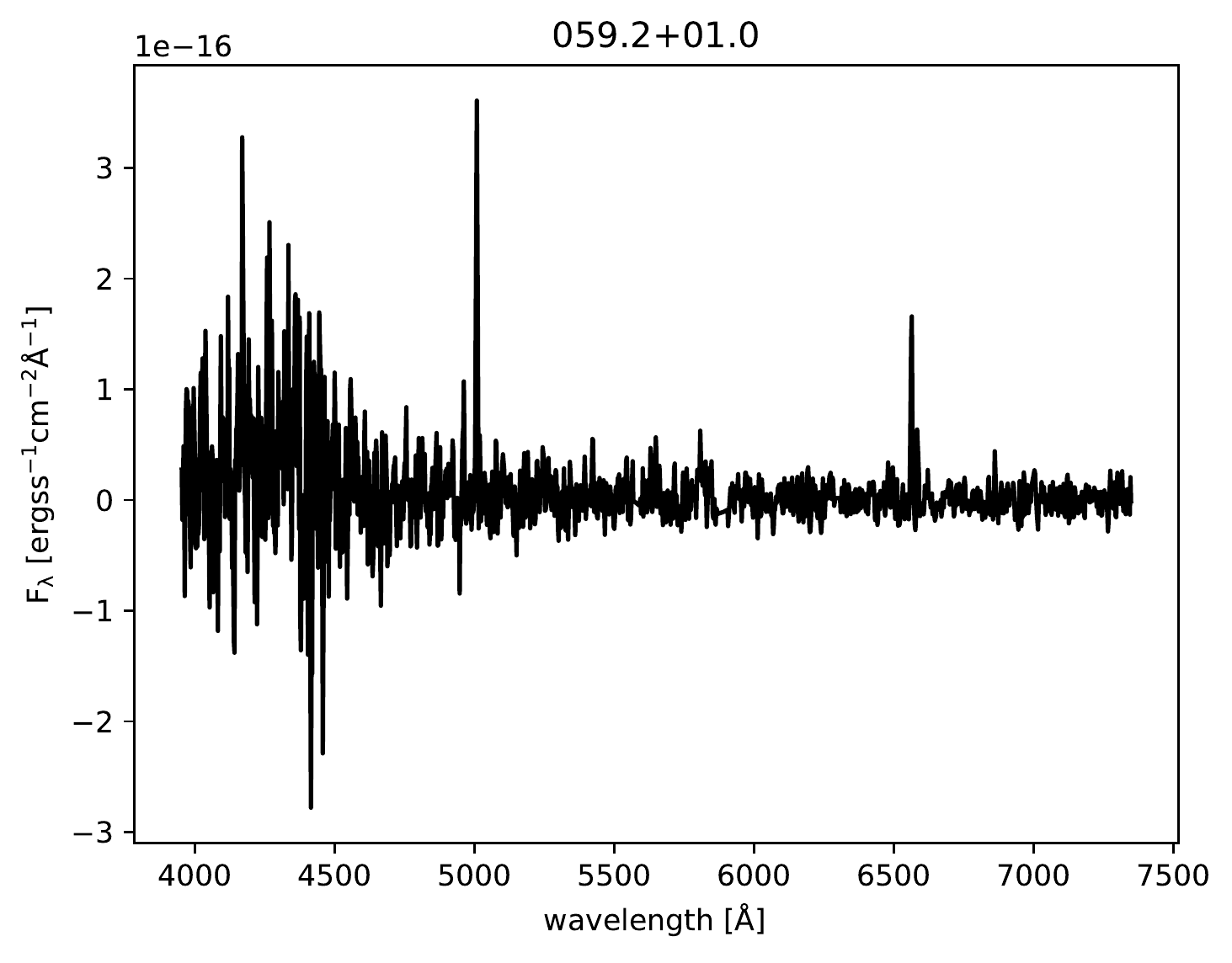}& \includegraphics[width=0.48\textwidth]{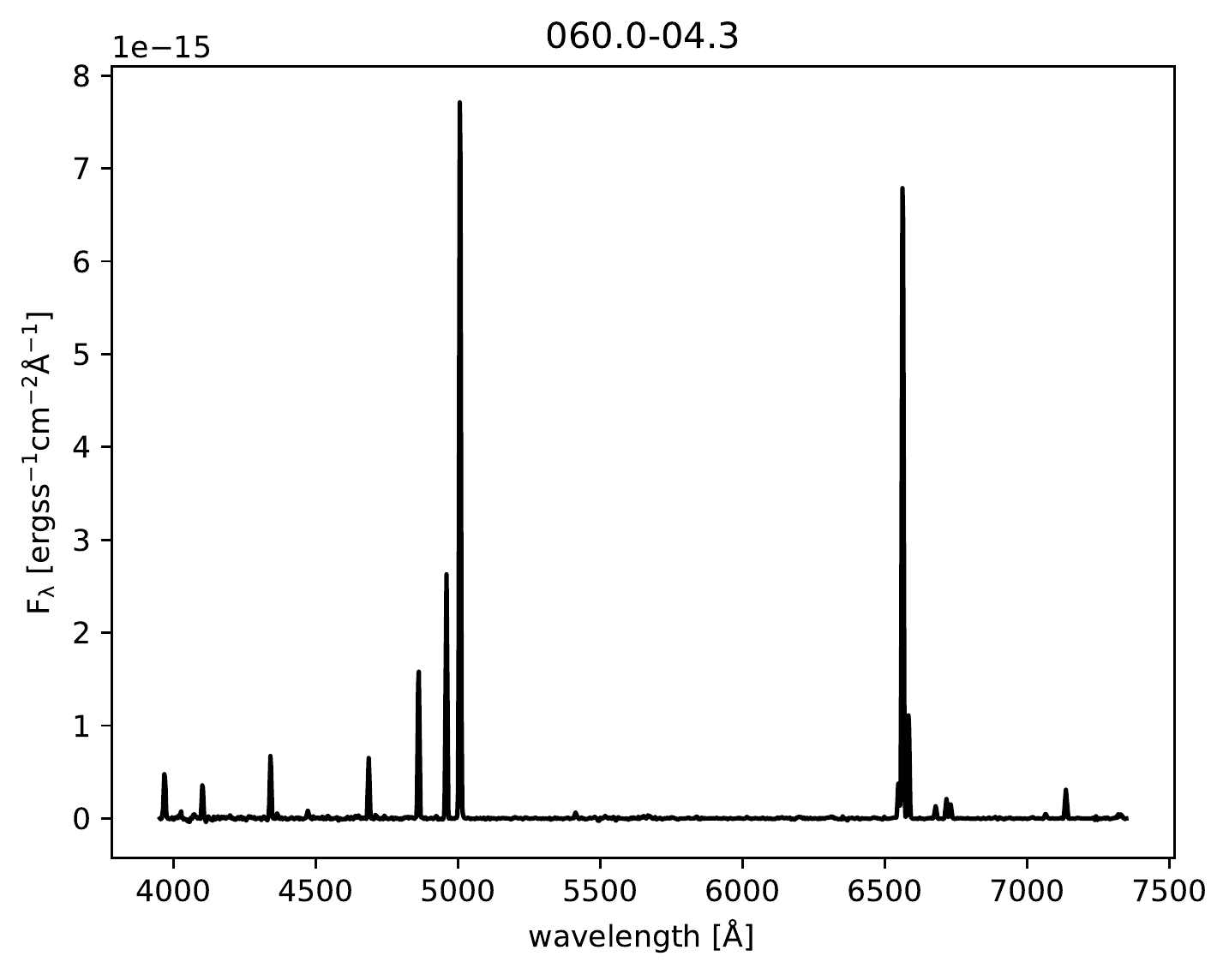}\\
 \includegraphics[width=0.48\textwidth]{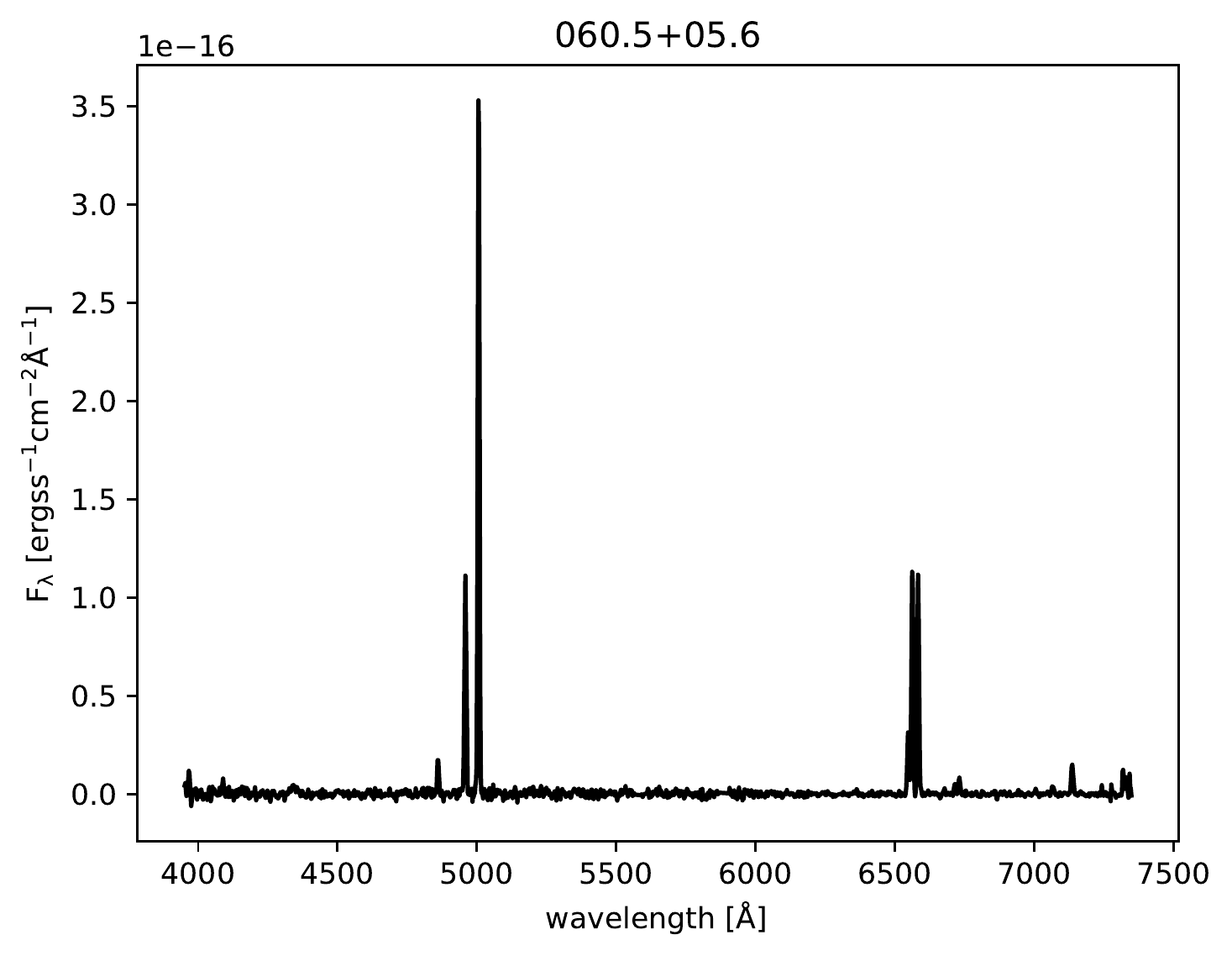}& \includegraphics[width=0.48\textwidth]{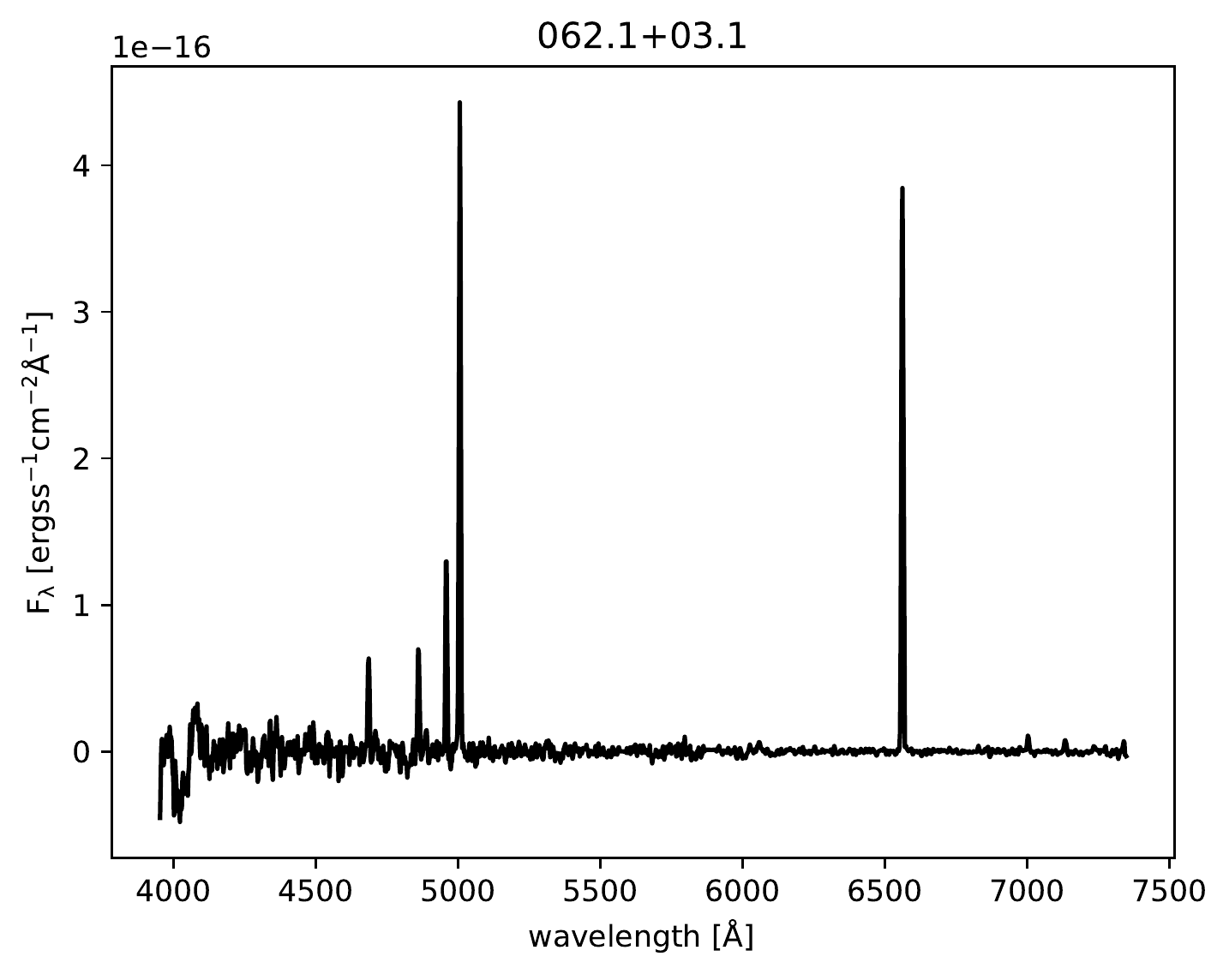}\\
 \includegraphics[width=0.48\textwidth]{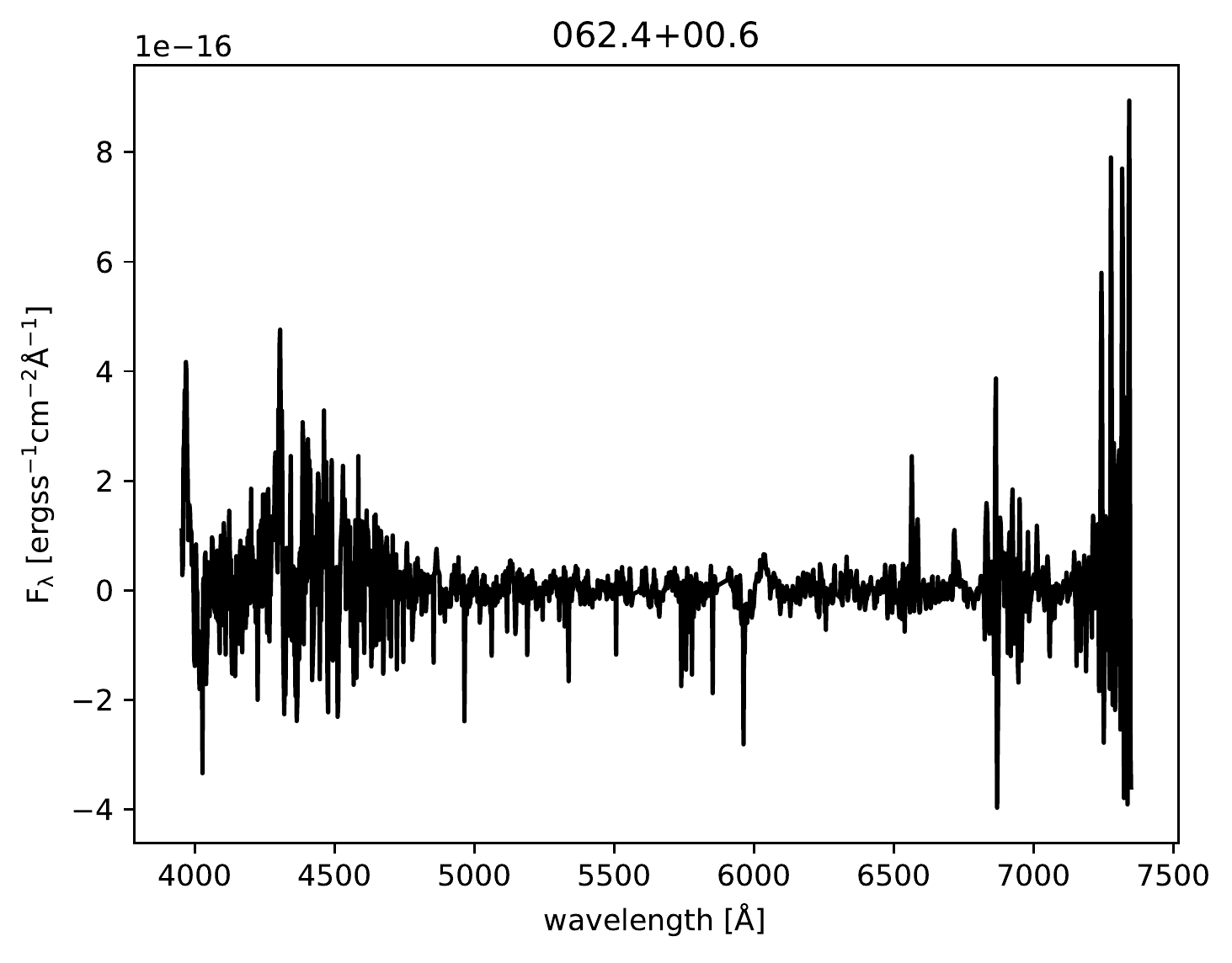}& \includegraphics[width=0.48\textwidth]{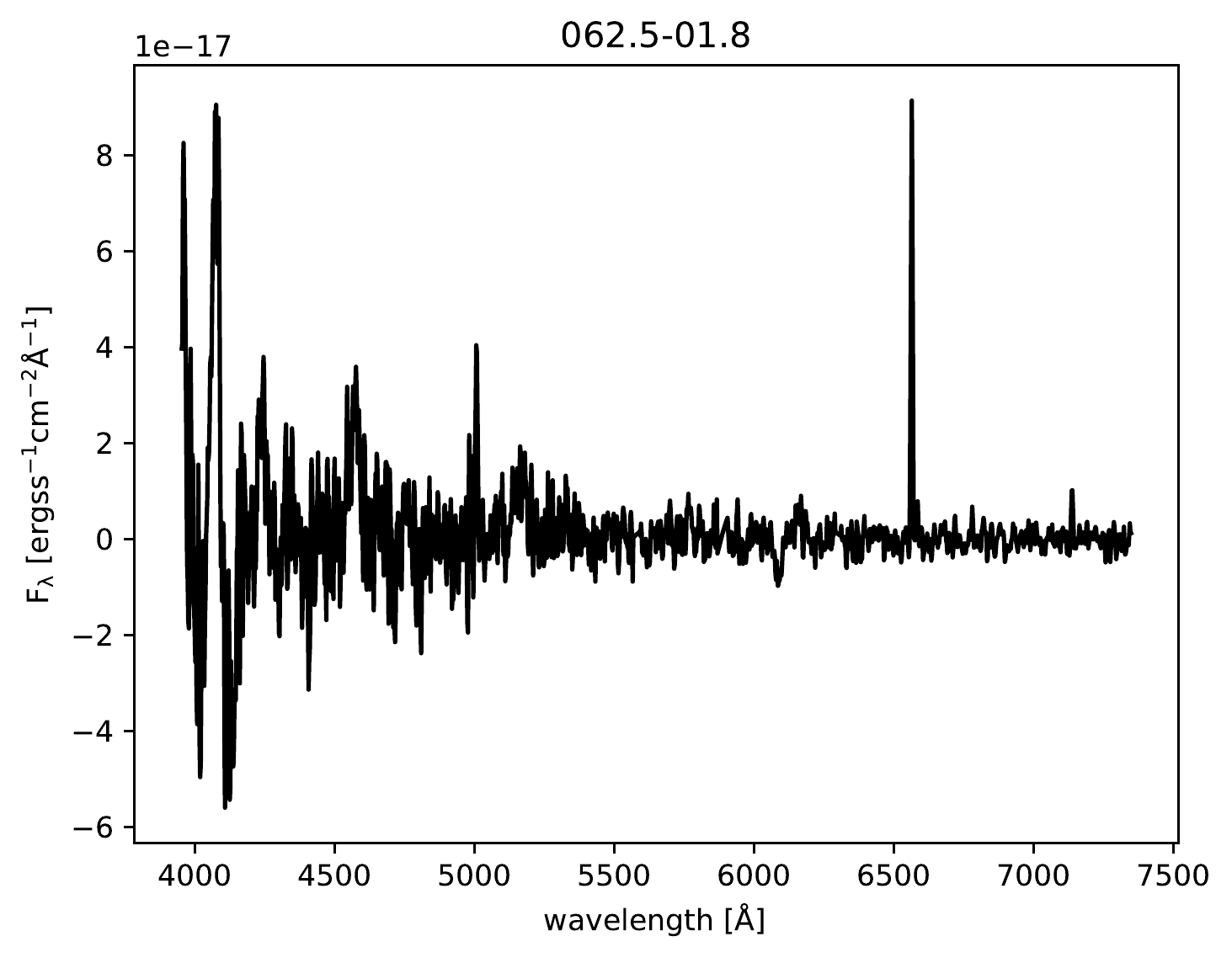}\\
 \includegraphics[width=0.48\textwidth]{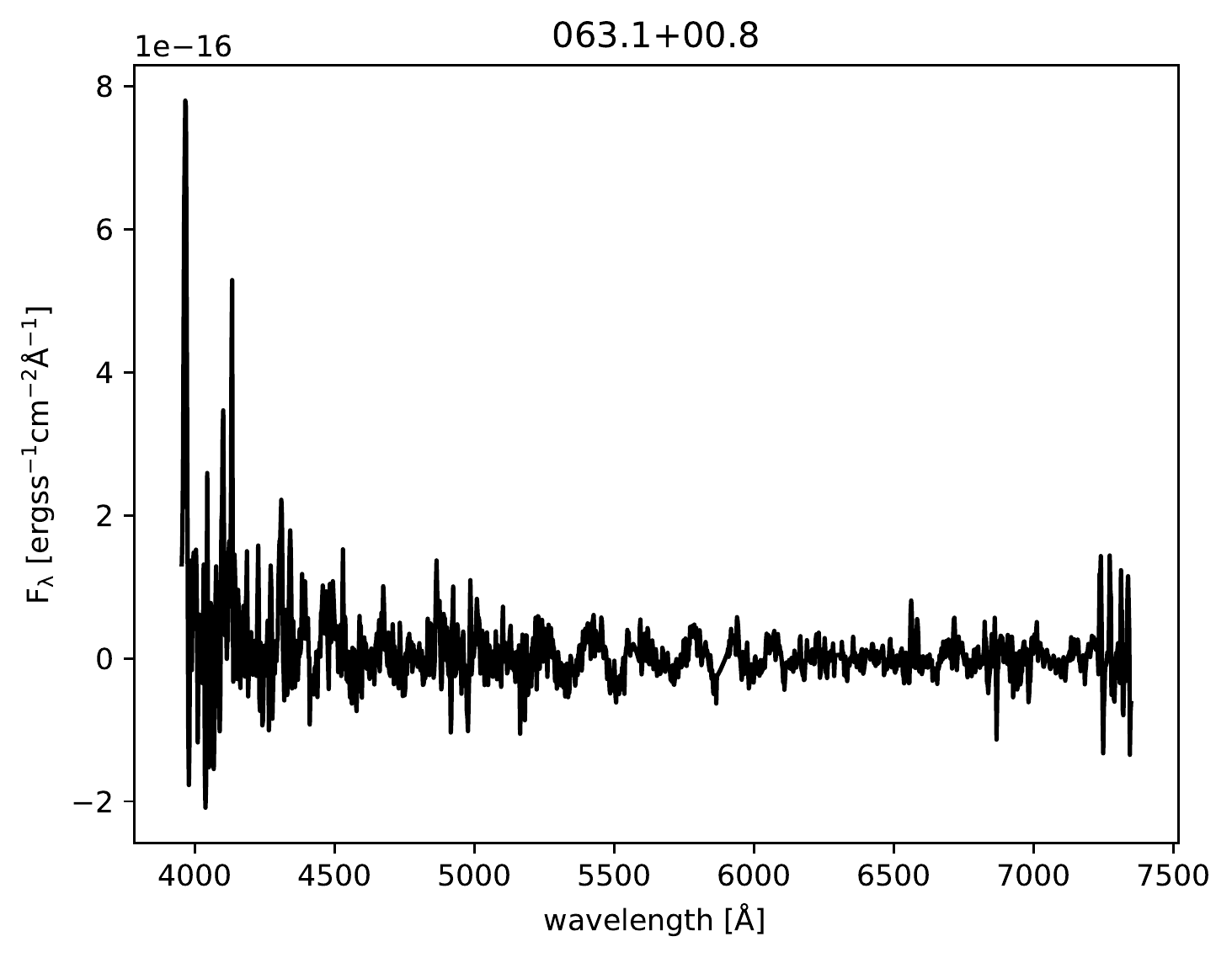}&\includegraphics[width=0.48\textwidth]{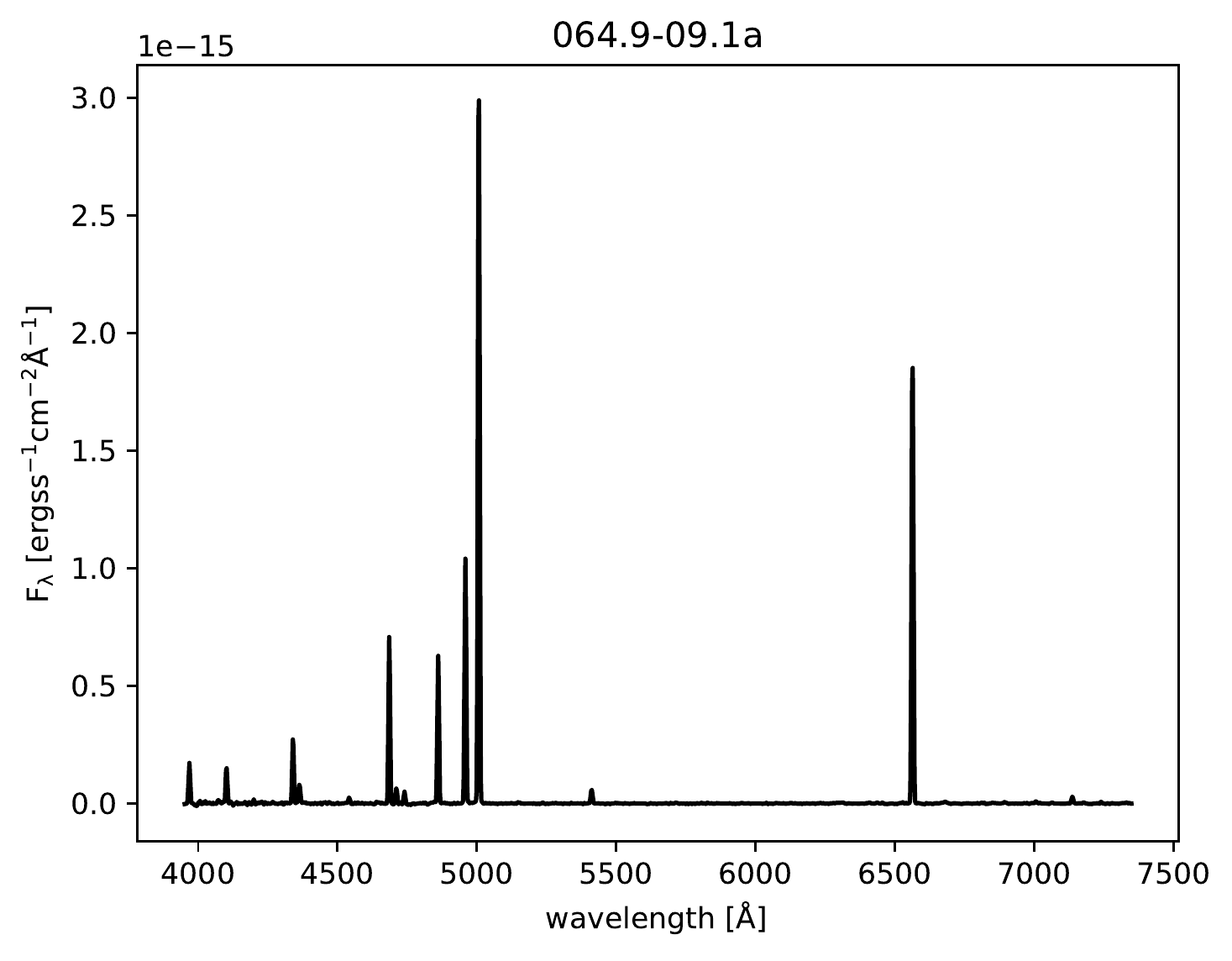}\\
 \includegraphics[width=0.48\textwidth]{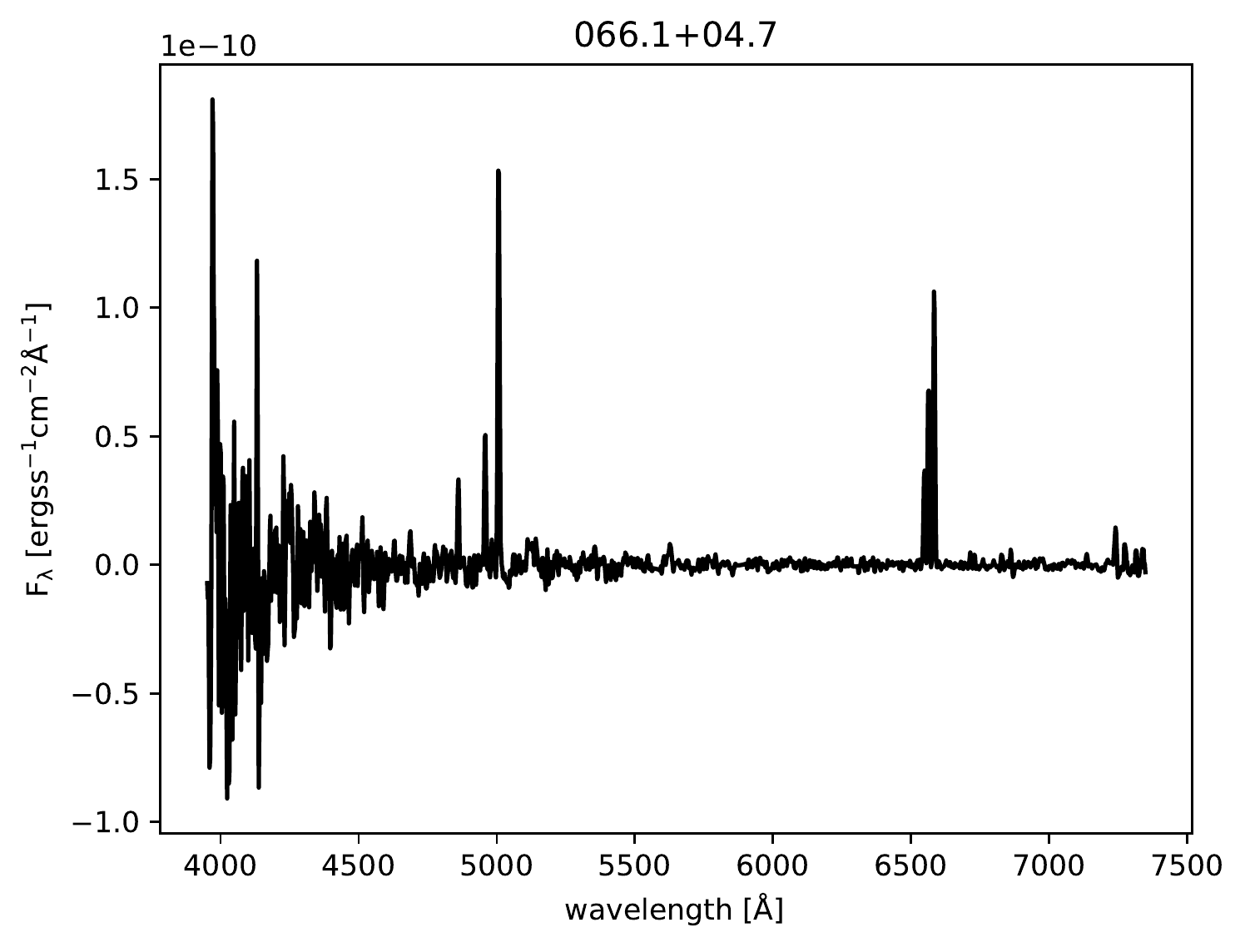}&\includegraphics[width=0.48\textwidth]{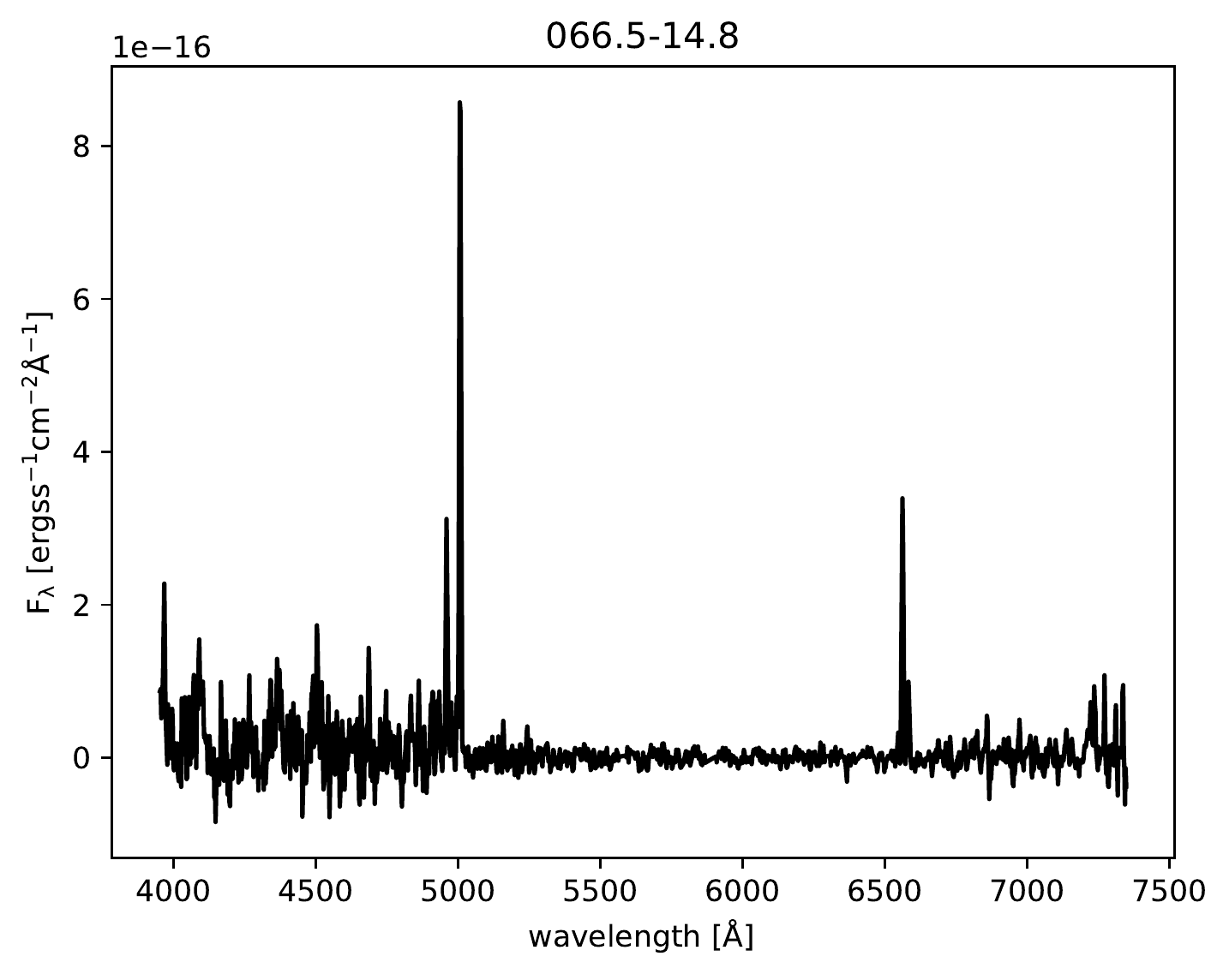}\\
\includegraphics[width=0.48\textwidth]{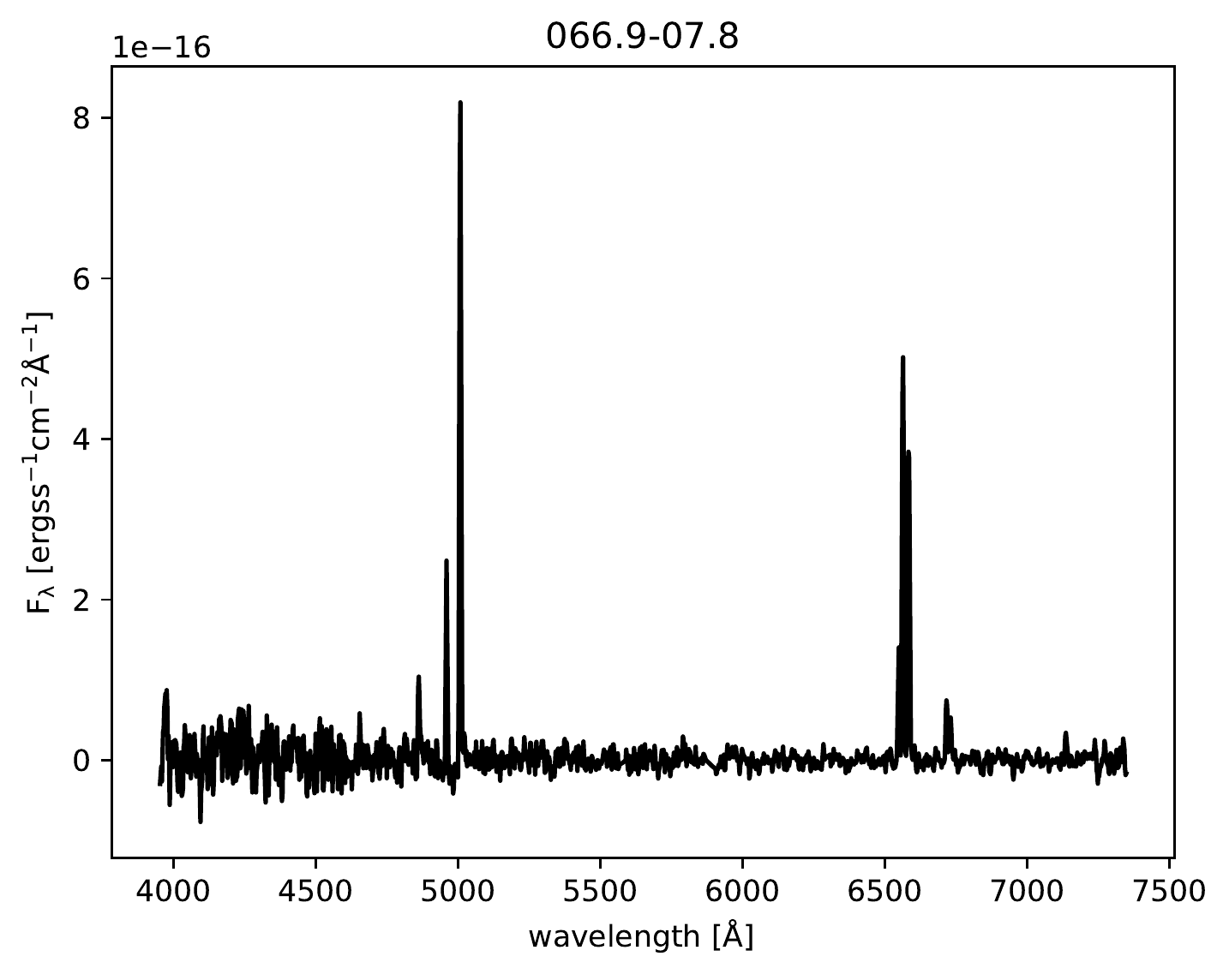}& \includegraphics[width=0.48\textwidth]{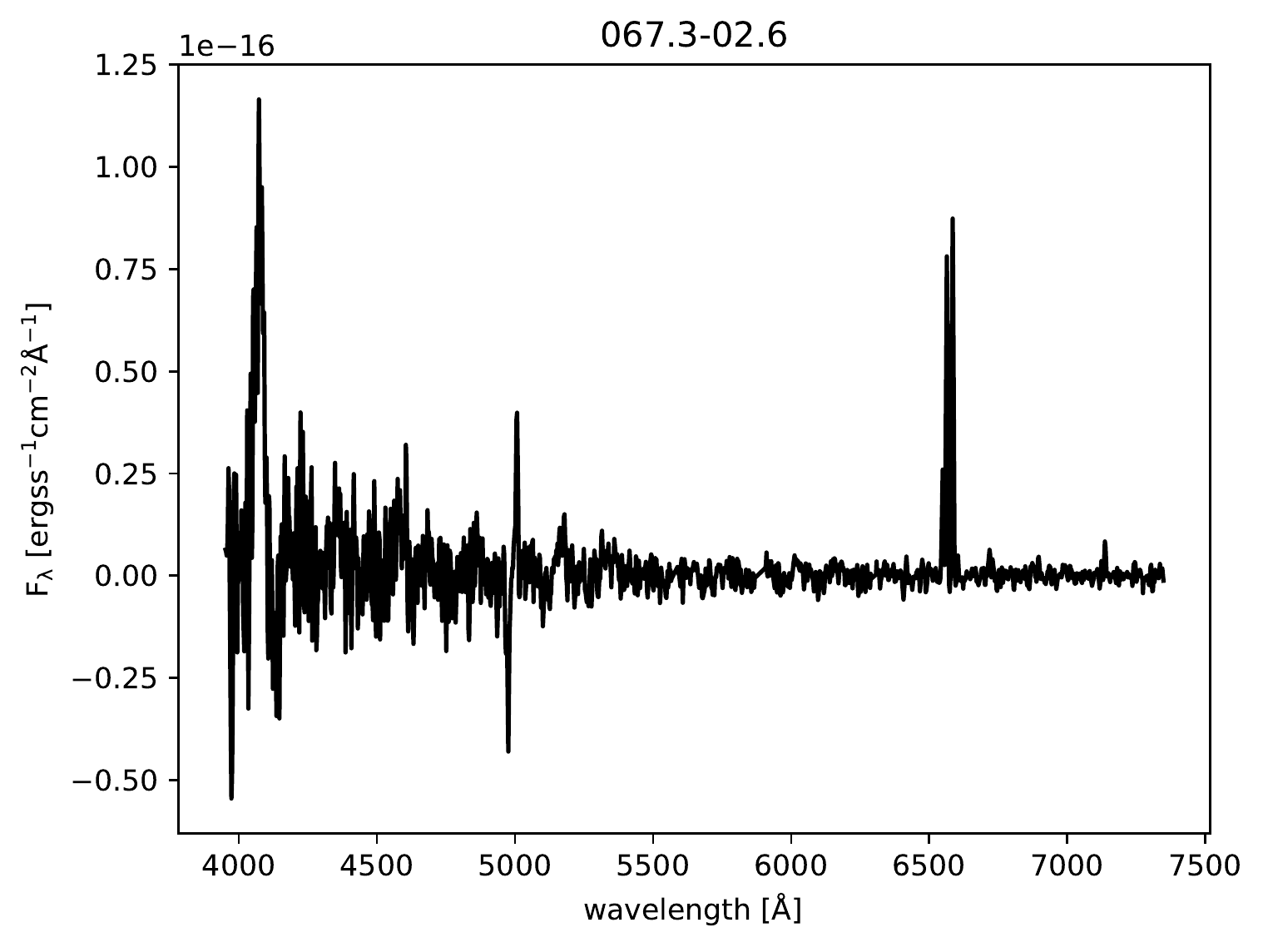}\\
\includegraphics[width=0.48\textwidth]{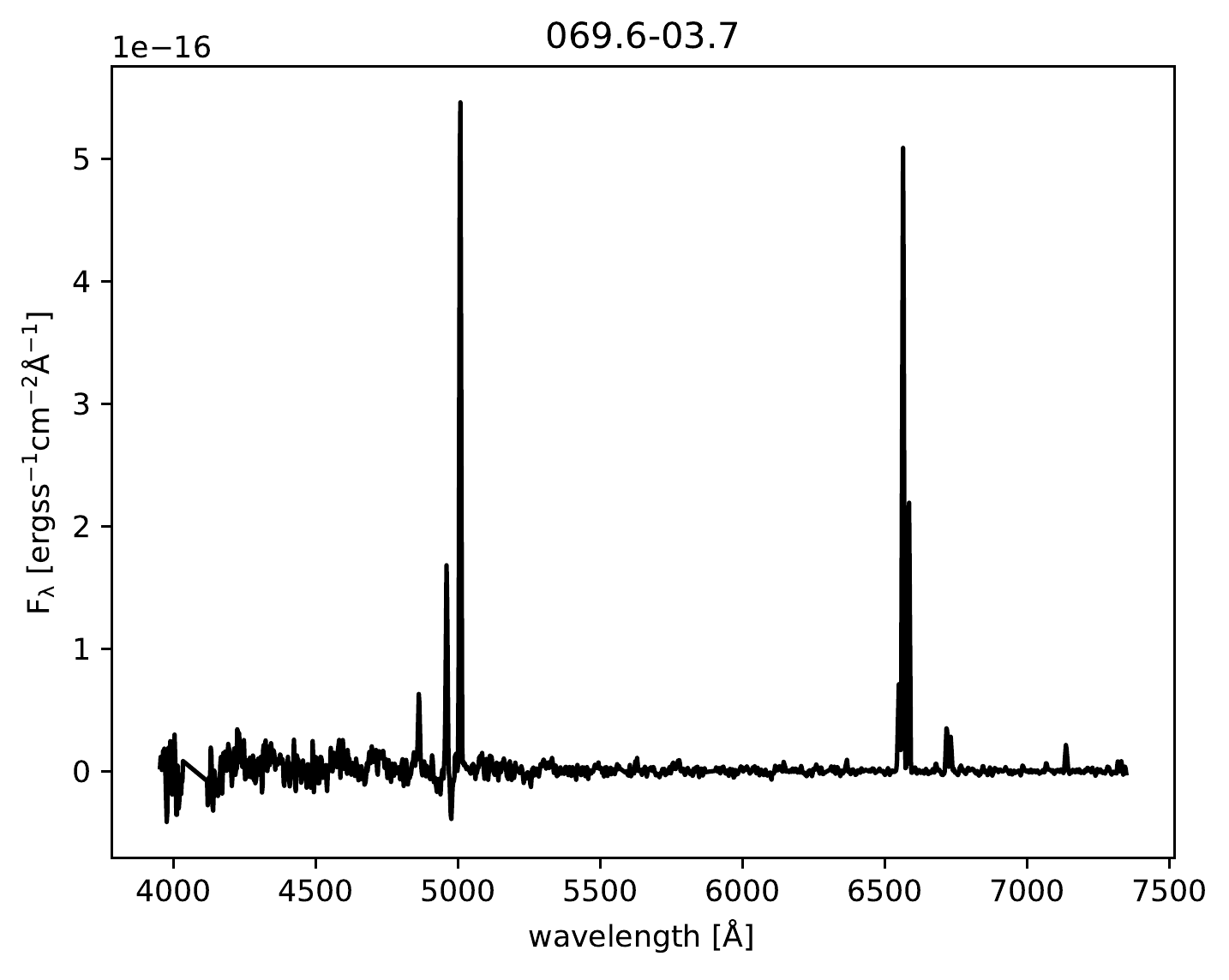}& \includegraphics[width=0.48\textwidth]{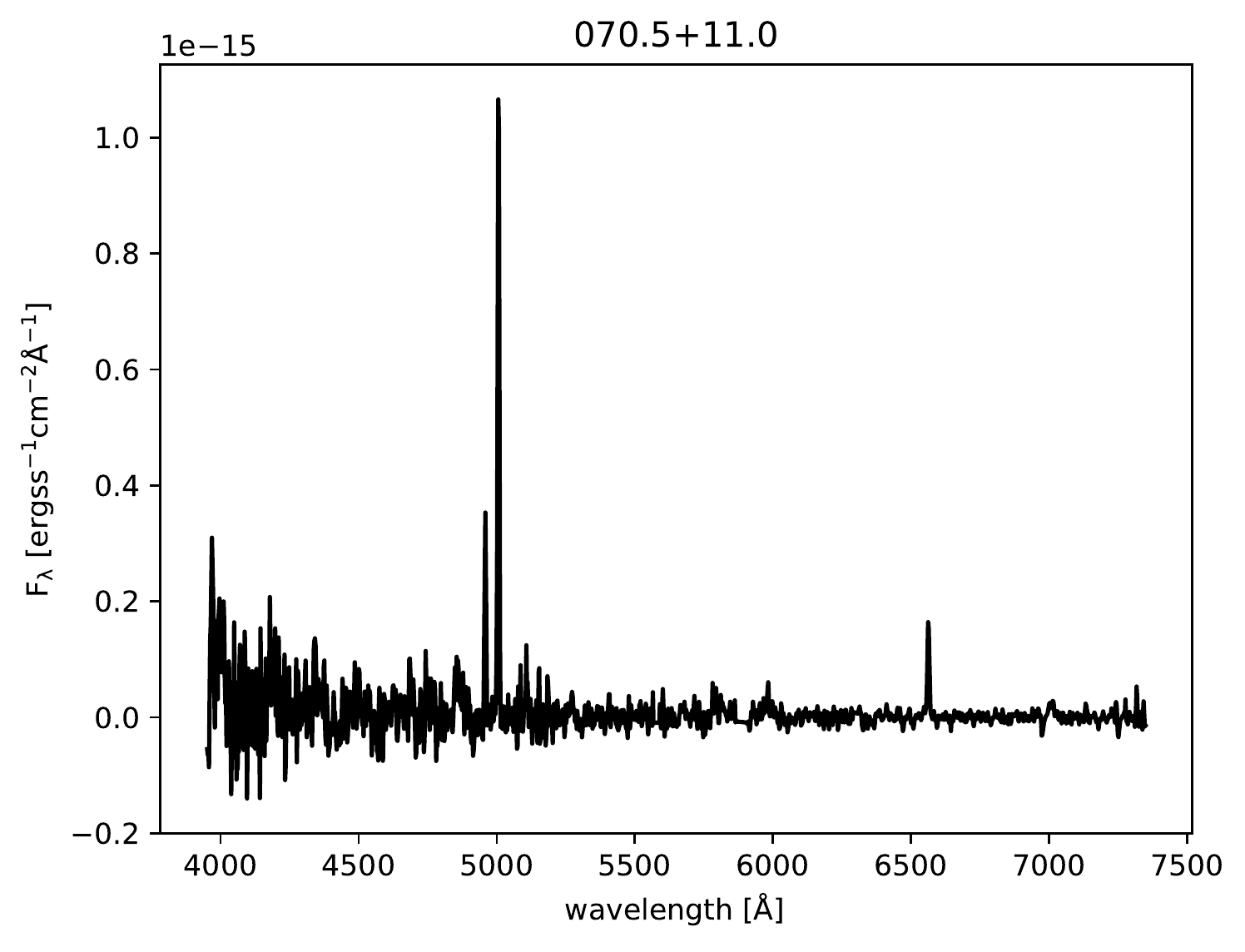}\\
 \includegraphics[width=0.48\textwidth]{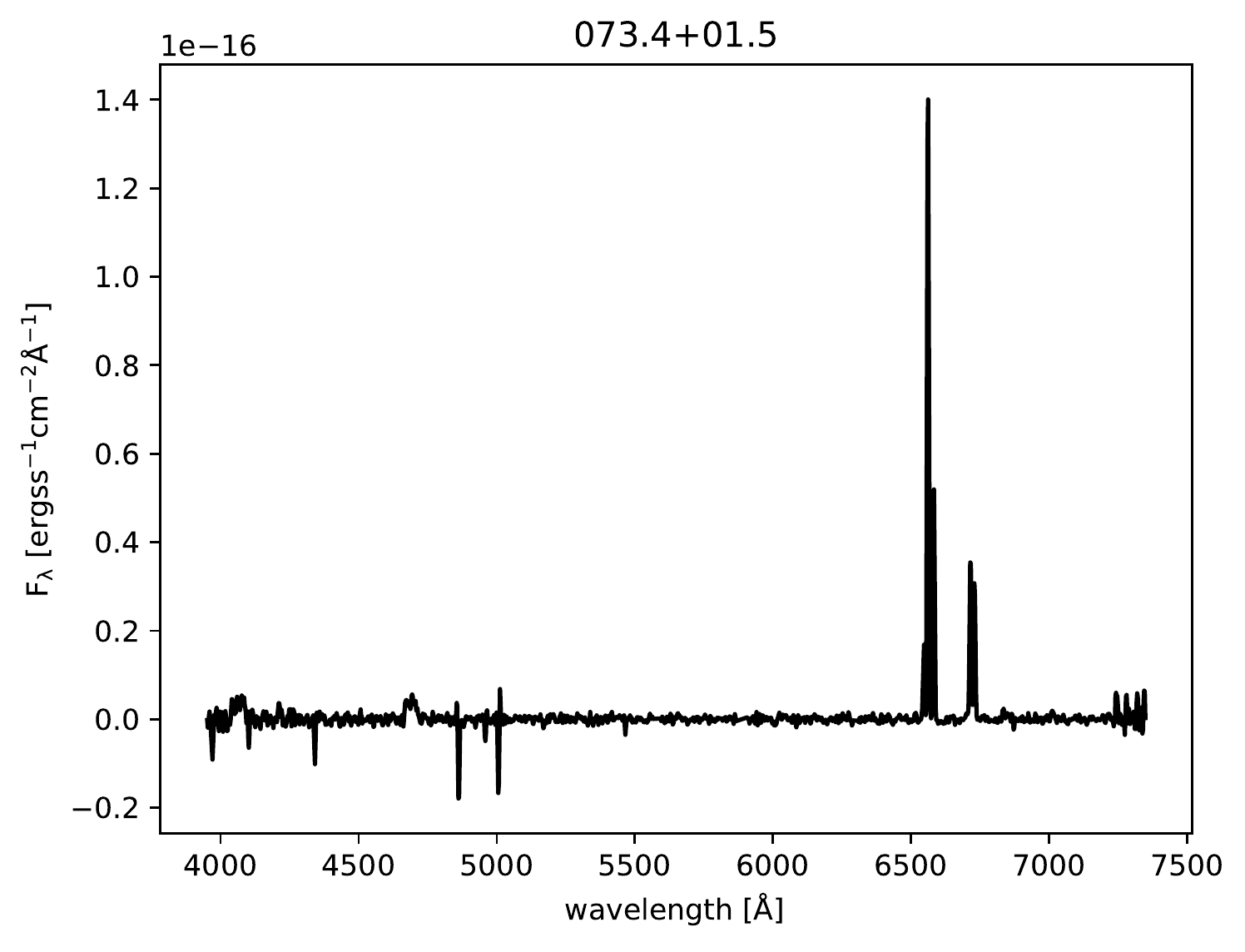}&\includegraphics[width=0.48\textwidth]{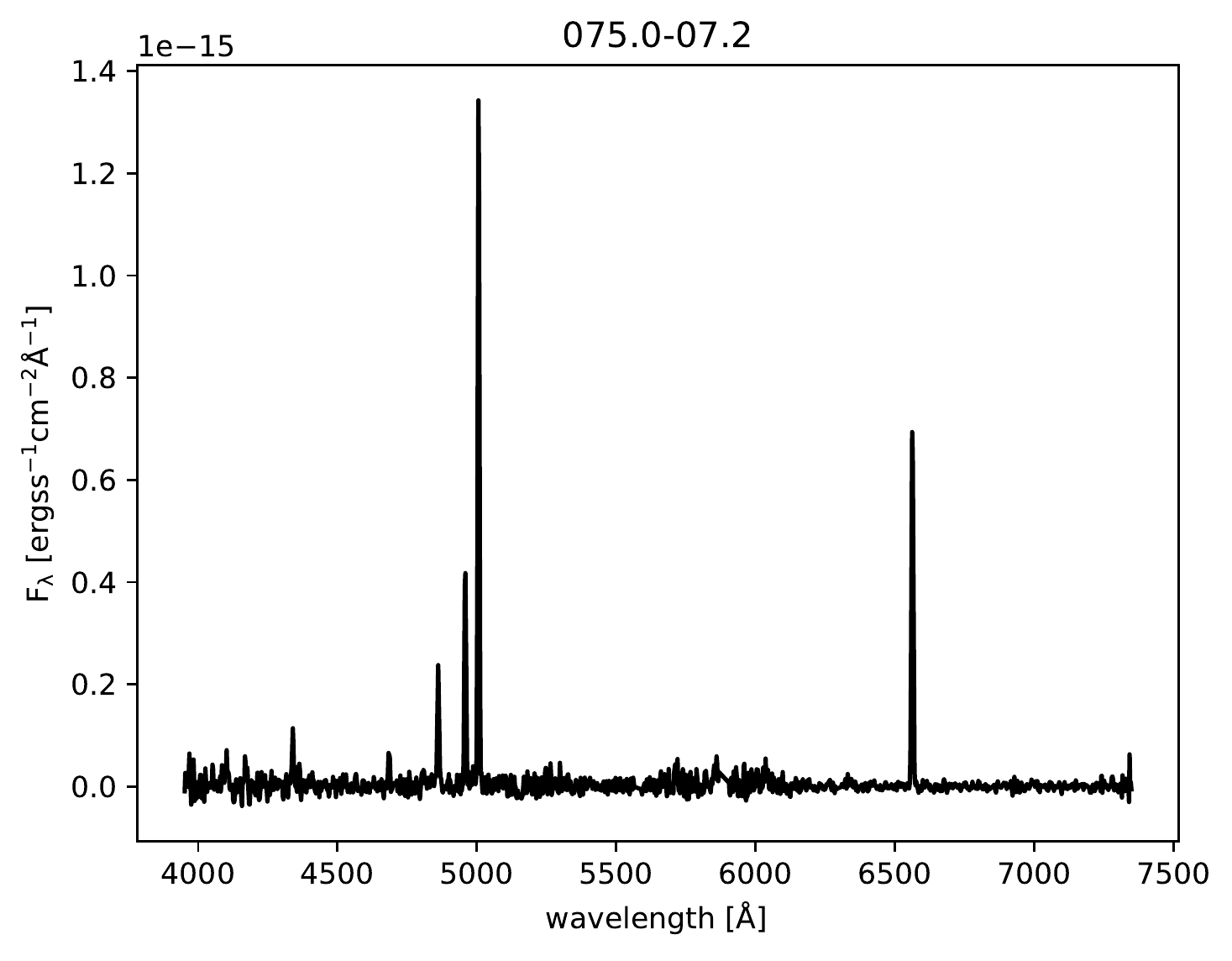}\\
 \includegraphics[width=0.48\textwidth]{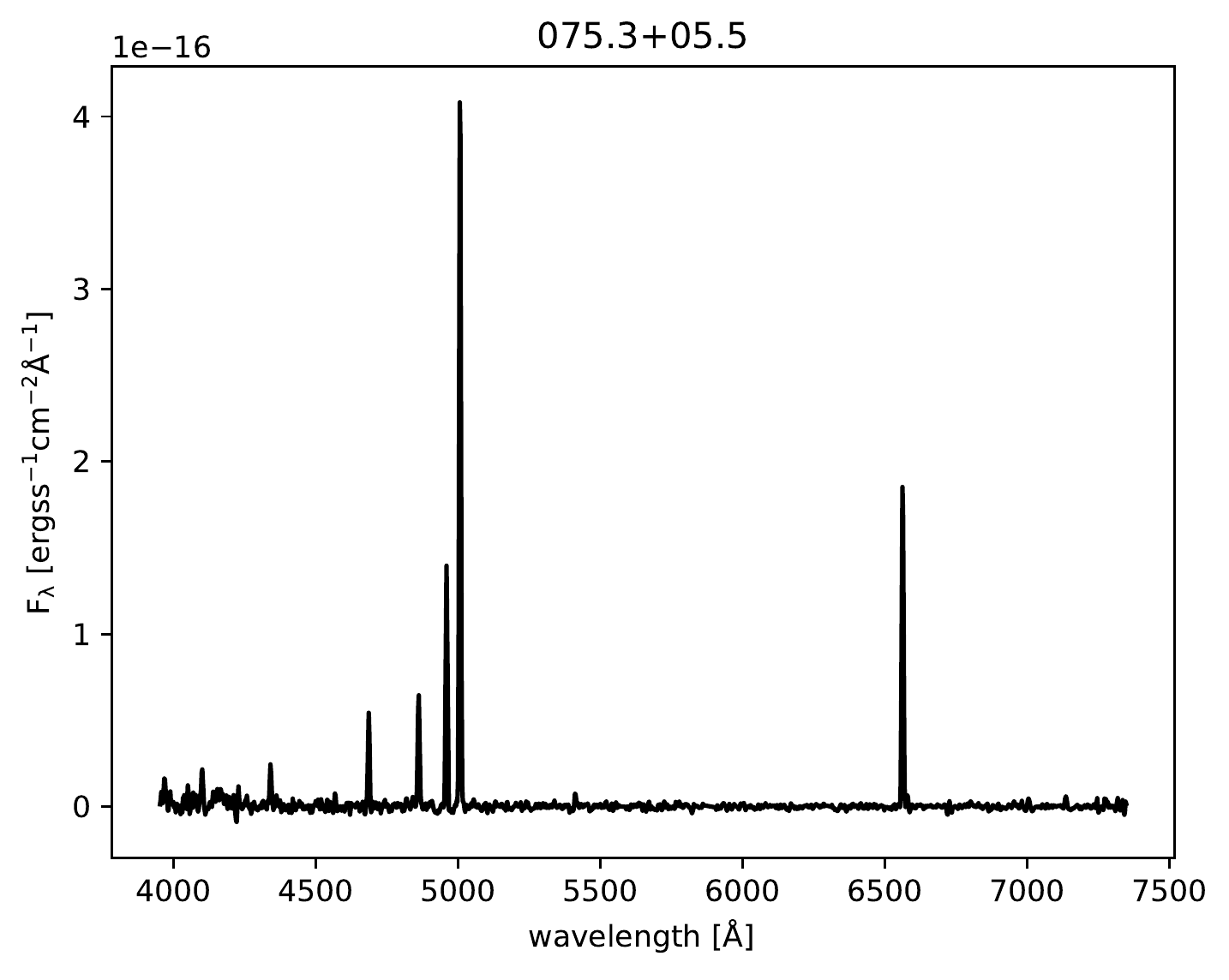}& \includegraphics[width=0.48\textwidth]{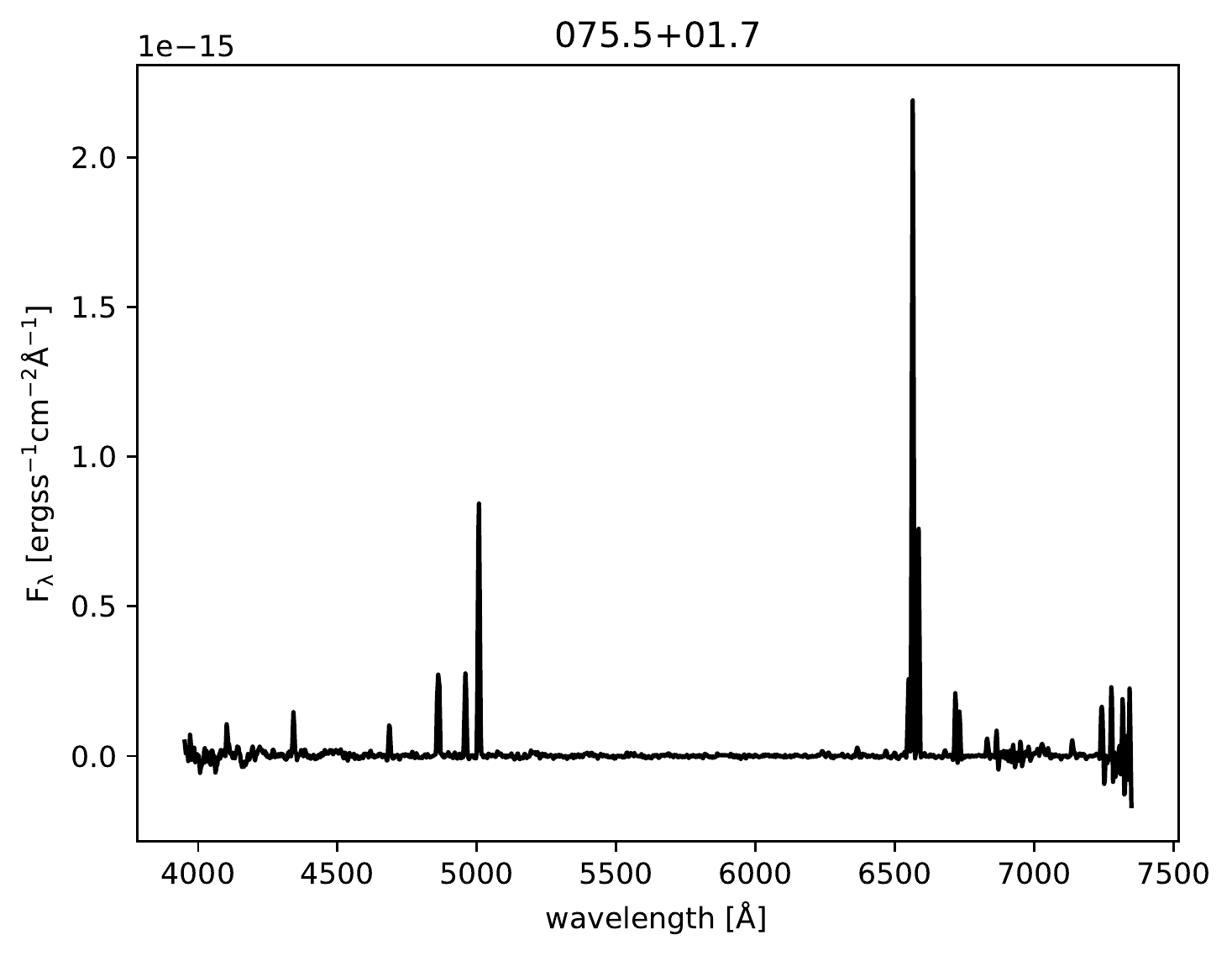}\\
\includegraphics[width=0.48\textwidth]{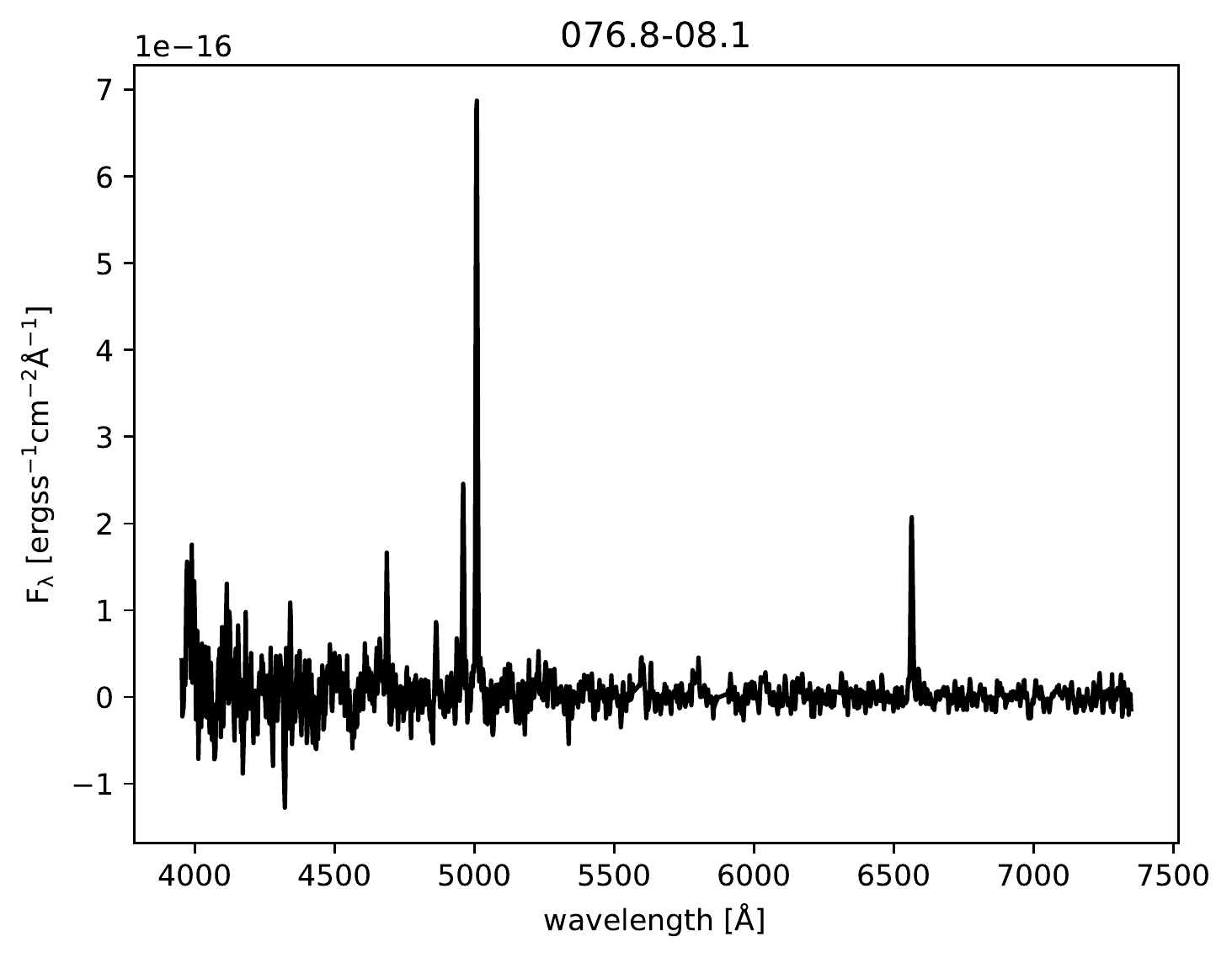}& \includegraphics[width=0.48\textwidth]{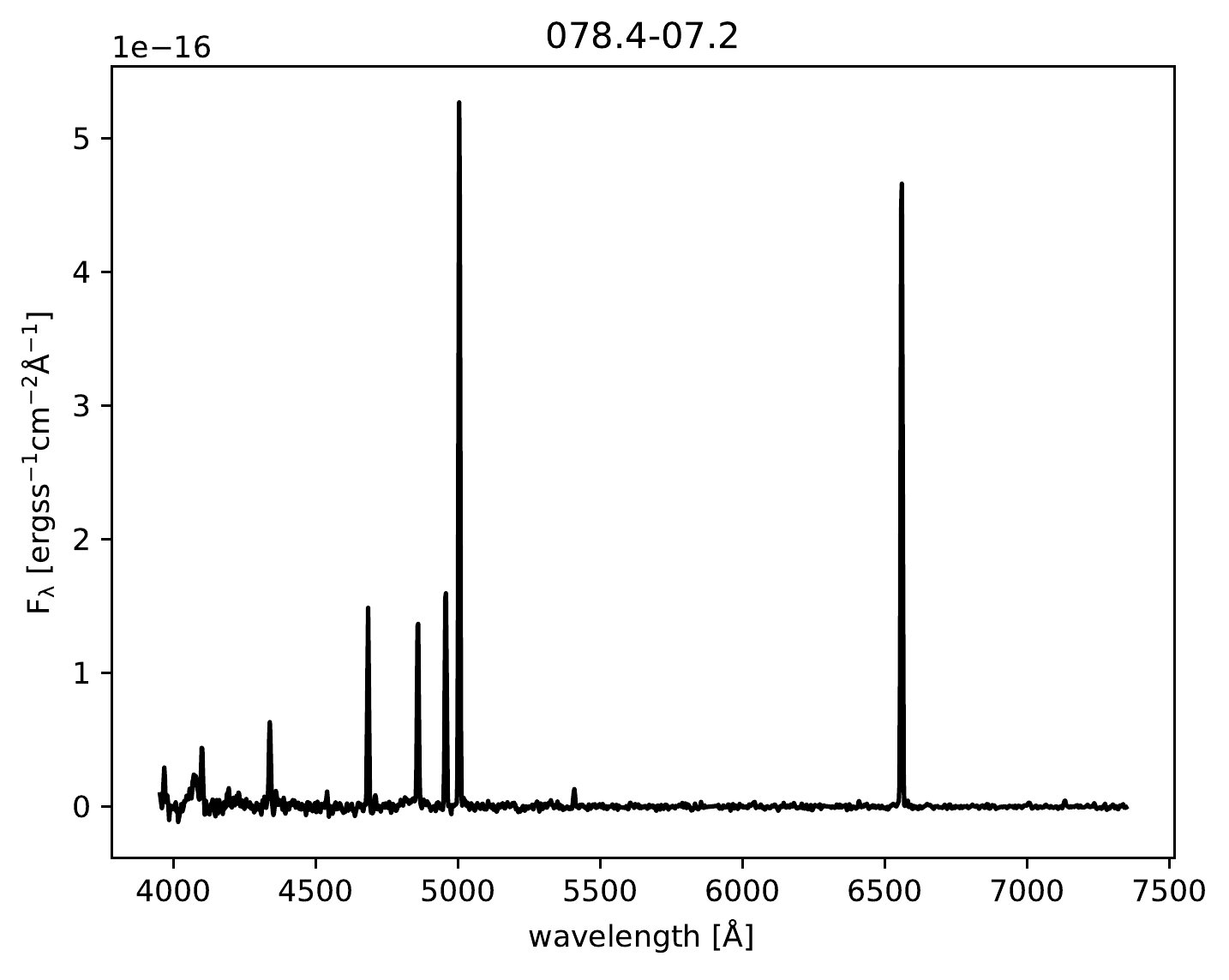}\\
\includegraphics[width=0.48\textwidth]{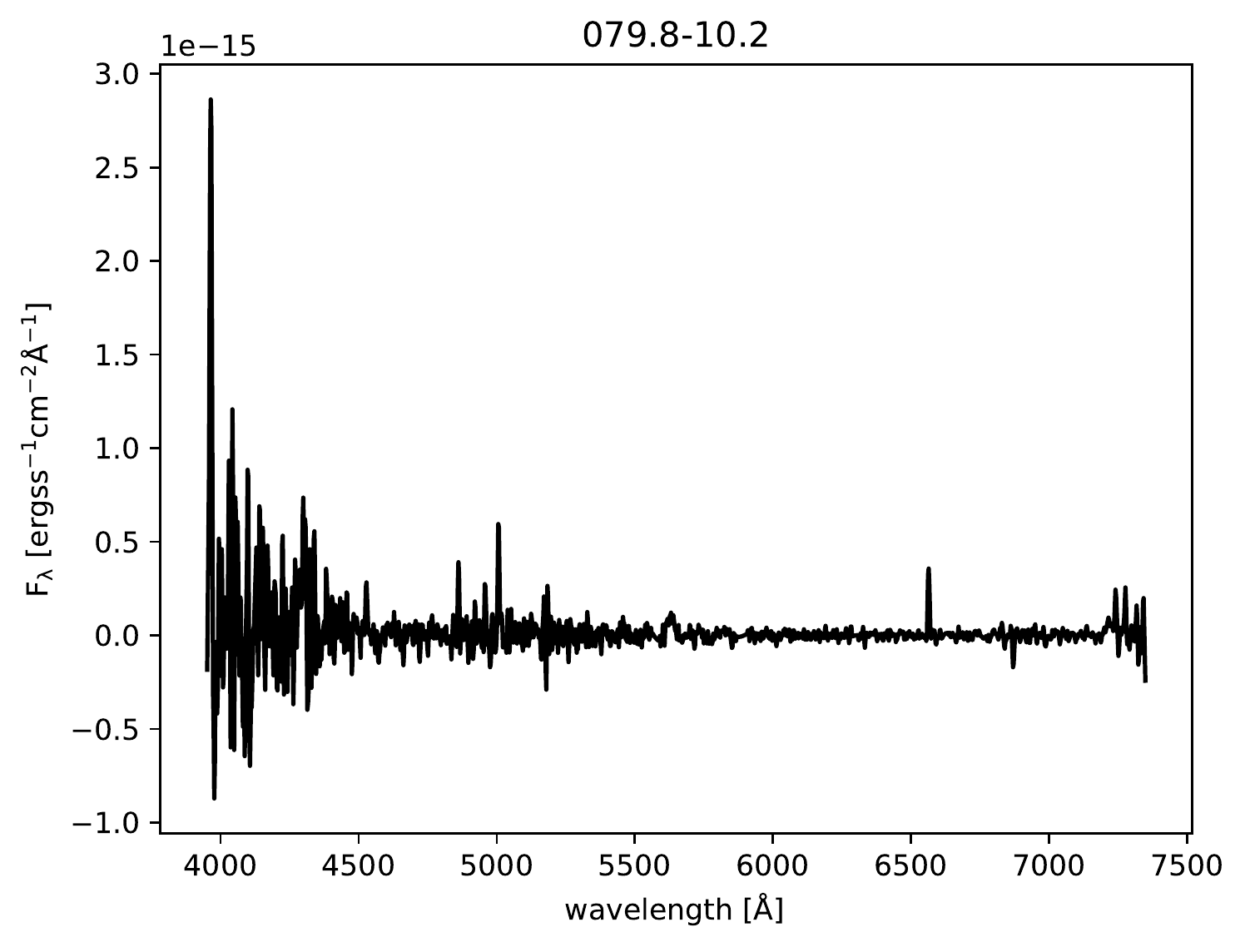}& \includegraphics[width=0.48\textwidth]{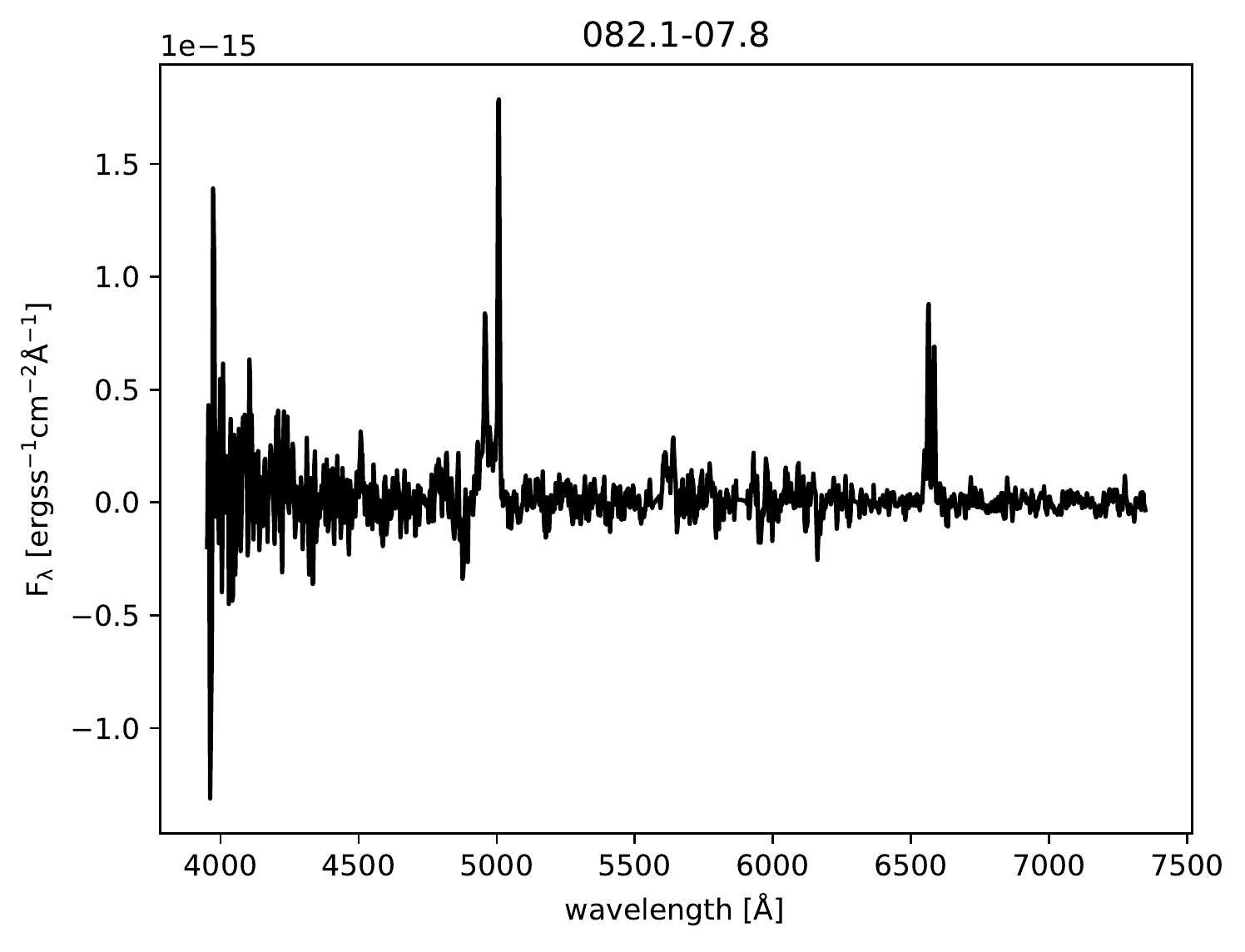}\\
 \includegraphics[width=0.48\textwidth]{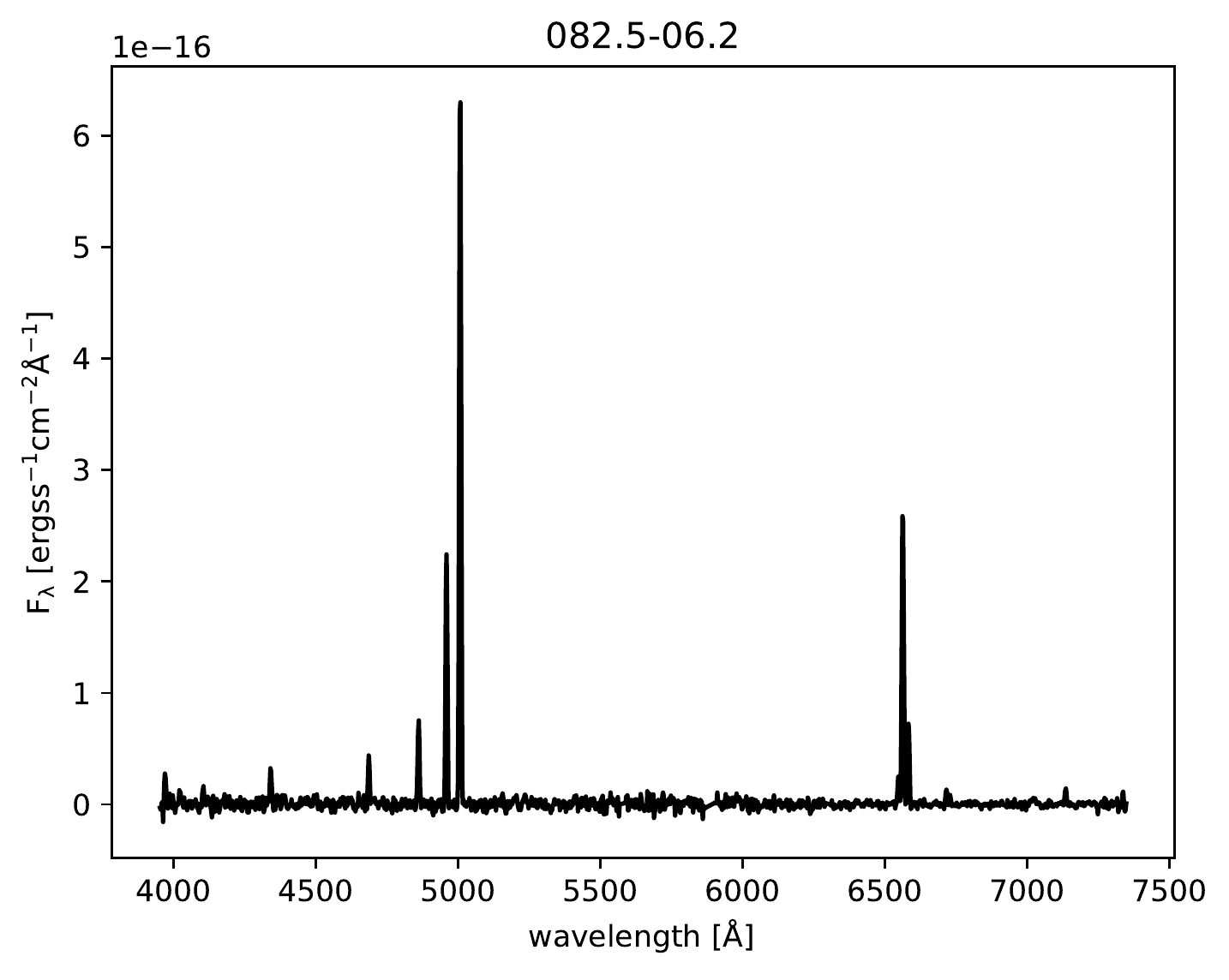}& \includegraphics[width=0.48\textwidth]{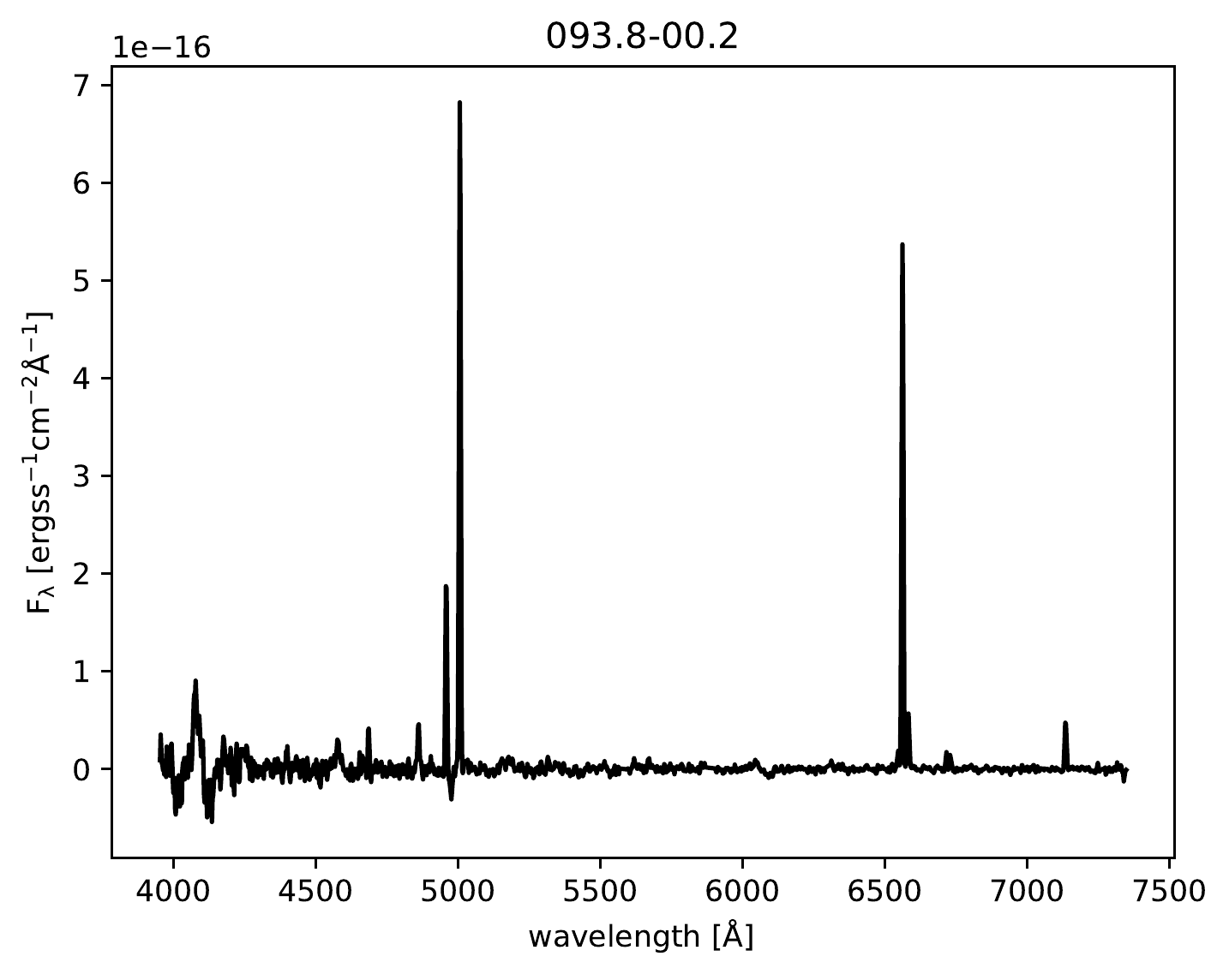}\\
 \includegraphics[width=0.48\textwidth]{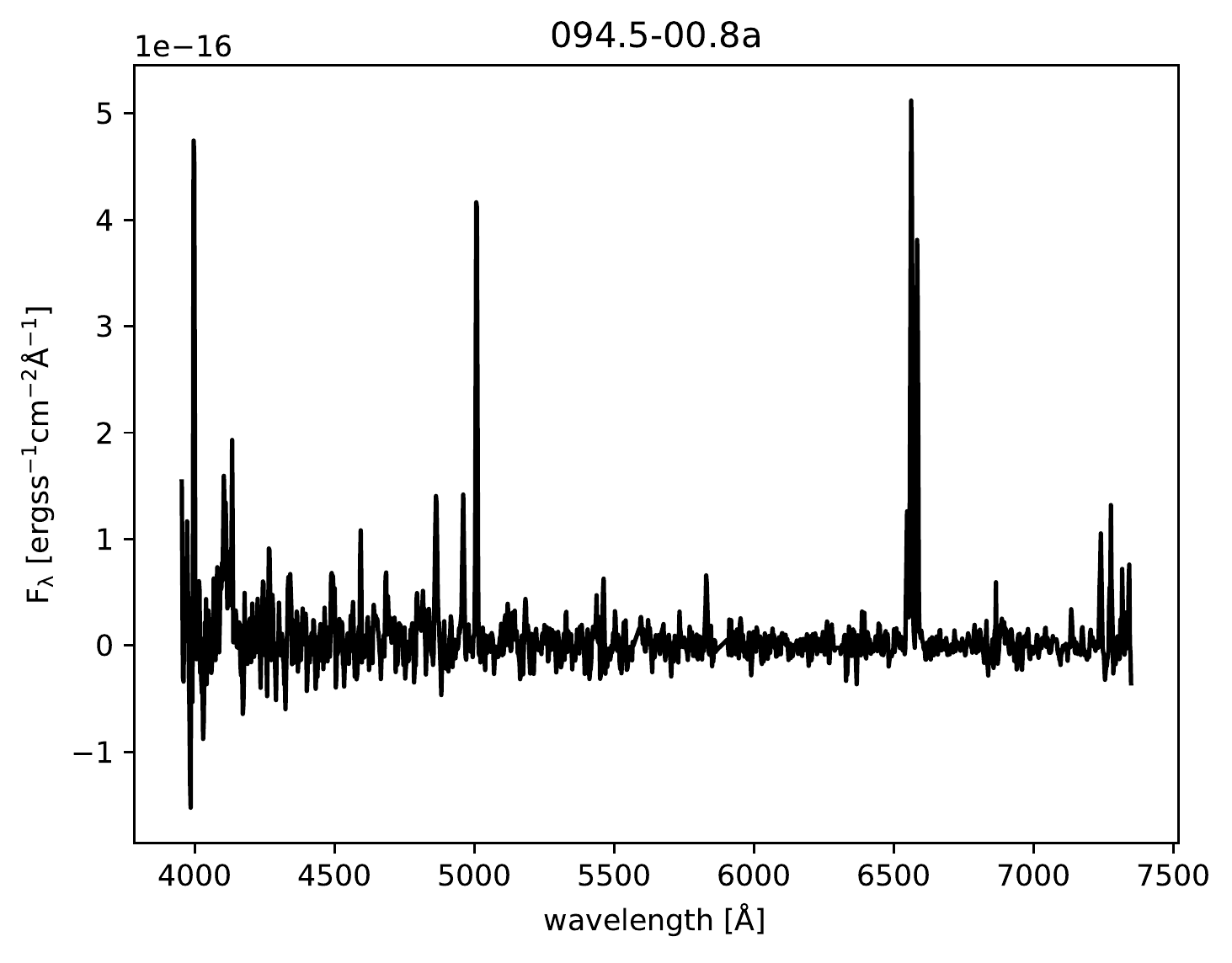}&\includegraphics[width=0.48\textwidth]{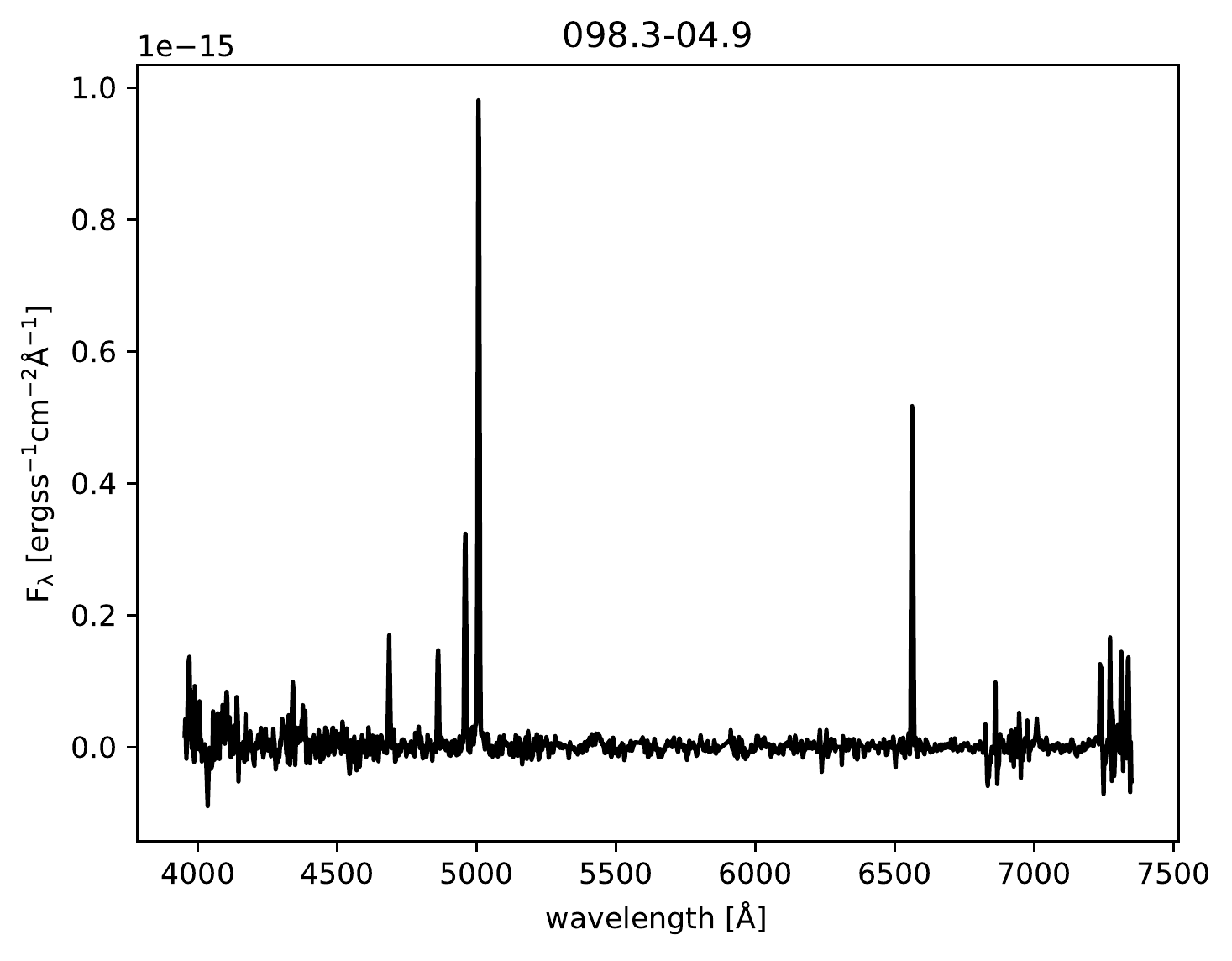}\\
\includegraphics[width=0.48\textwidth]{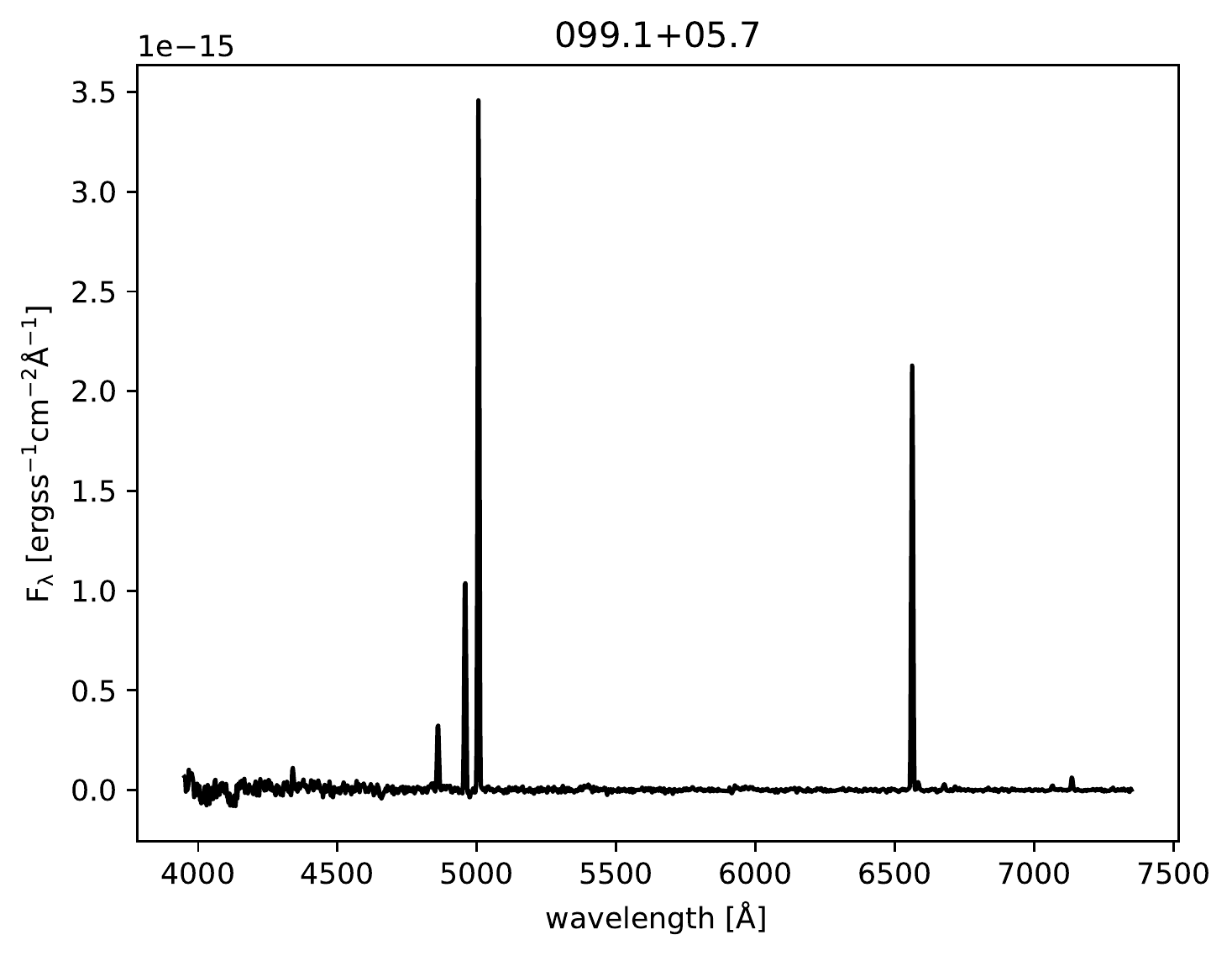}&\includegraphics[width=0.48\textwidth]{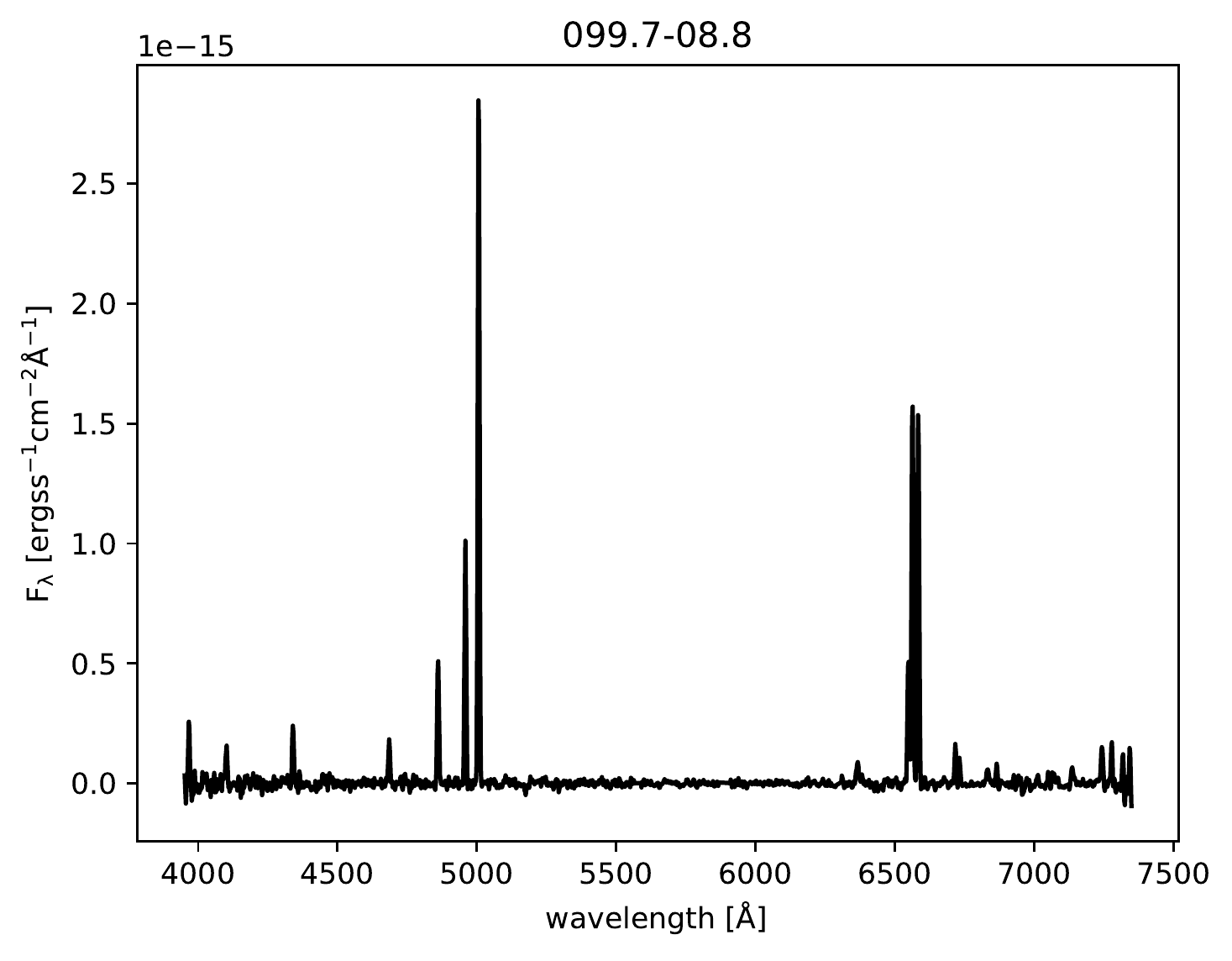}\\
\includegraphics[width=0.48\textwidth]{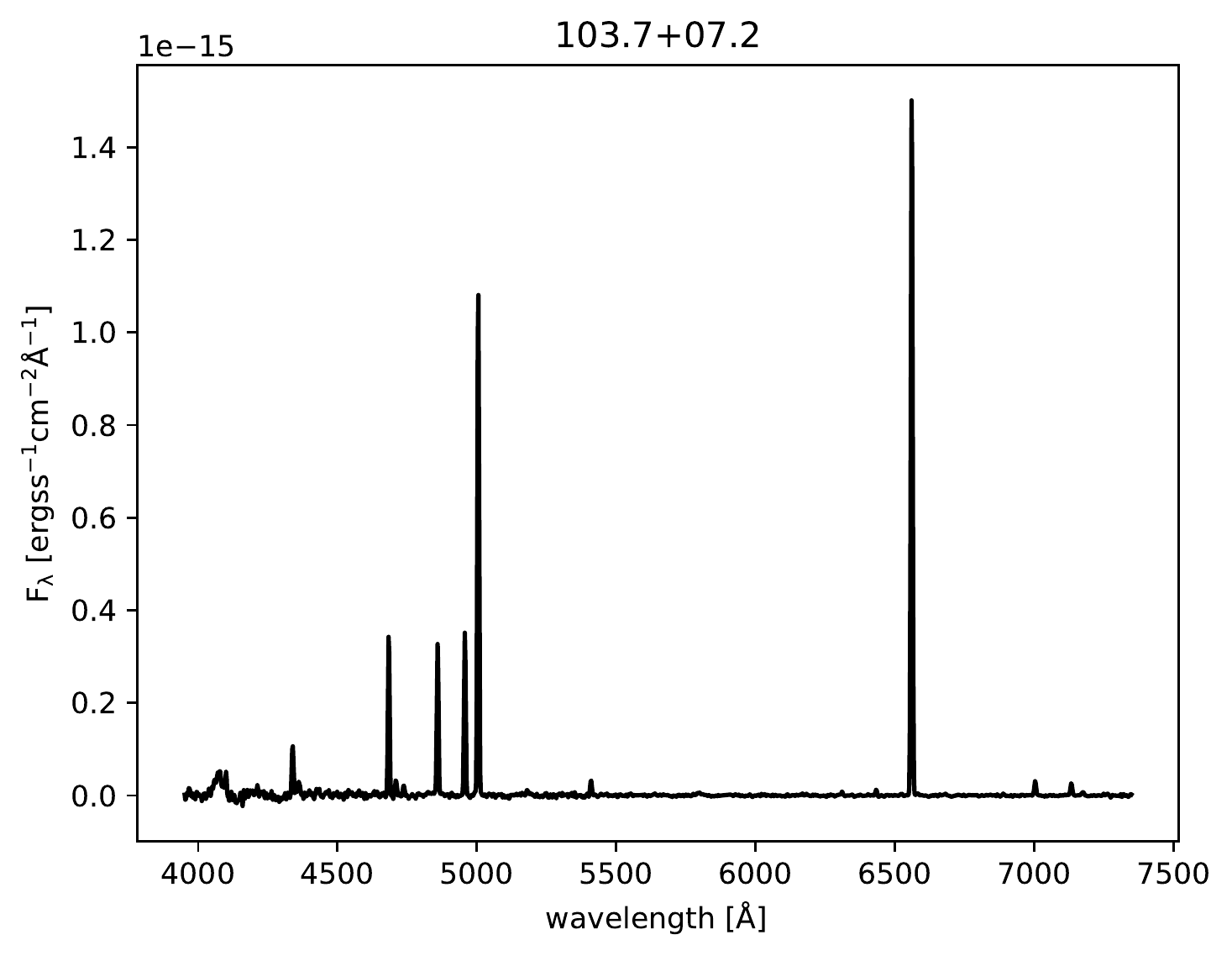}&\includegraphics[width=0.48\textwidth]{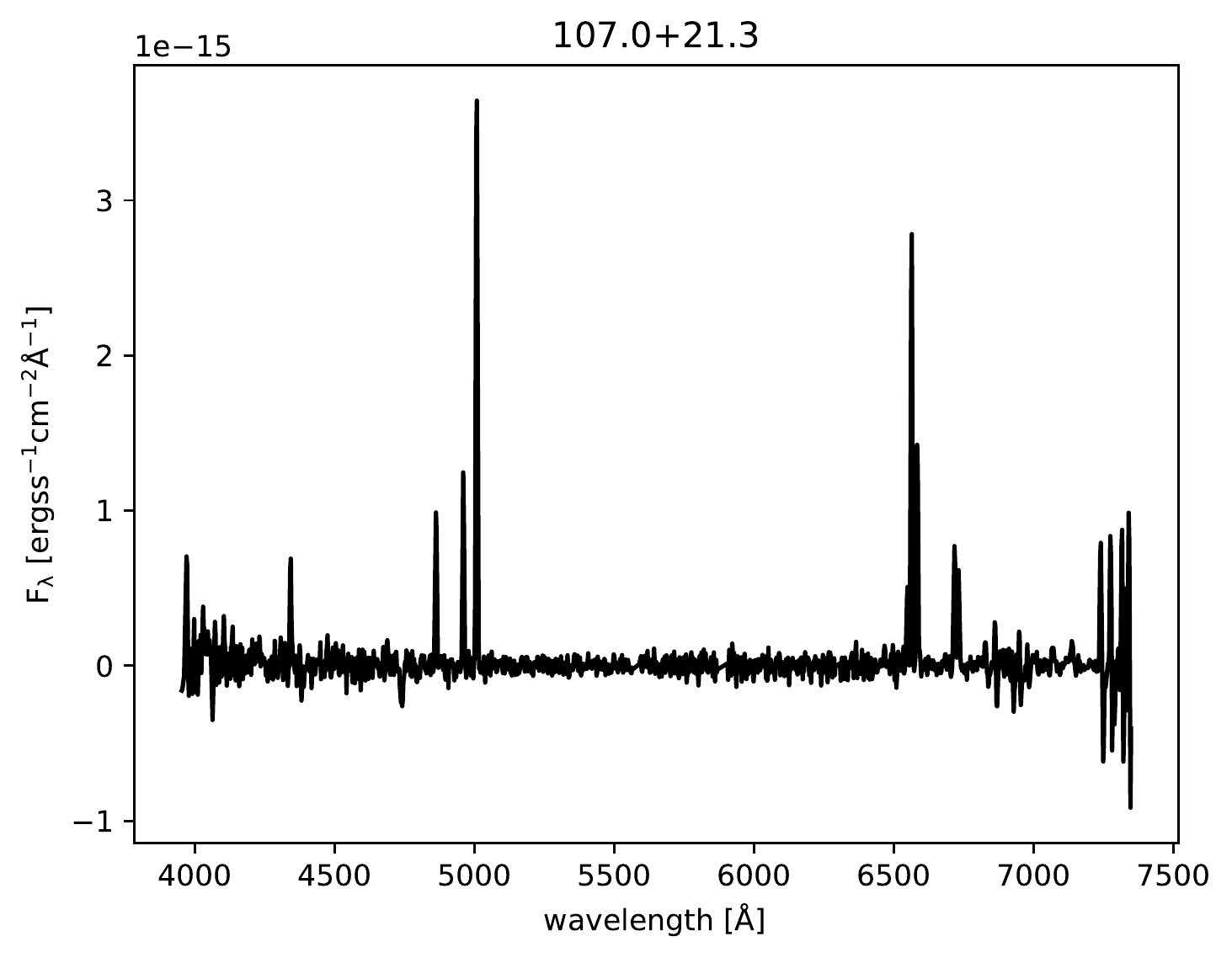}\\
\includegraphics[width=0.48\textwidth]{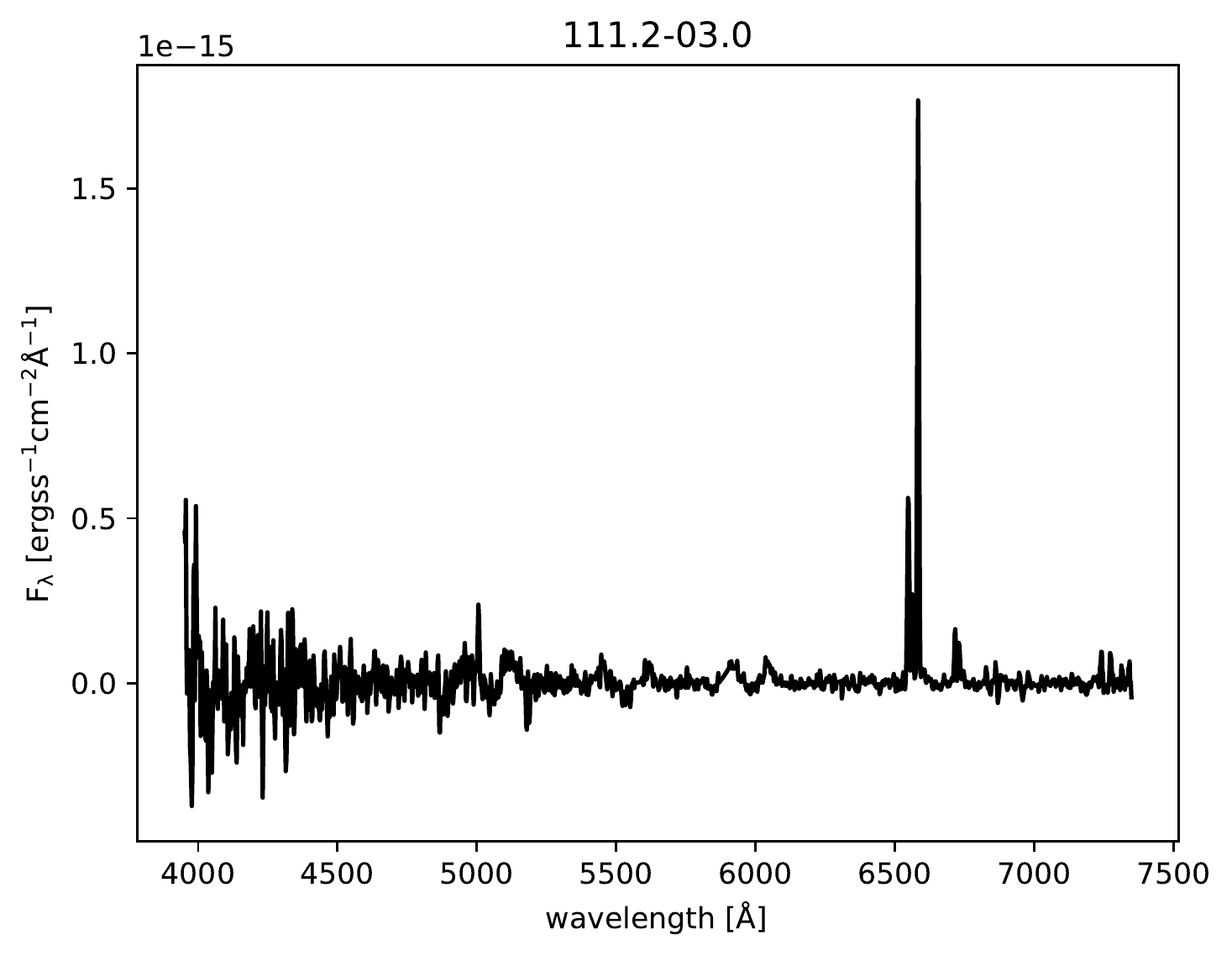}&\includegraphics[width=0.48\textwidth]{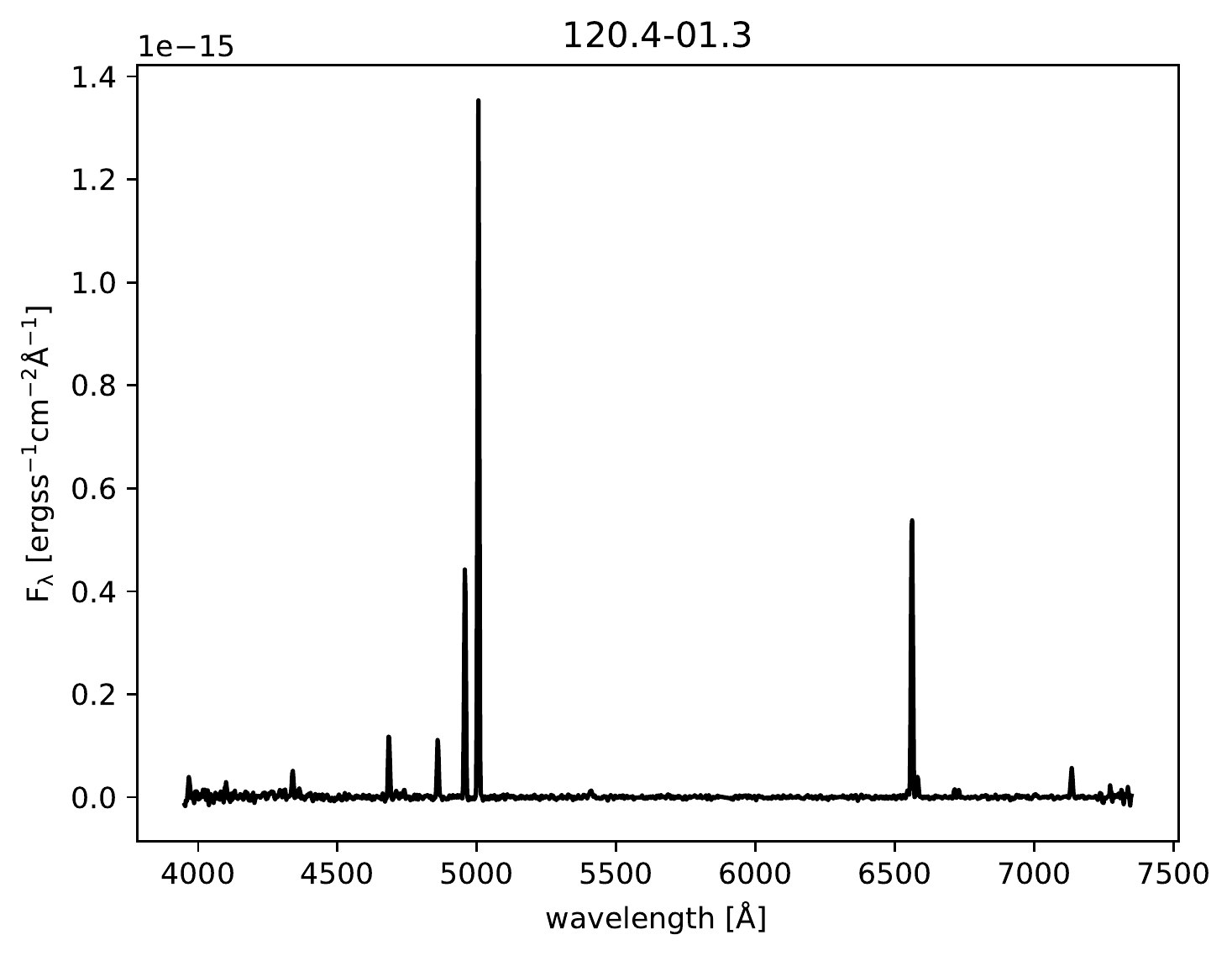}\\
\includegraphics[width=0.48\textwidth]{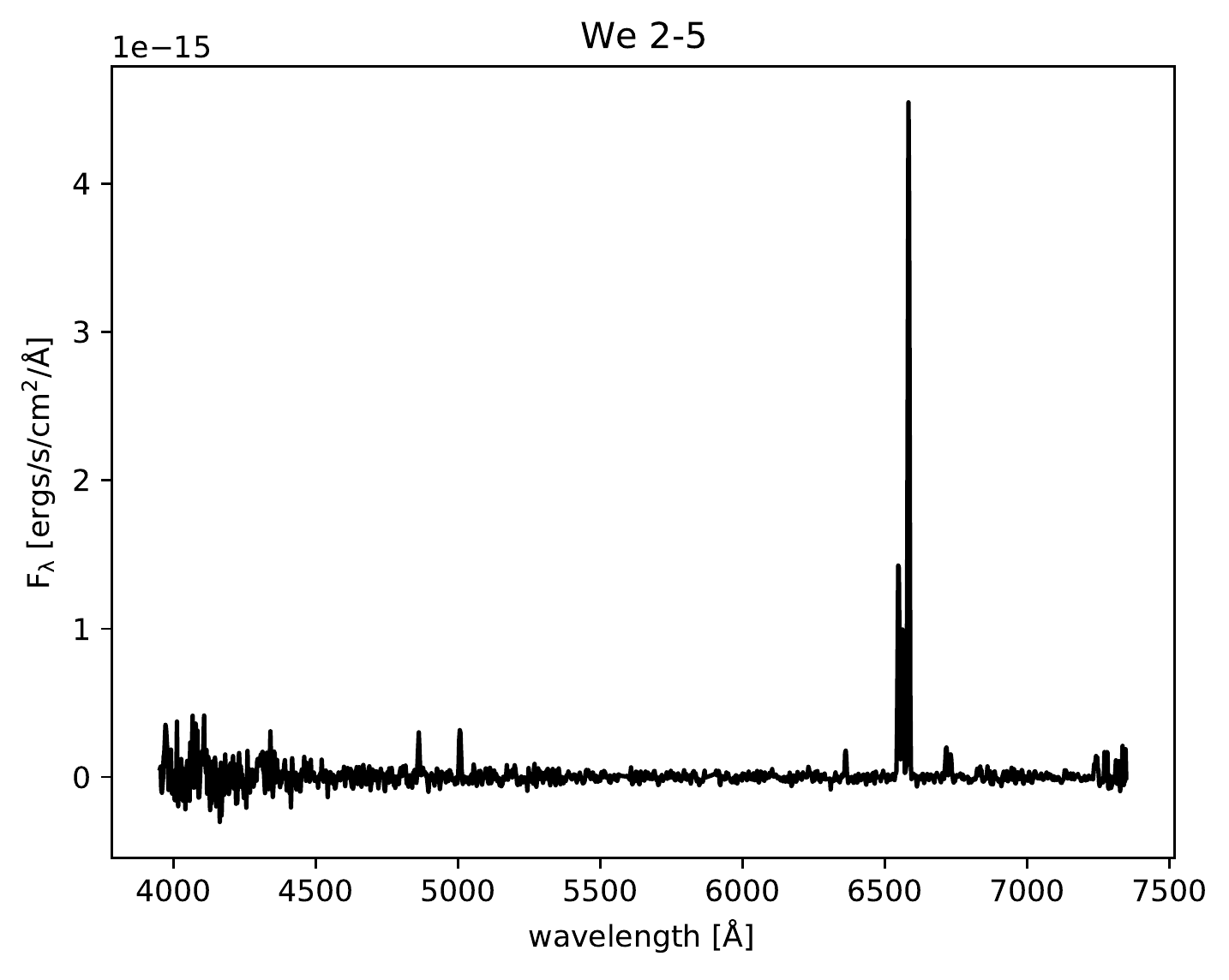}&\includegraphics[width=0.48\textwidth]{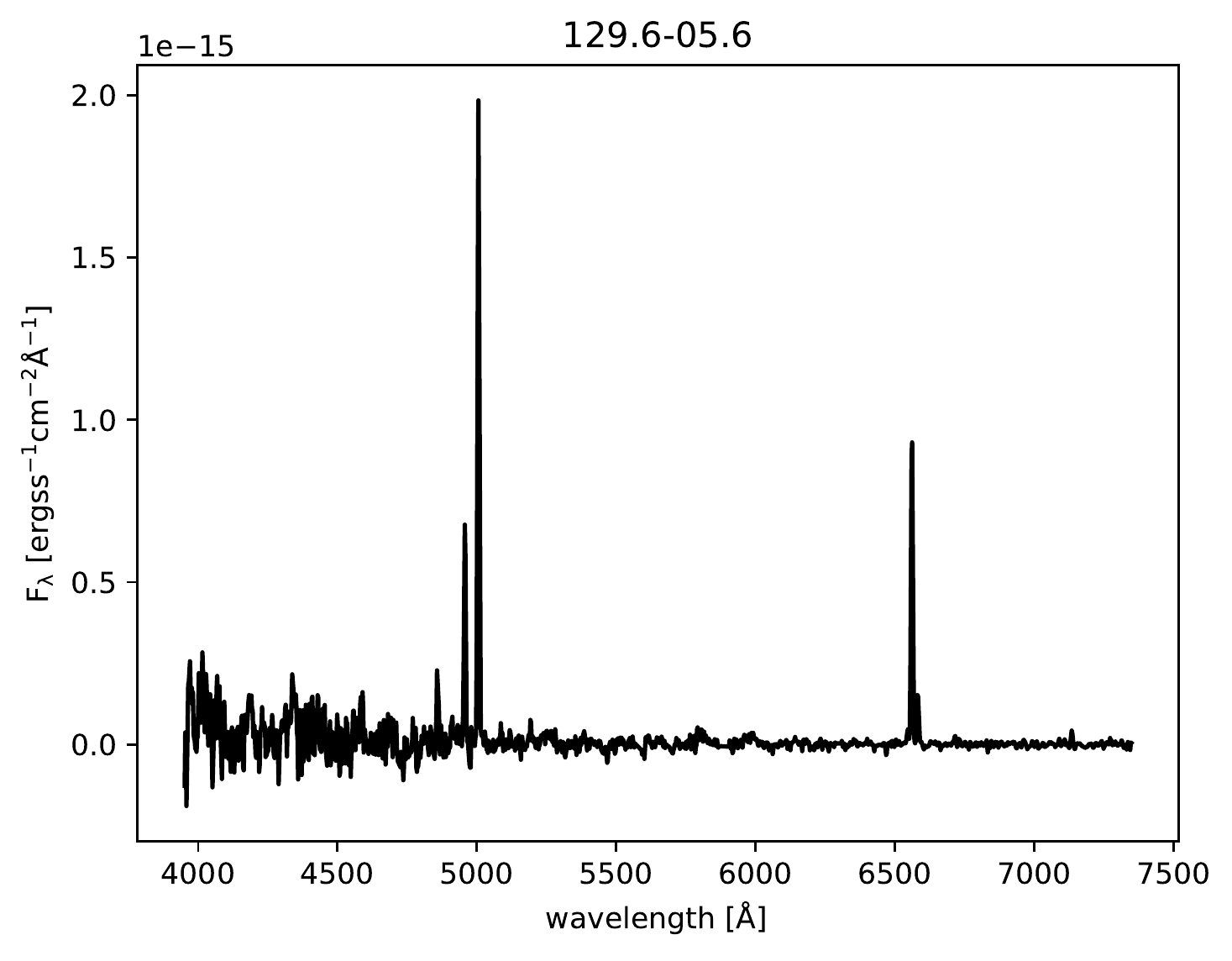}\\
\includegraphics[width=0.48\textwidth]{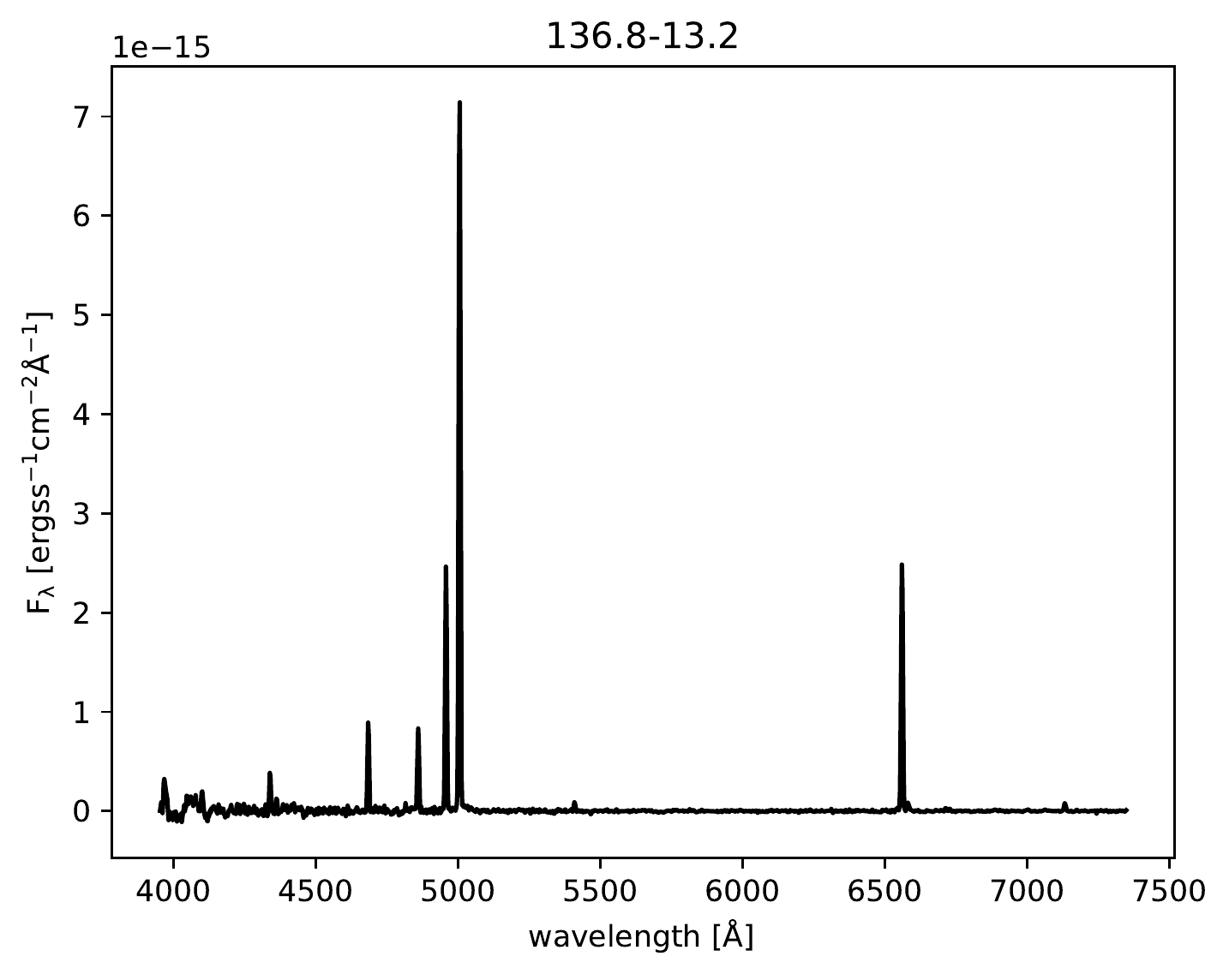}&\includegraphics[width=0.48\textwidth]{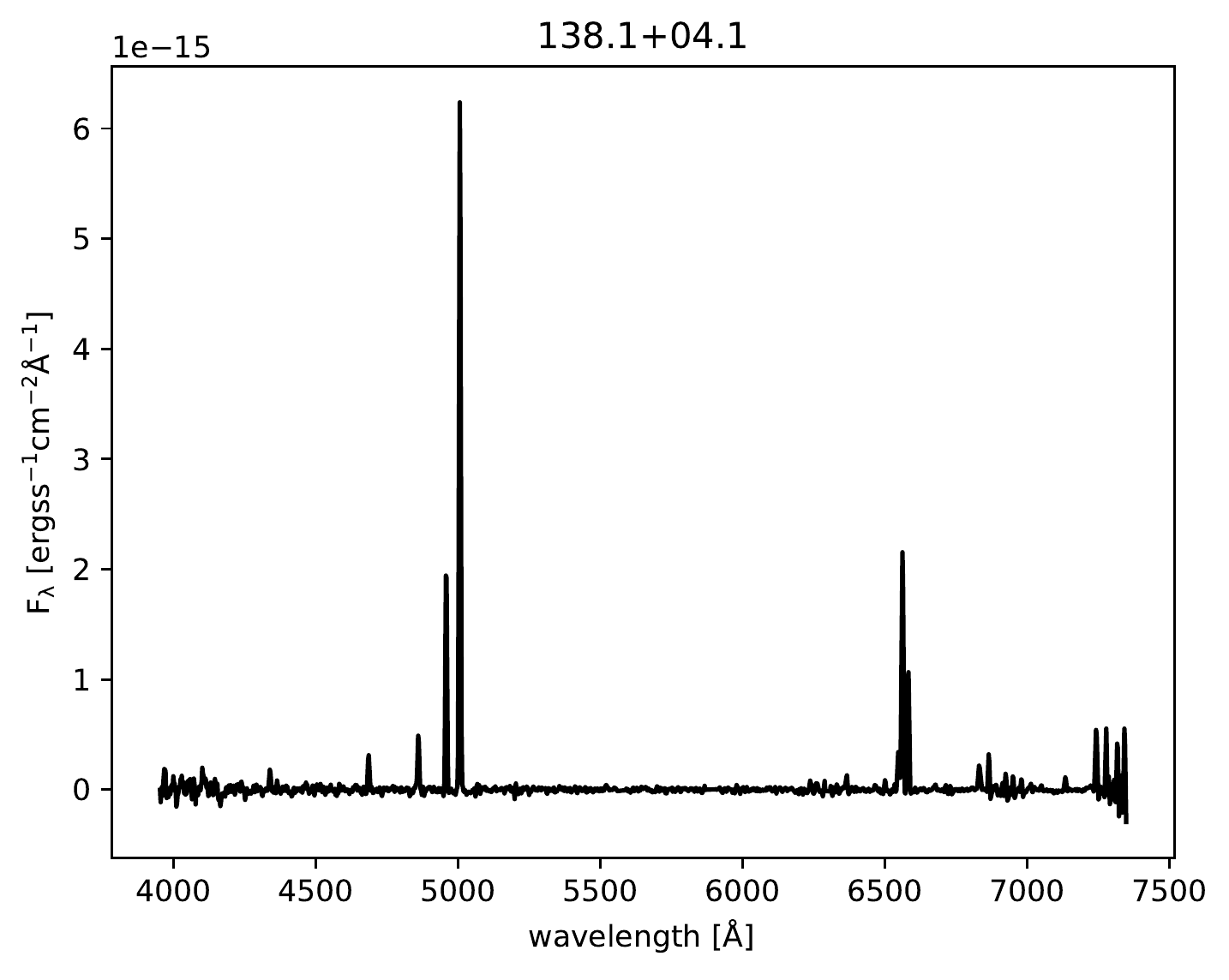}\\
\includegraphics[width=0.48\textwidth]{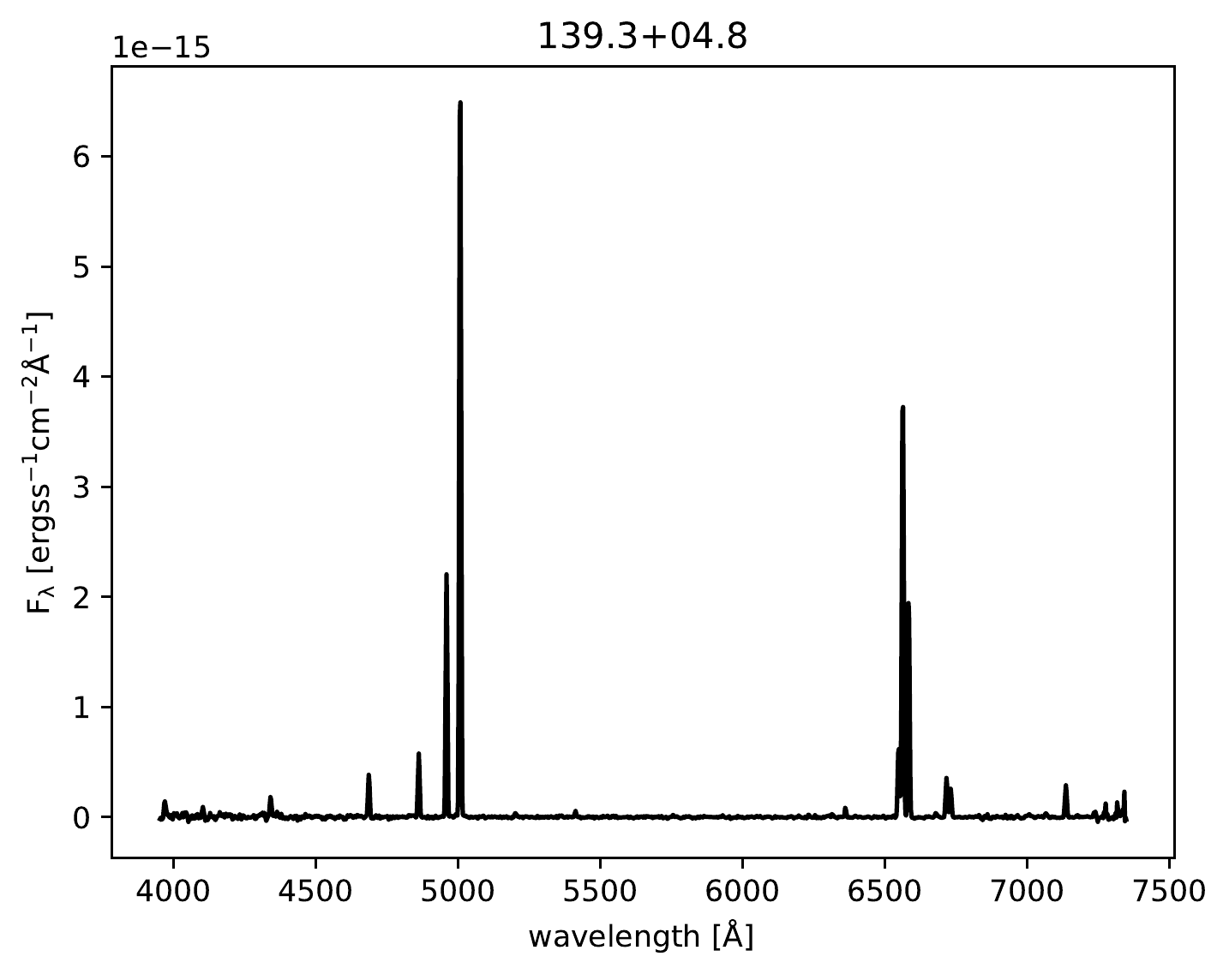}&\includegraphics[width=0.48\textwidth]{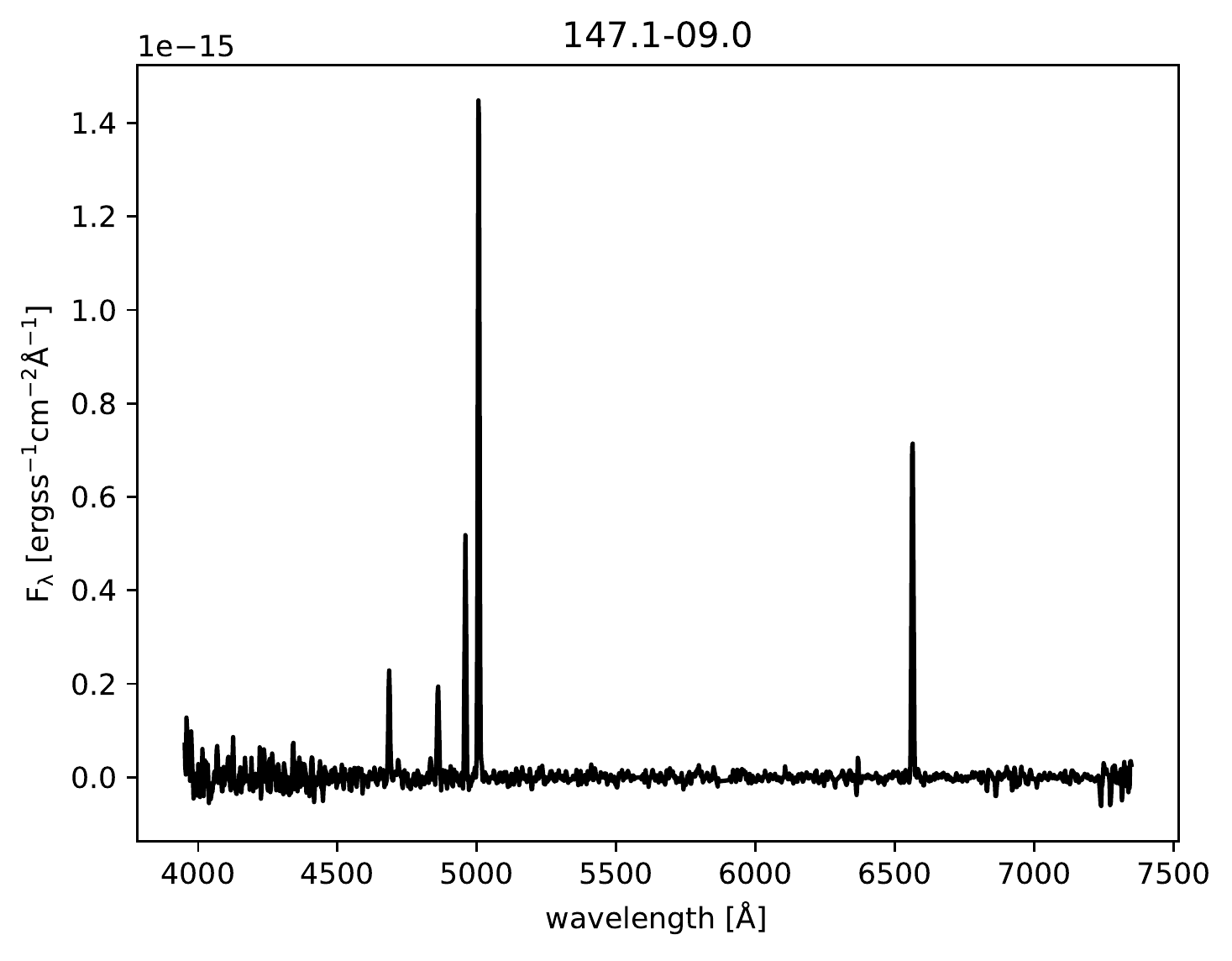}\\
\includegraphics[width=0.48\textwidth]{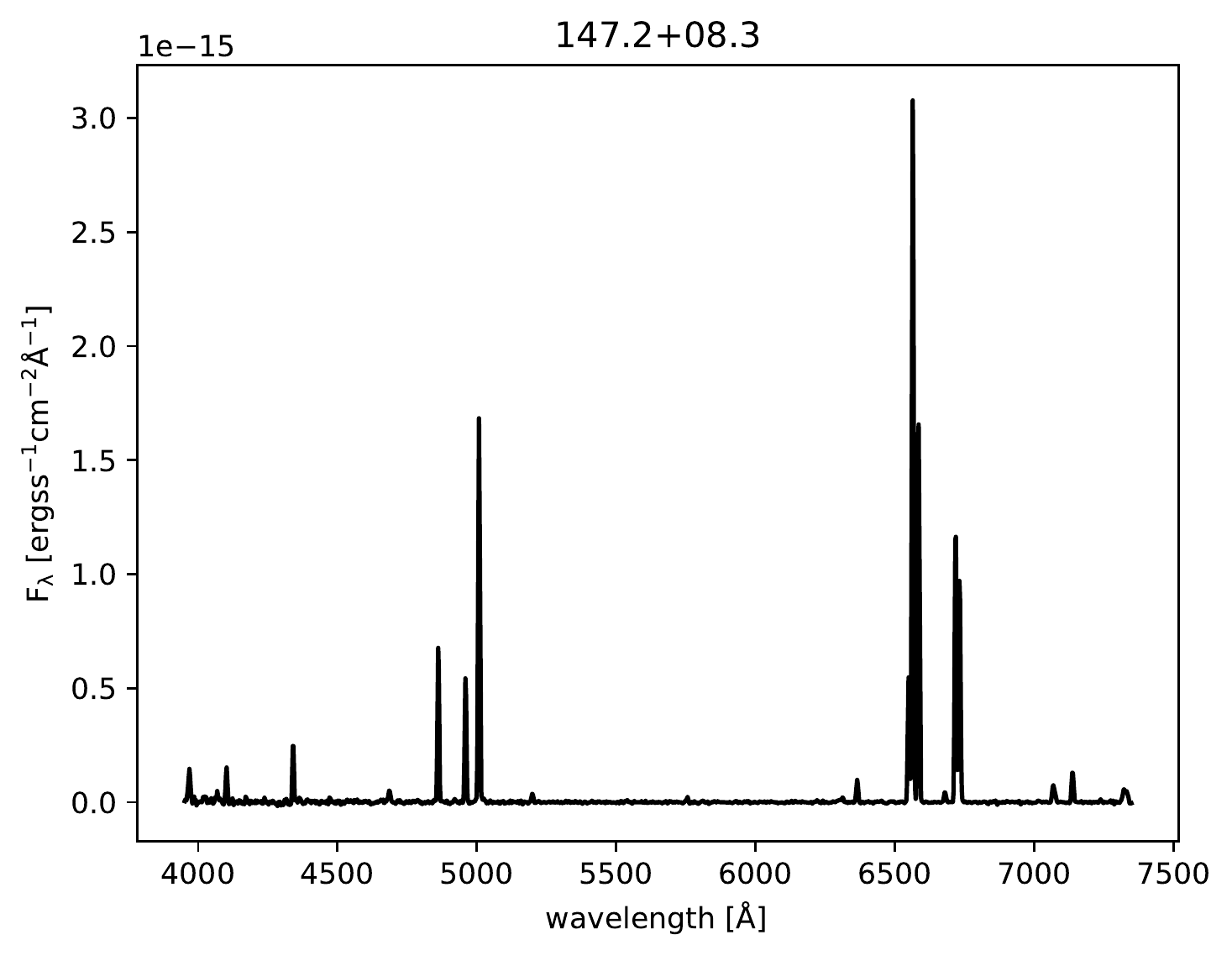}&\includegraphics[width=0.48\textwidth]{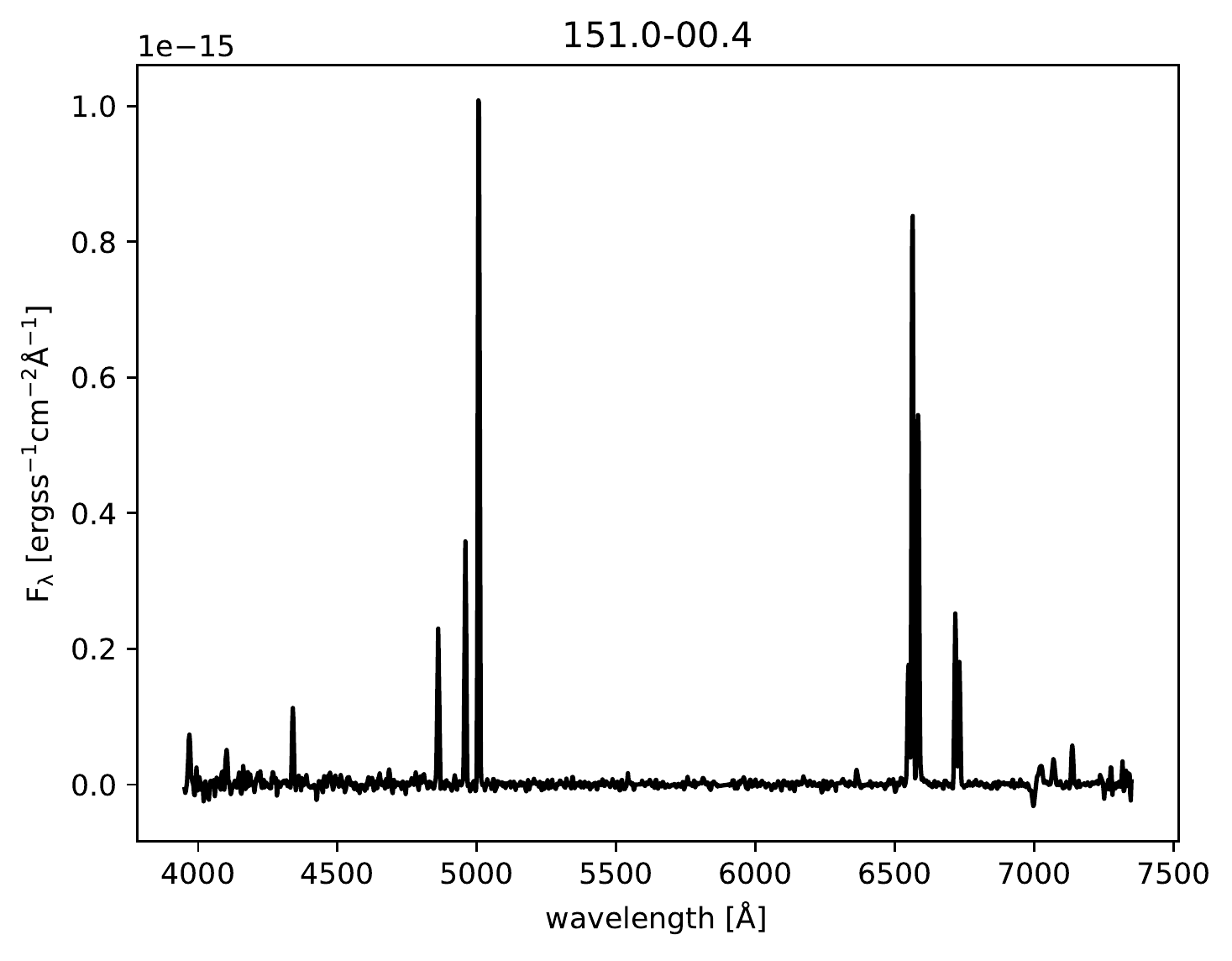}\\
\includegraphics[width=0.48\textwidth]{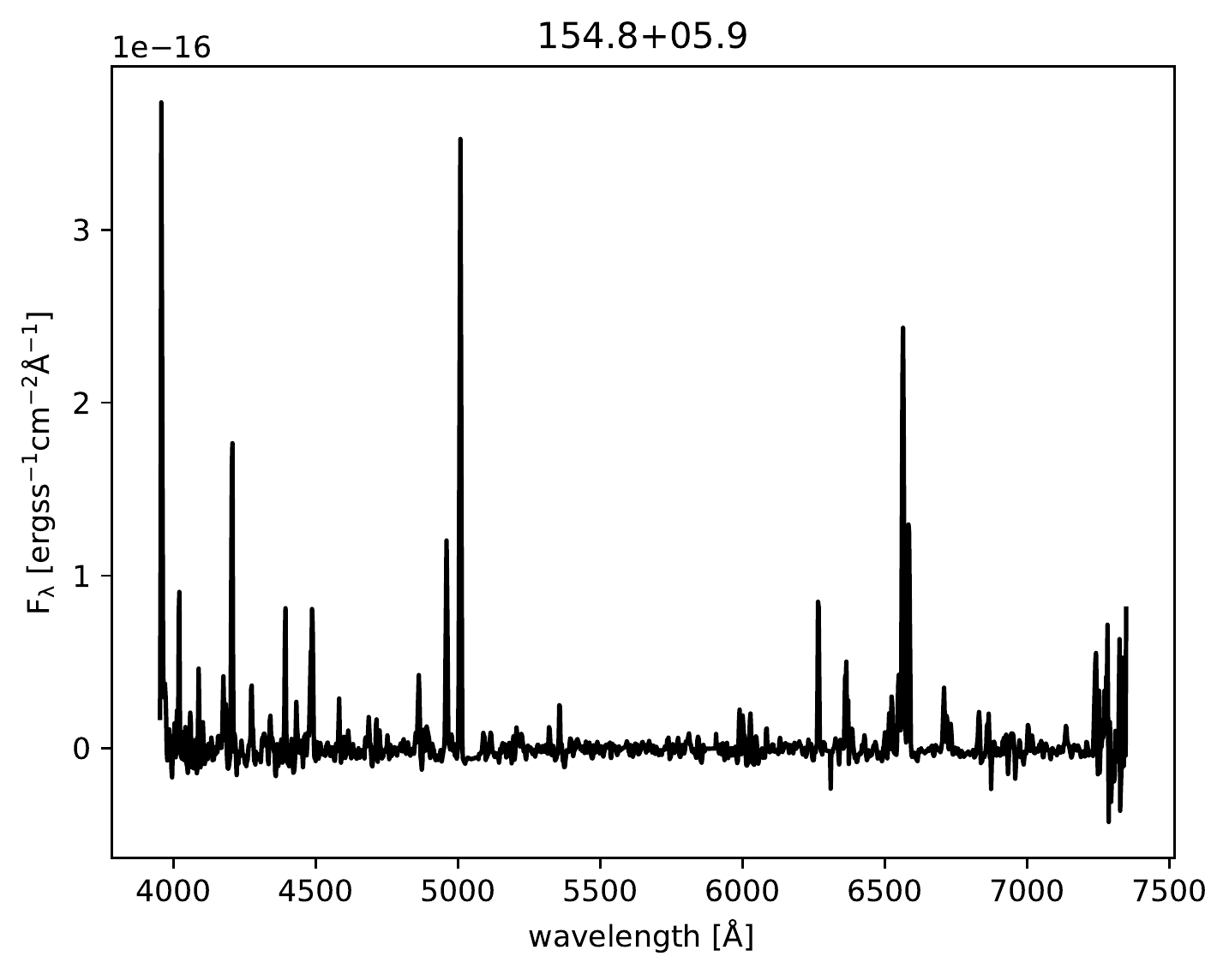}&\includegraphics[width=0.48\textwidth]{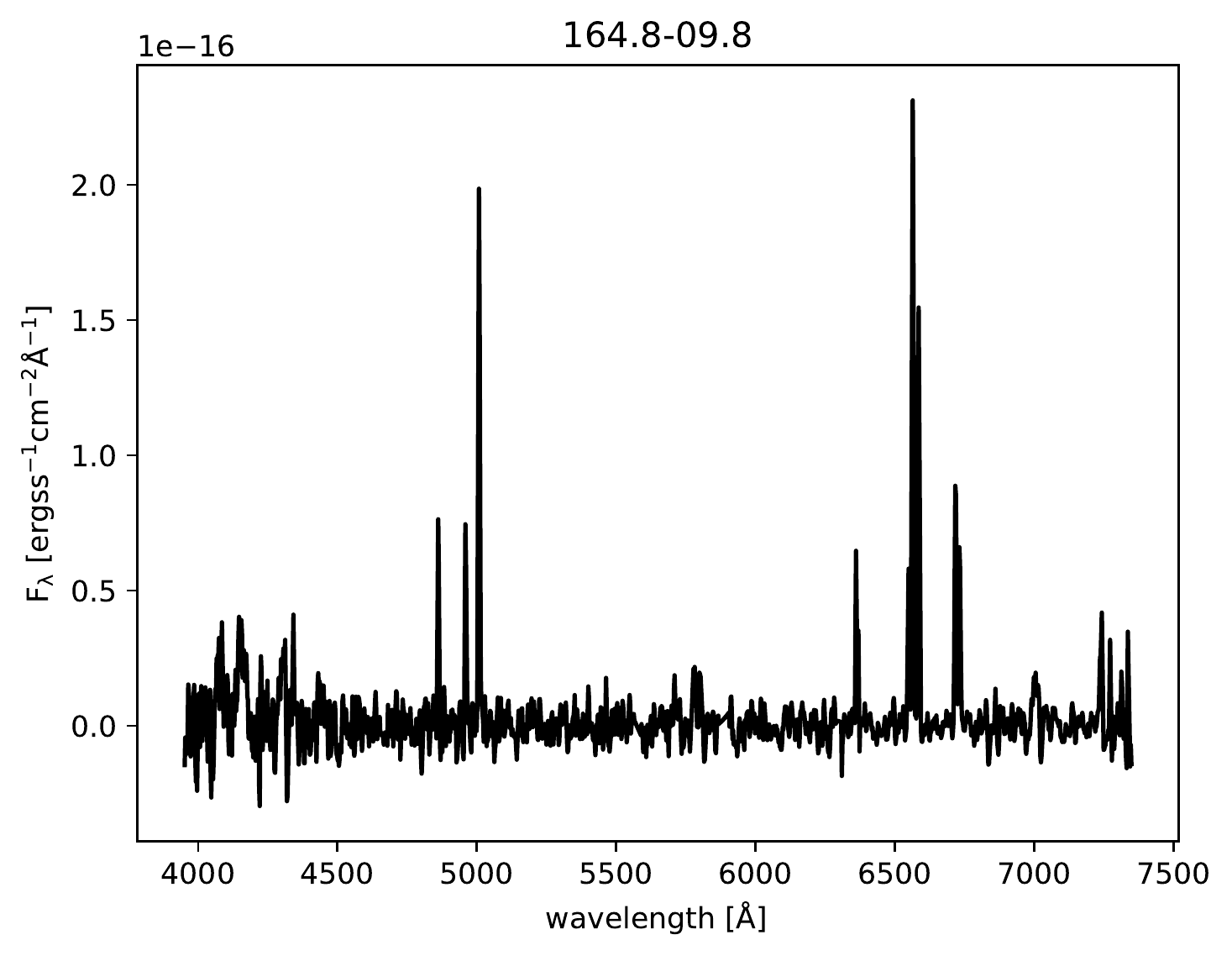}\\
\includegraphics[width=0.48\textwidth]{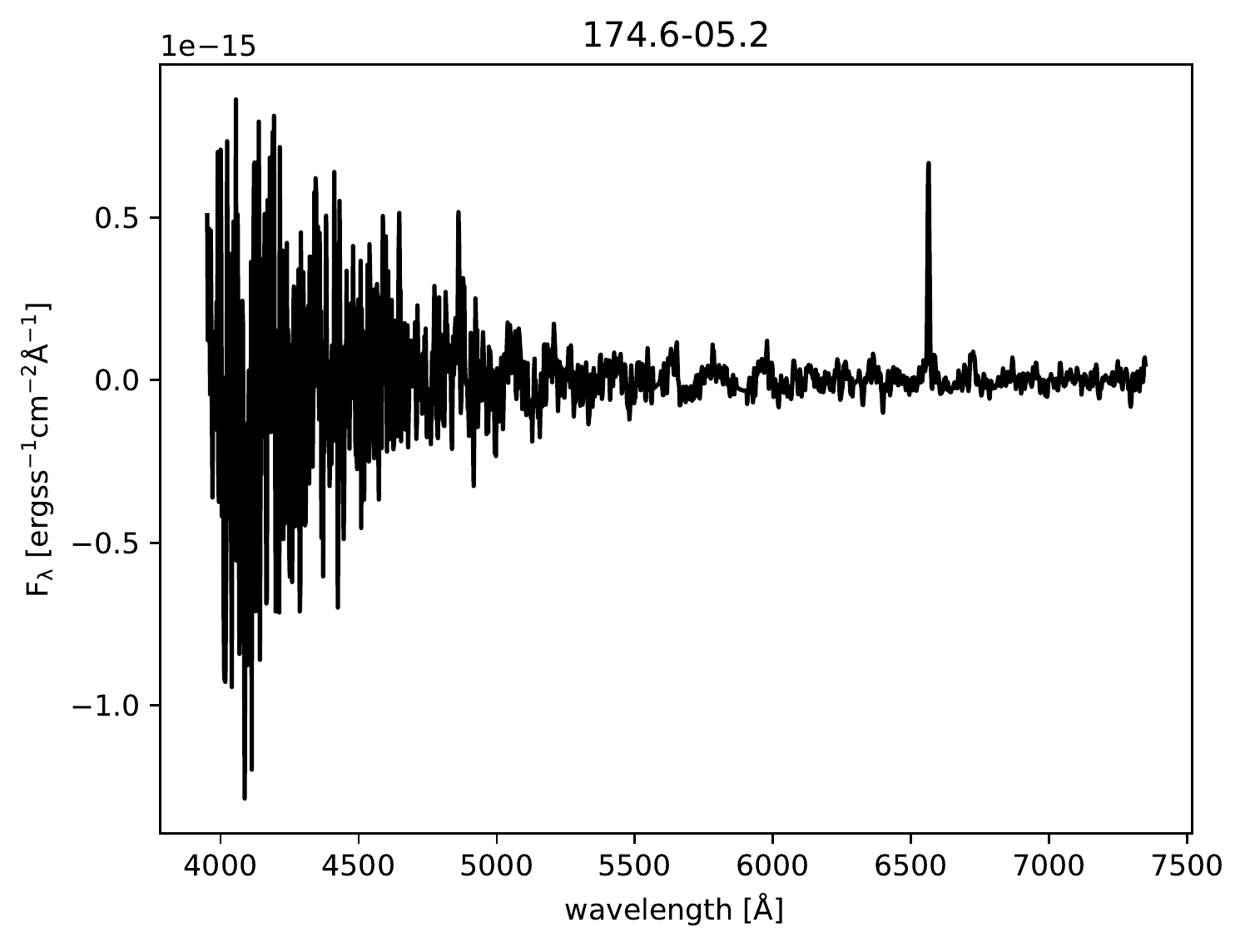}&\includegraphics[width=0.48\textwidth]{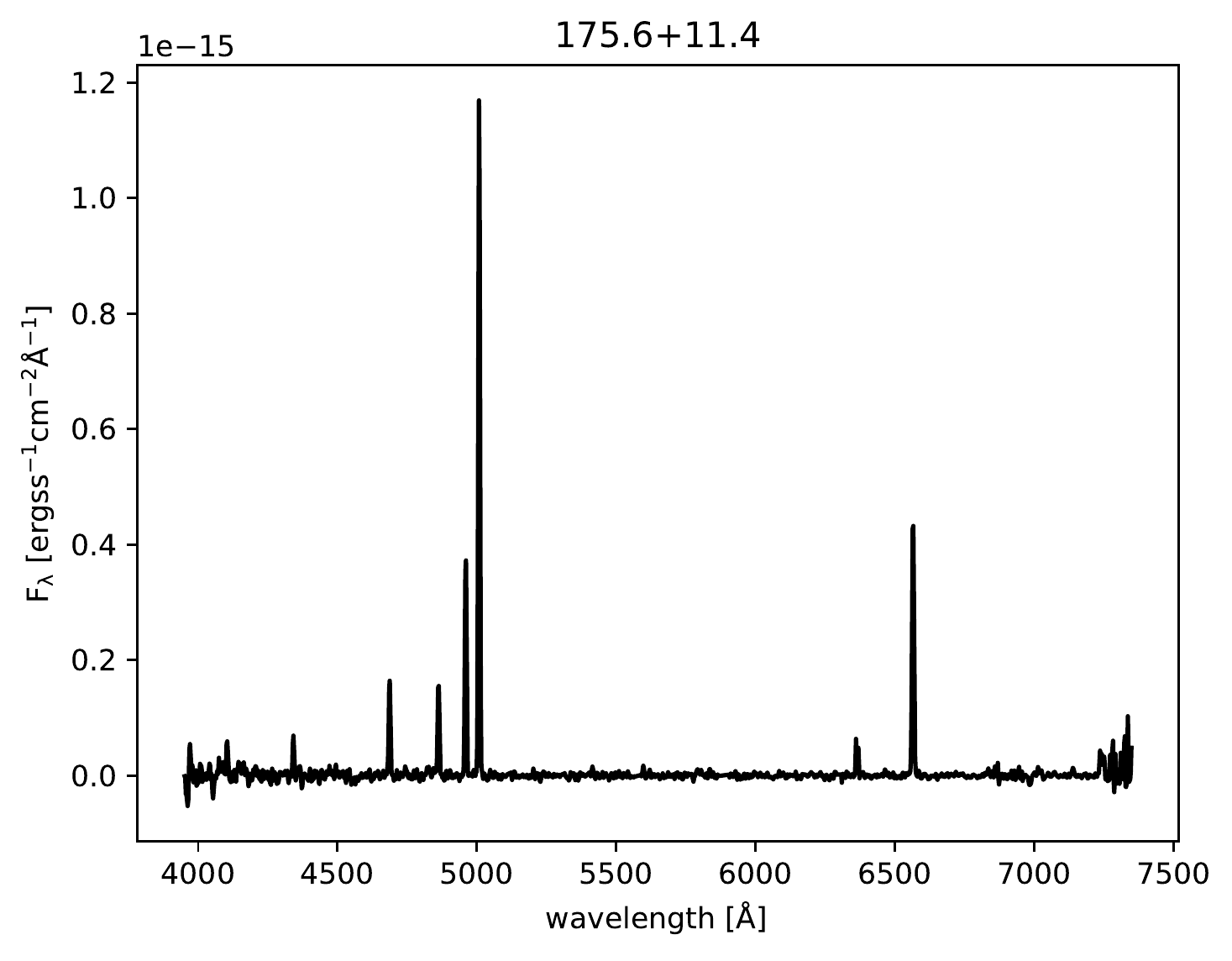}\\
\includegraphics[width=0.48\textwidth]{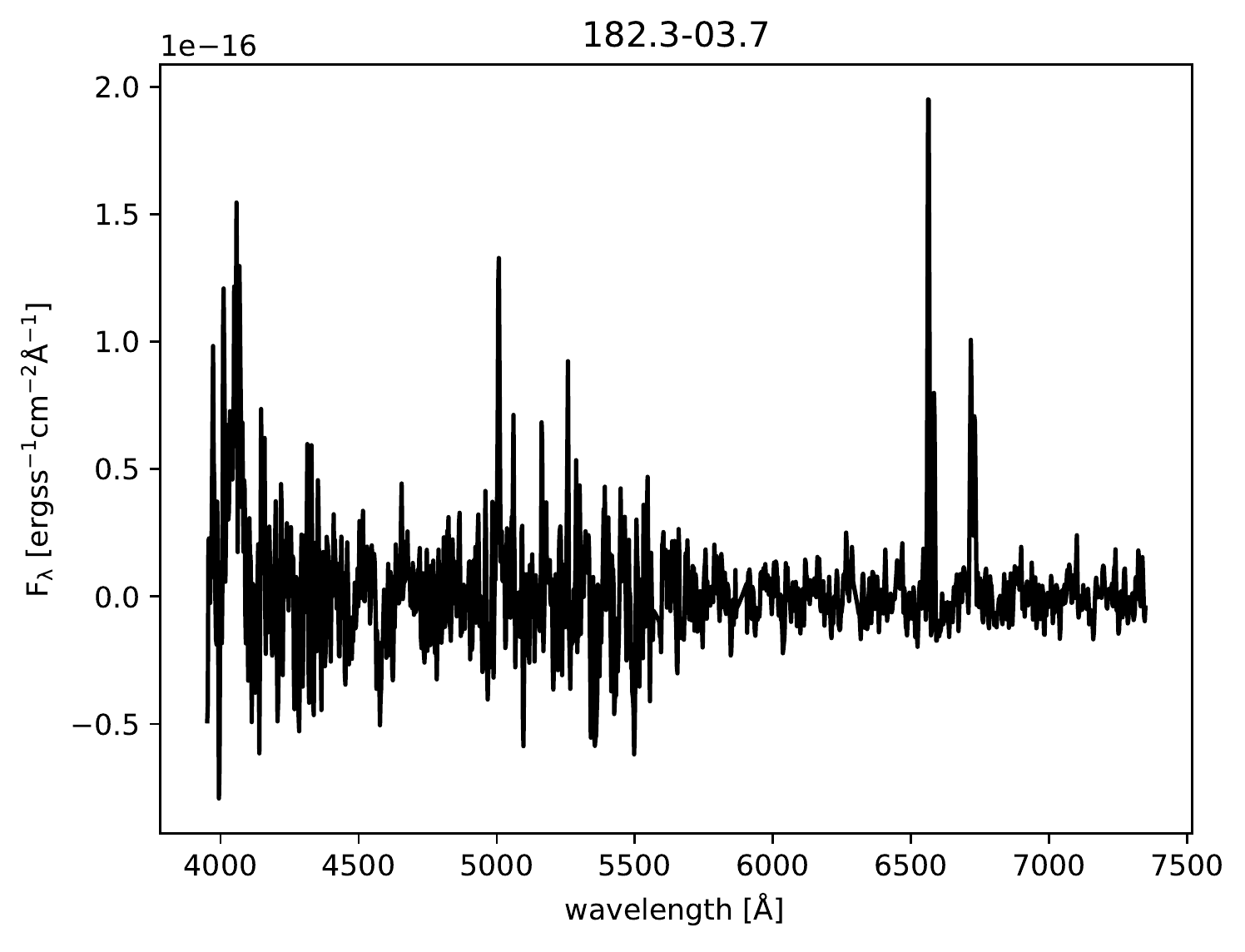}&
\end{longtable}
\begin{longtable}{ *{2}{l} }
    \caption{\edited{Spectra of the  previously observed candidates that required better spectra for final confirmation.}}\label{tab:spectra2}\\
    \endhead  
\includegraphics[width=0.48\textwidth]{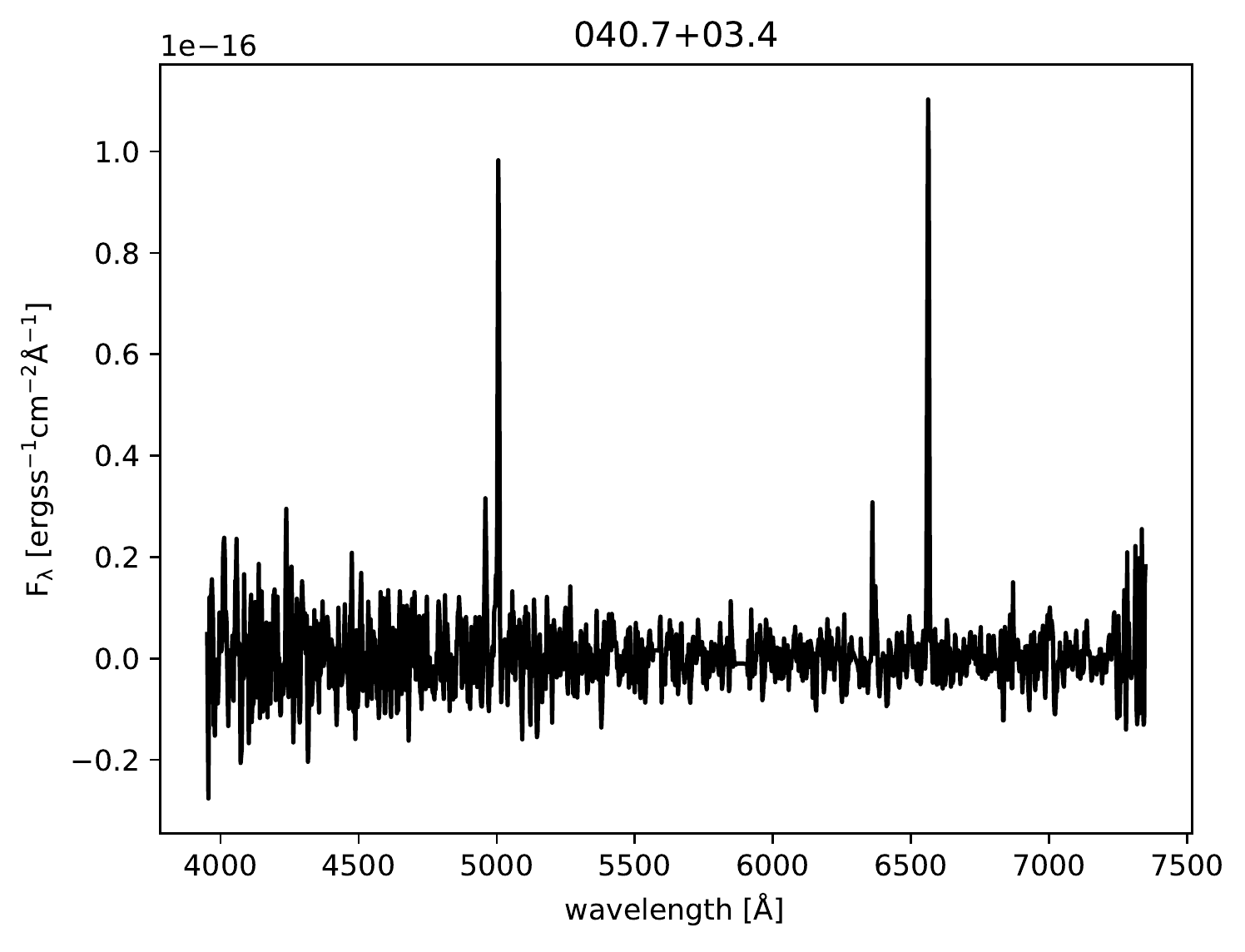}& \includegraphics[width=0.48\textwidth]{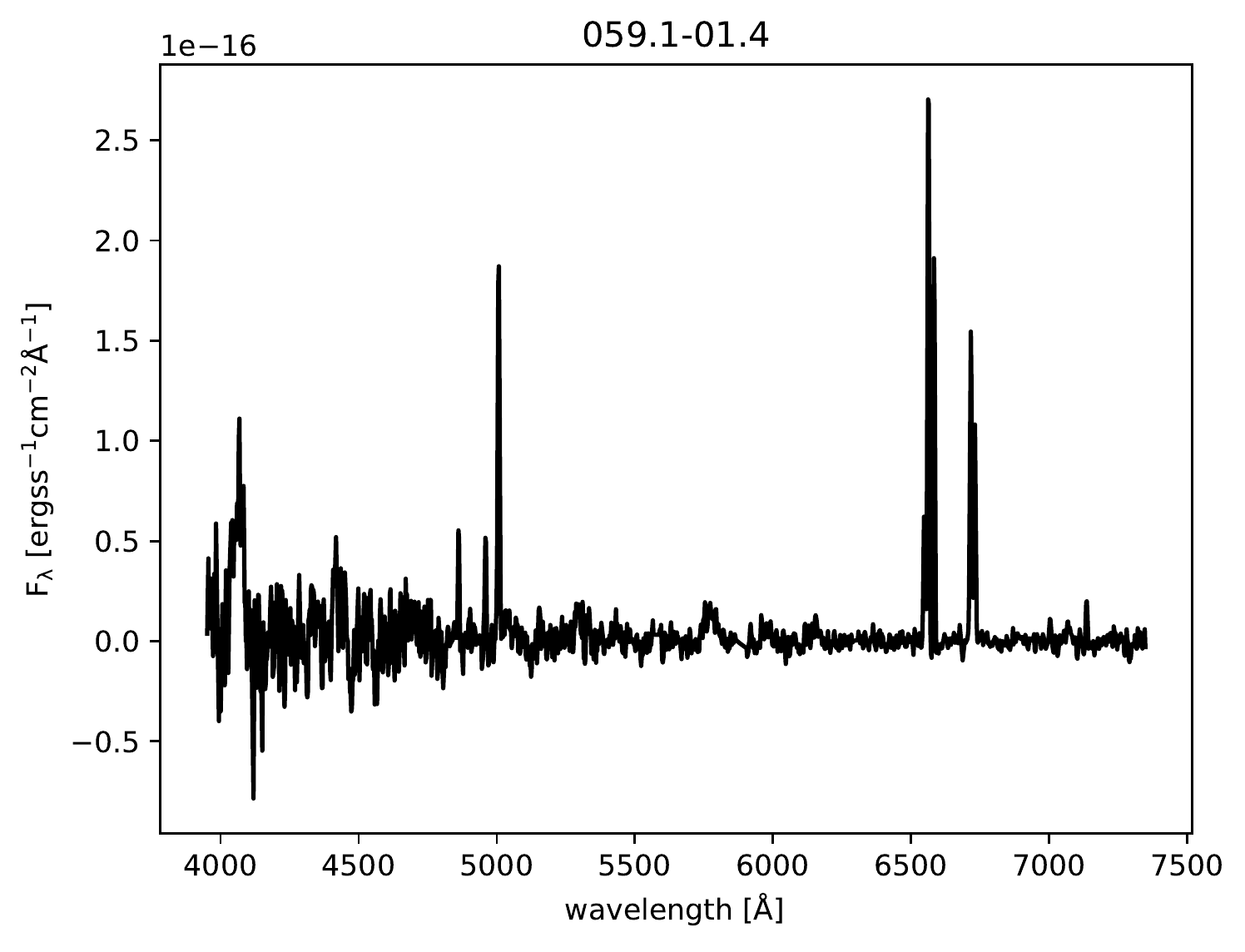}\\
 \includegraphics[width=0.48\textwidth]{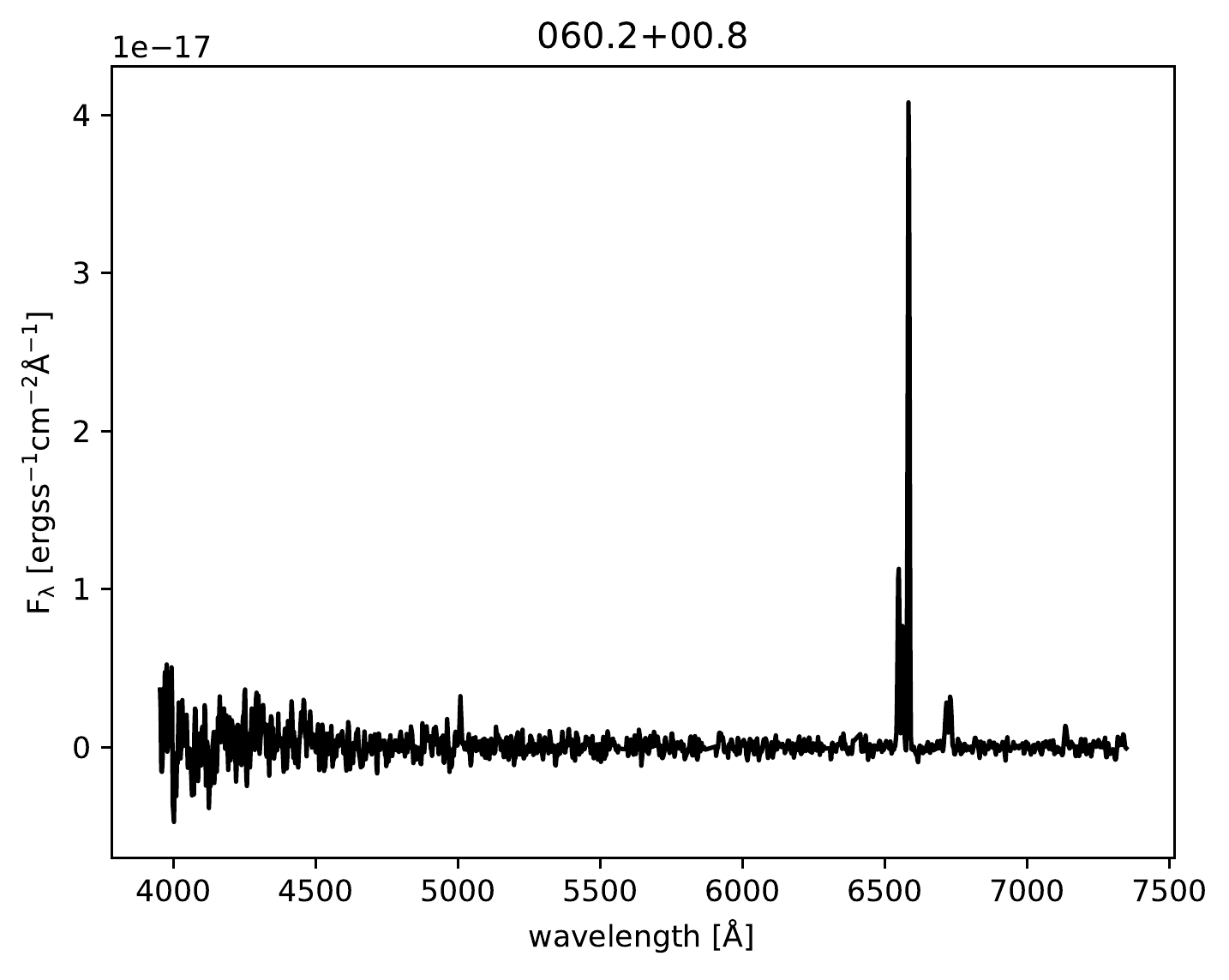}& \includegraphics[width=0.48\textwidth]{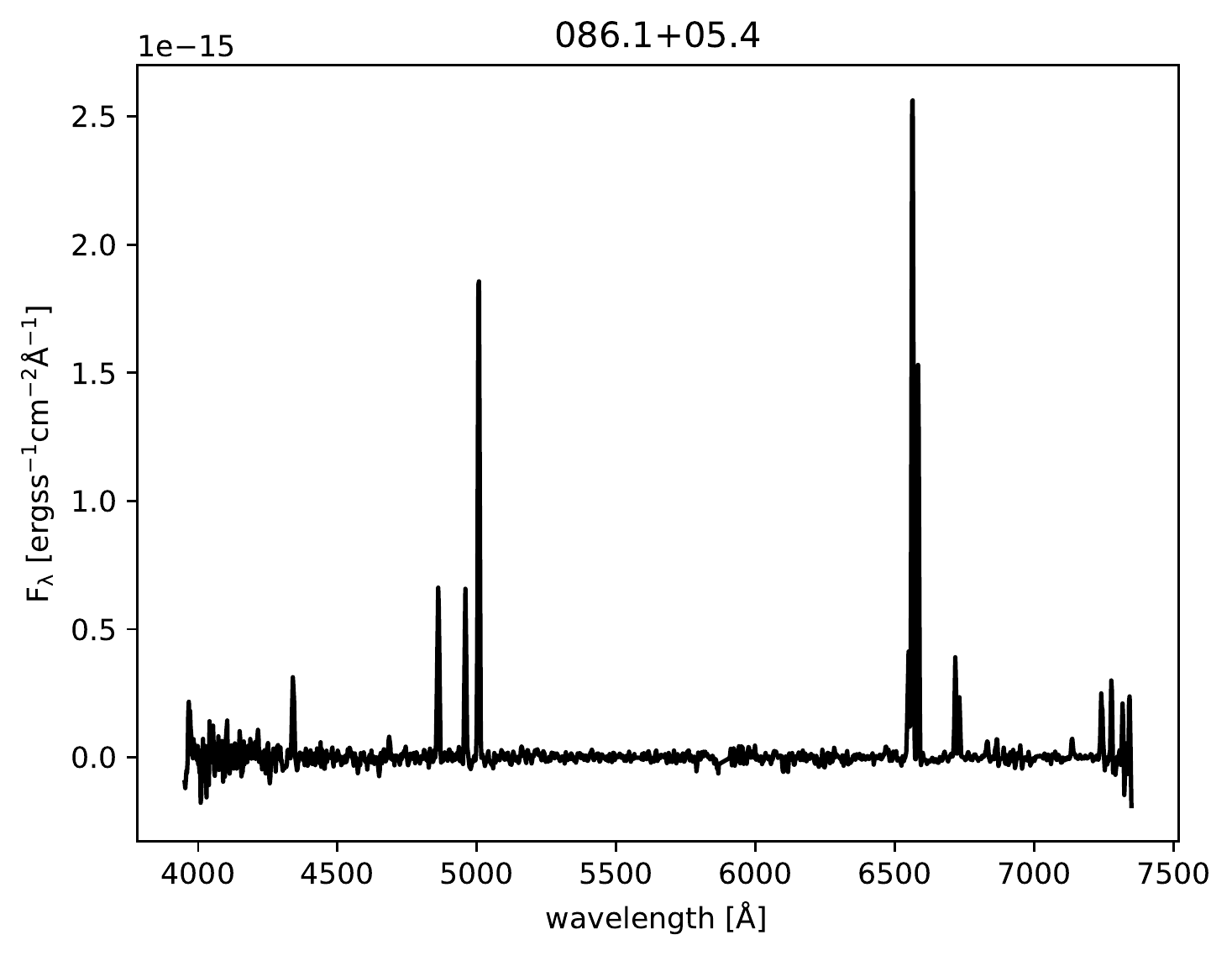}\\
 \includegraphics[width=0.48\textwidth]{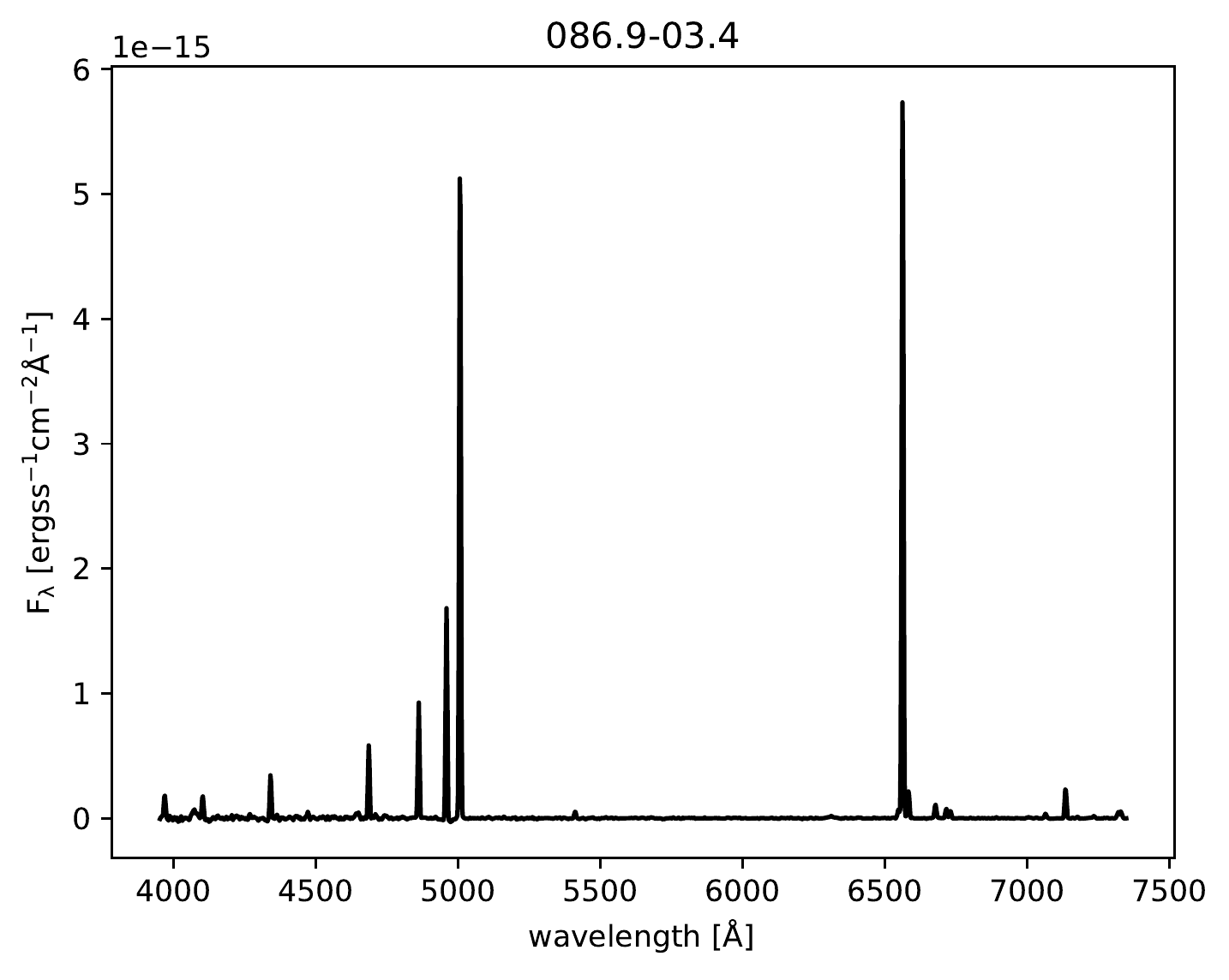}&\includegraphics[width=0.48\textwidth]{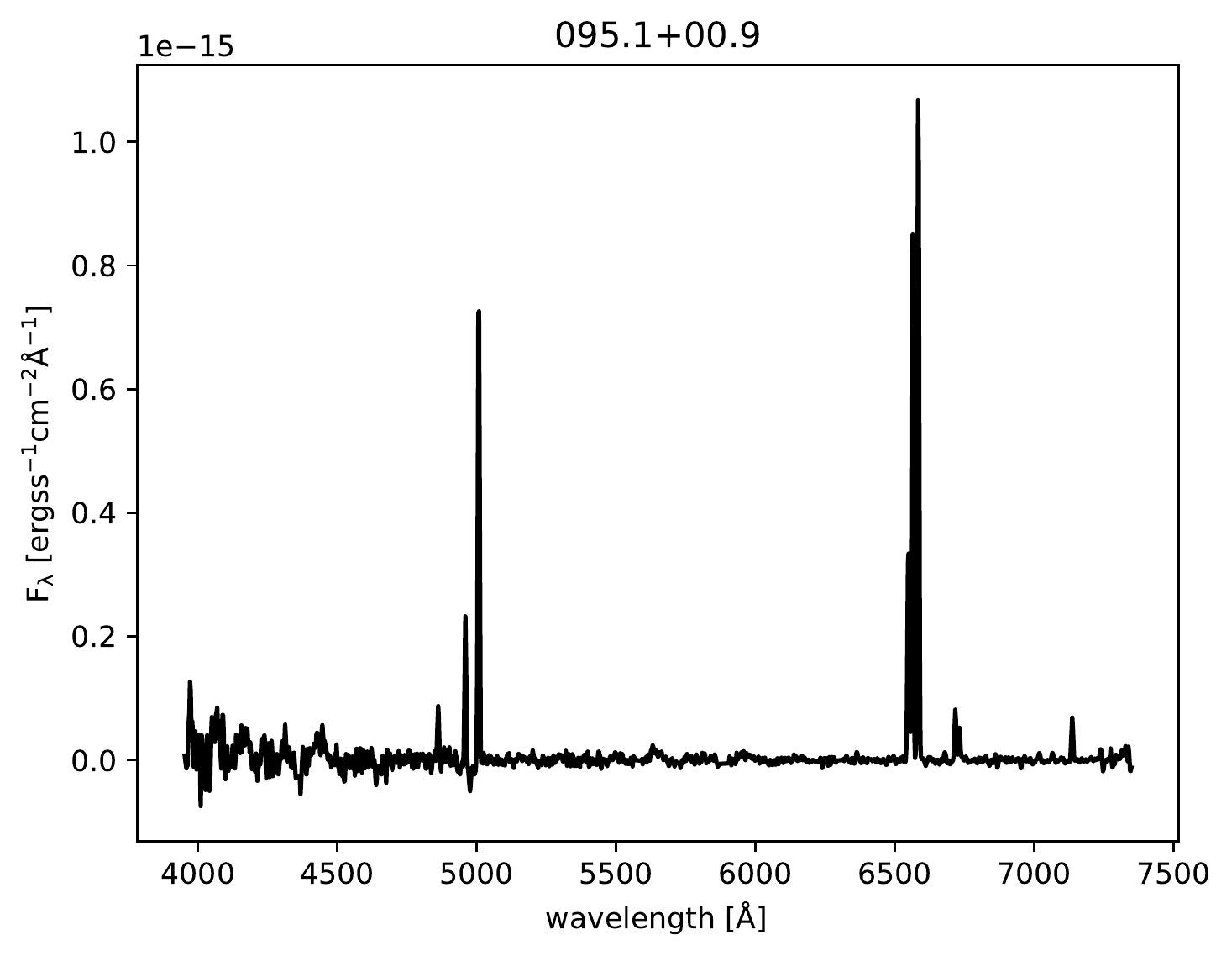}\\
\includegraphics[width=0.48\textwidth]{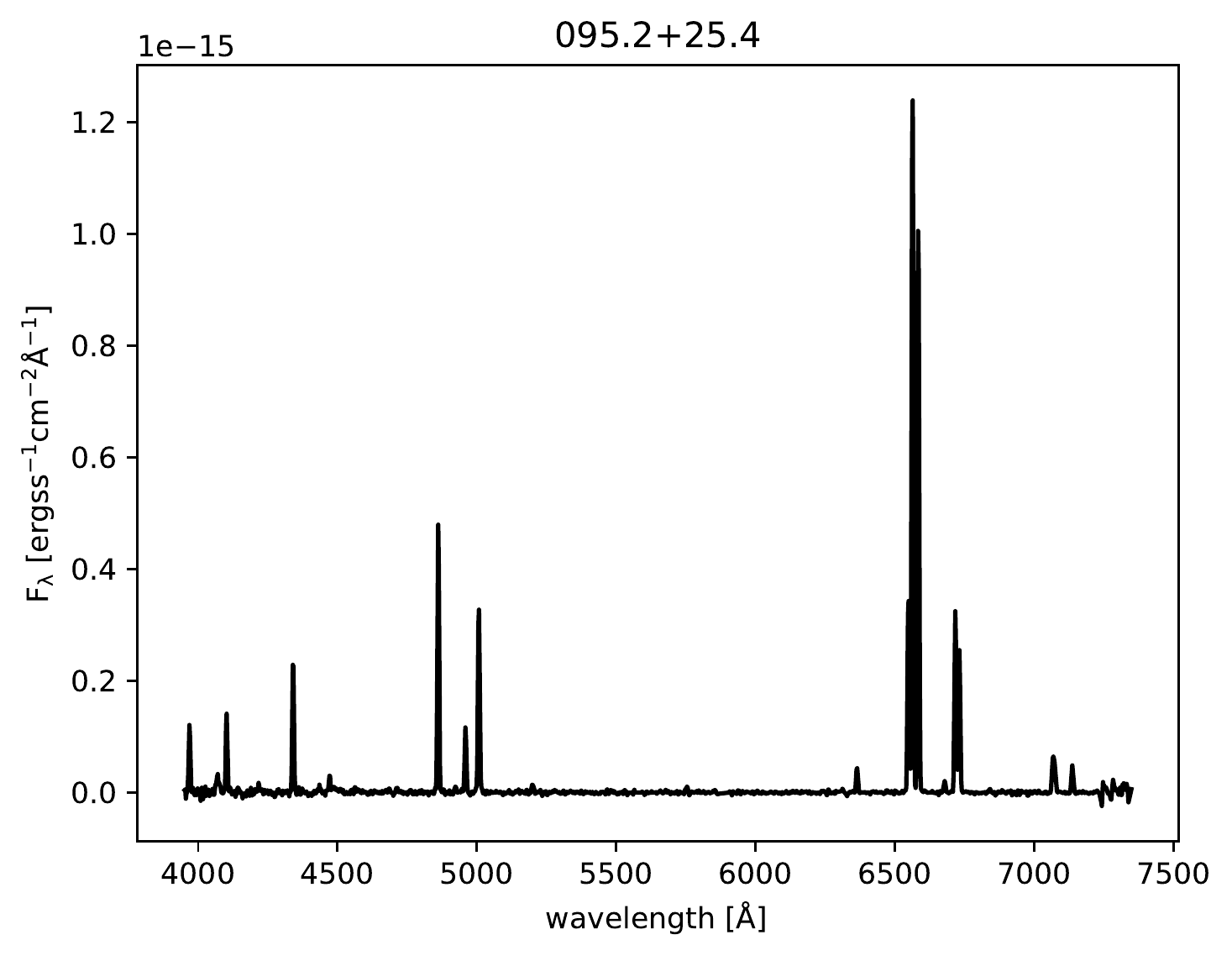}&\includegraphics[width=0.48\textwidth]{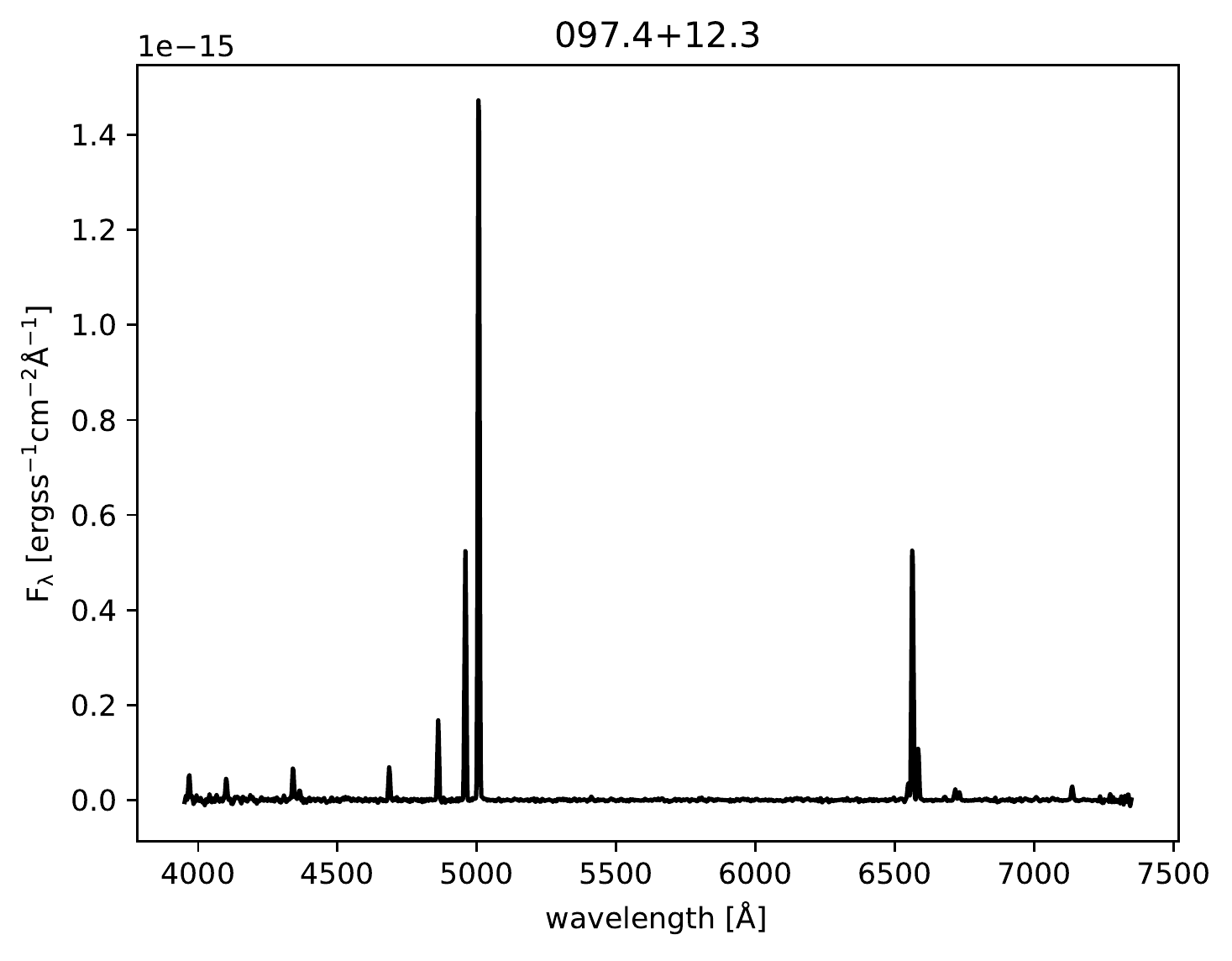}\\
 \includegraphics[width=0.48\textwidth]{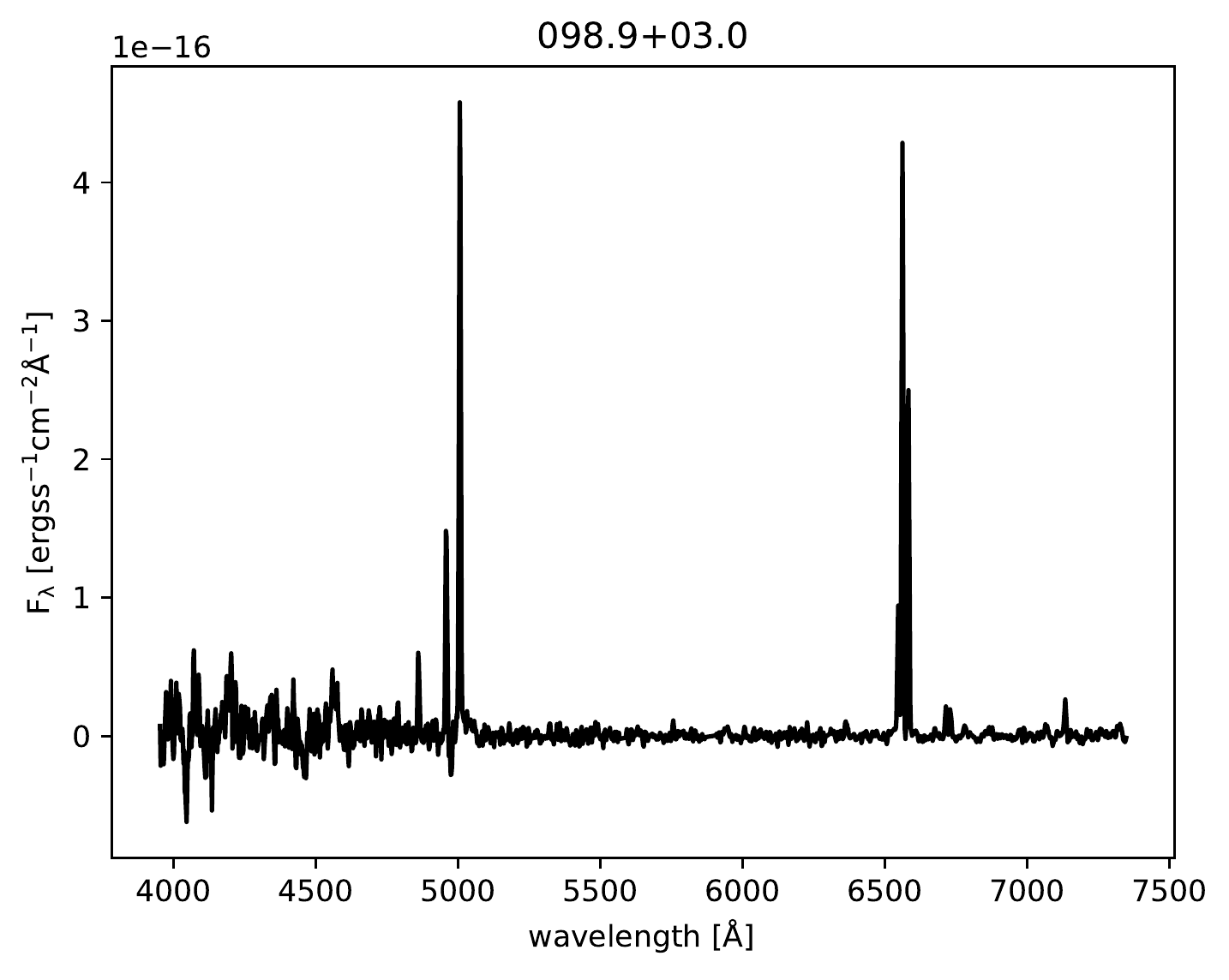}&\includegraphics[width=0.48\textwidth]{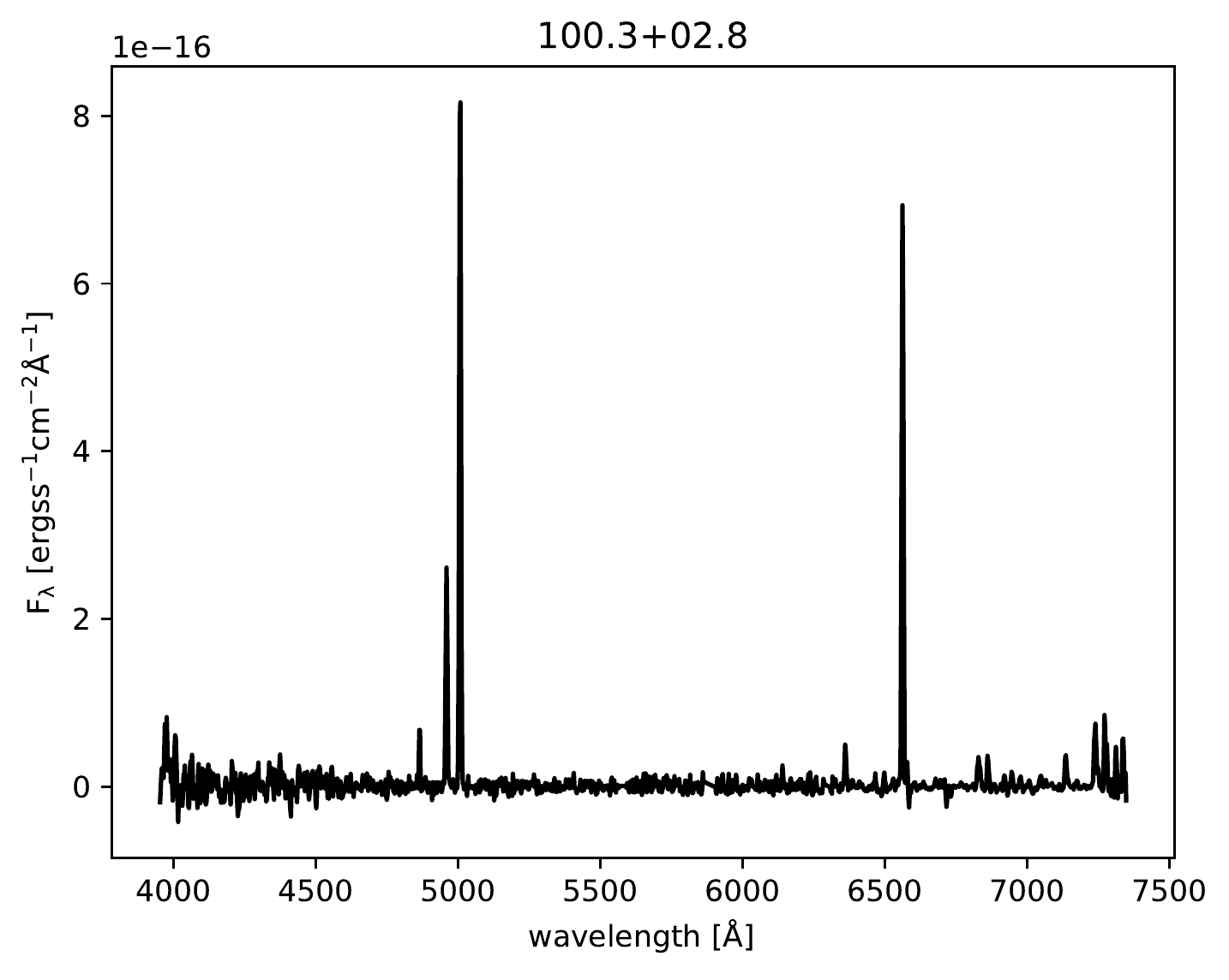}\\
\includegraphics[width=0.48\textwidth]{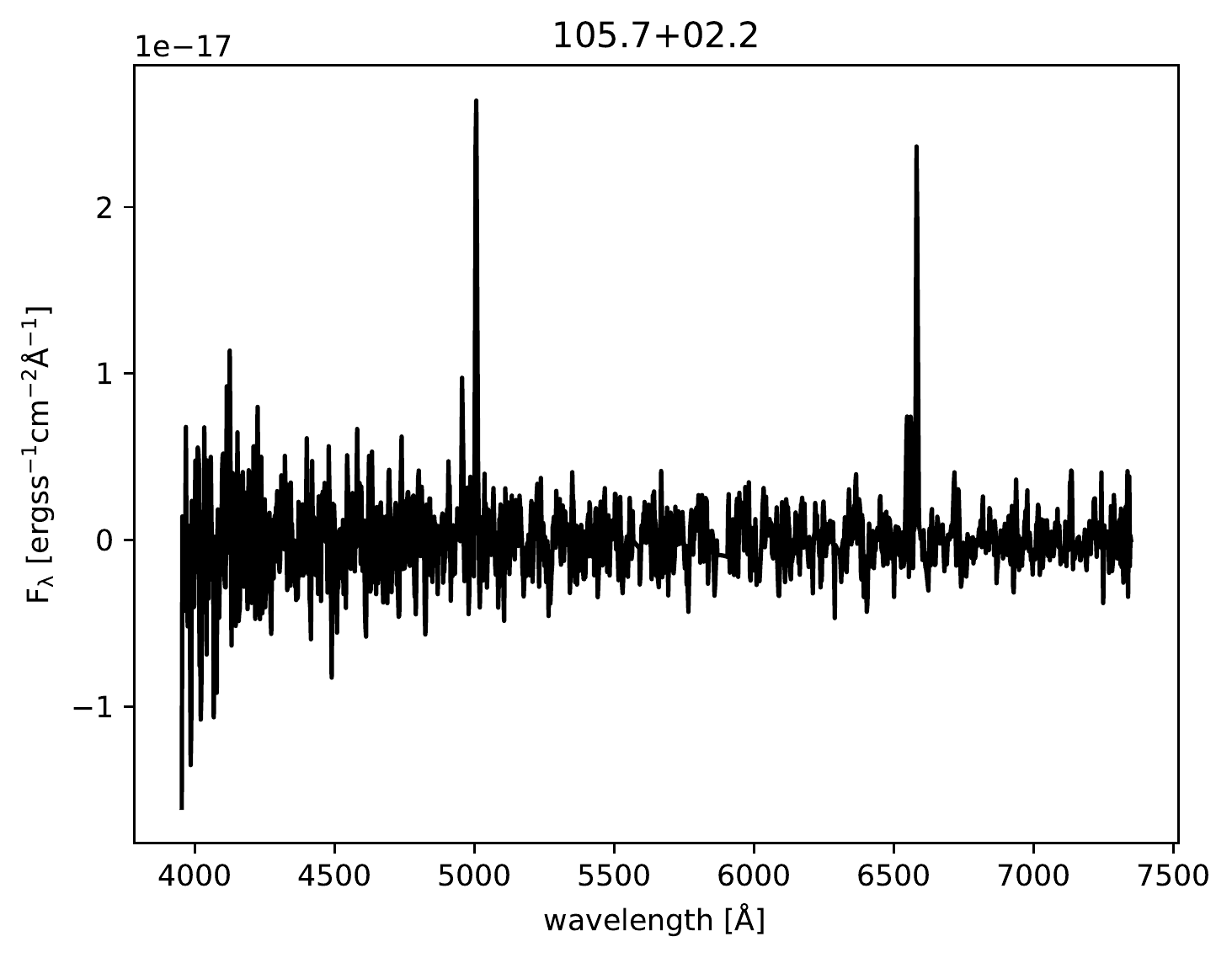}&\includegraphics[width=0.48\textwidth]{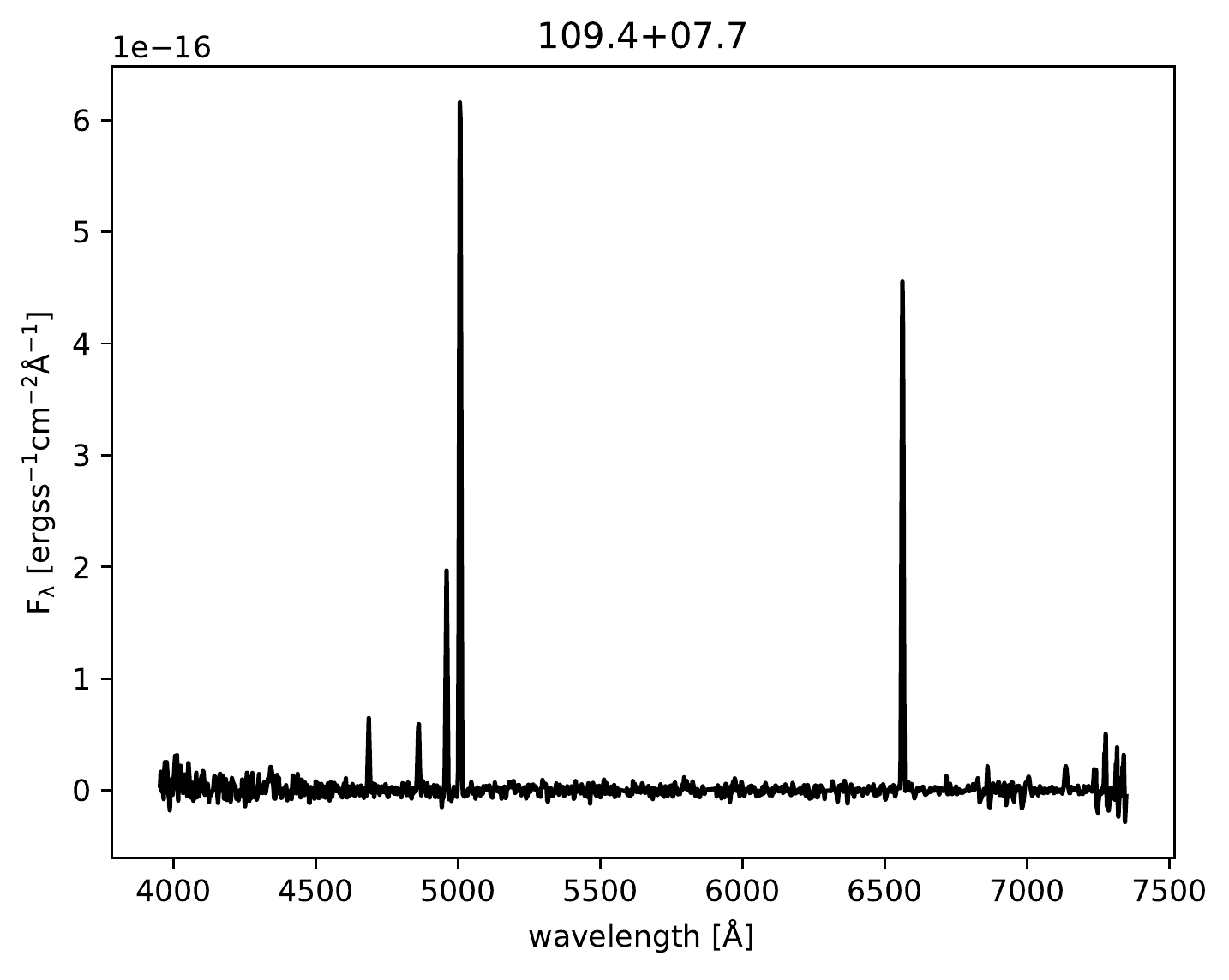}\\
\includegraphics[width=0.48\textwidth]{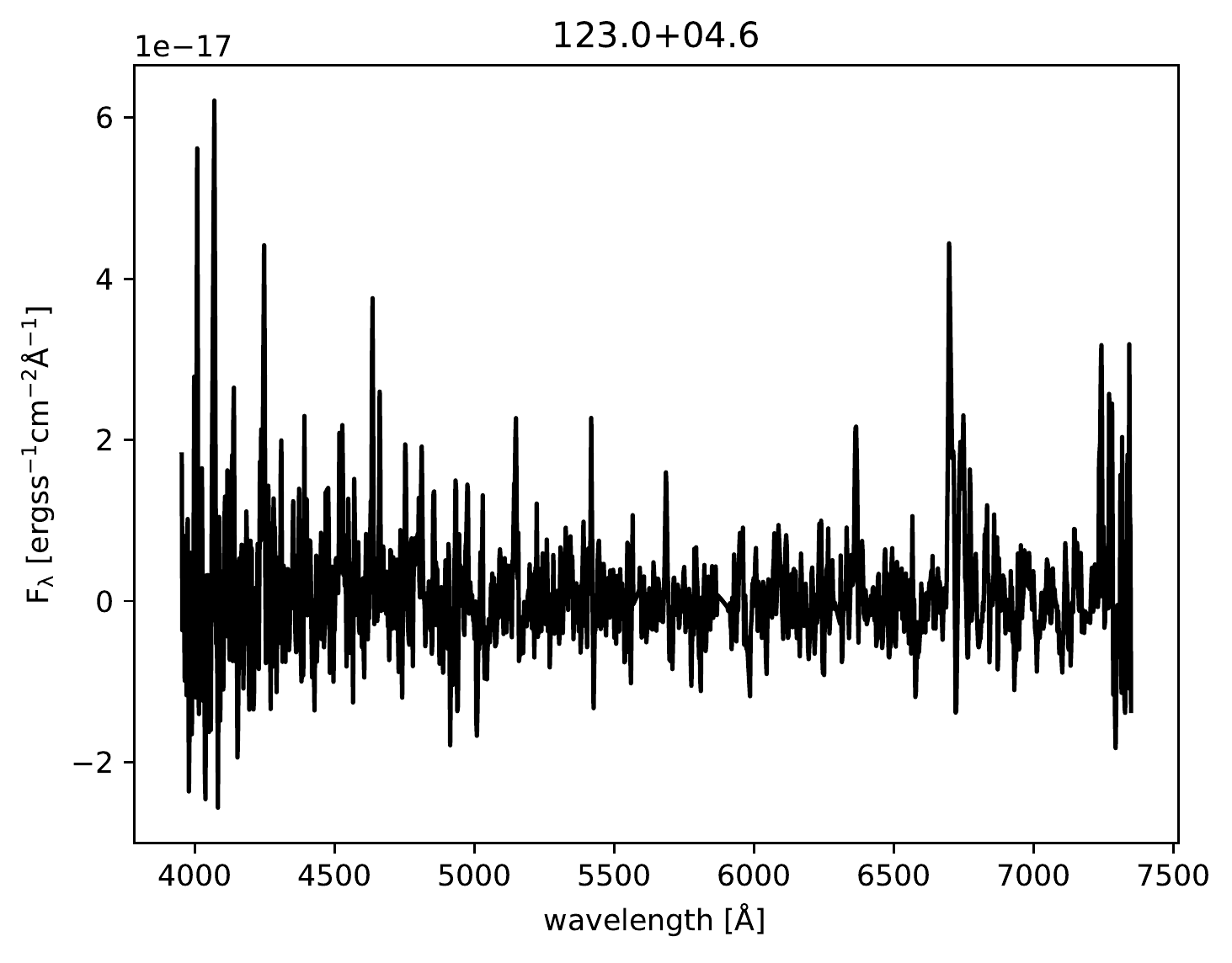}&\includegraphics[width=0.48\textwidth]{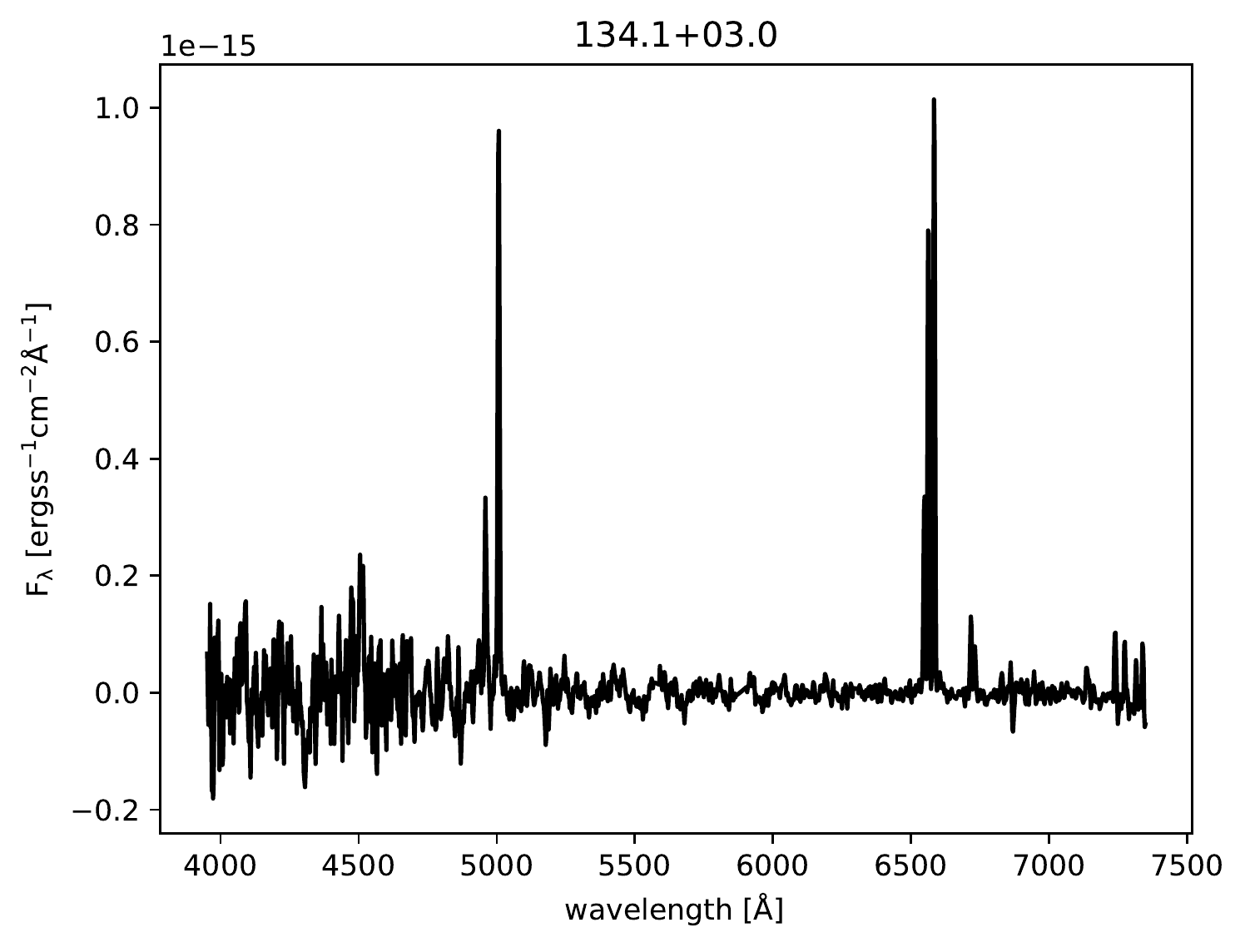}\\
\includegraphics[width=0.48\textwidth]{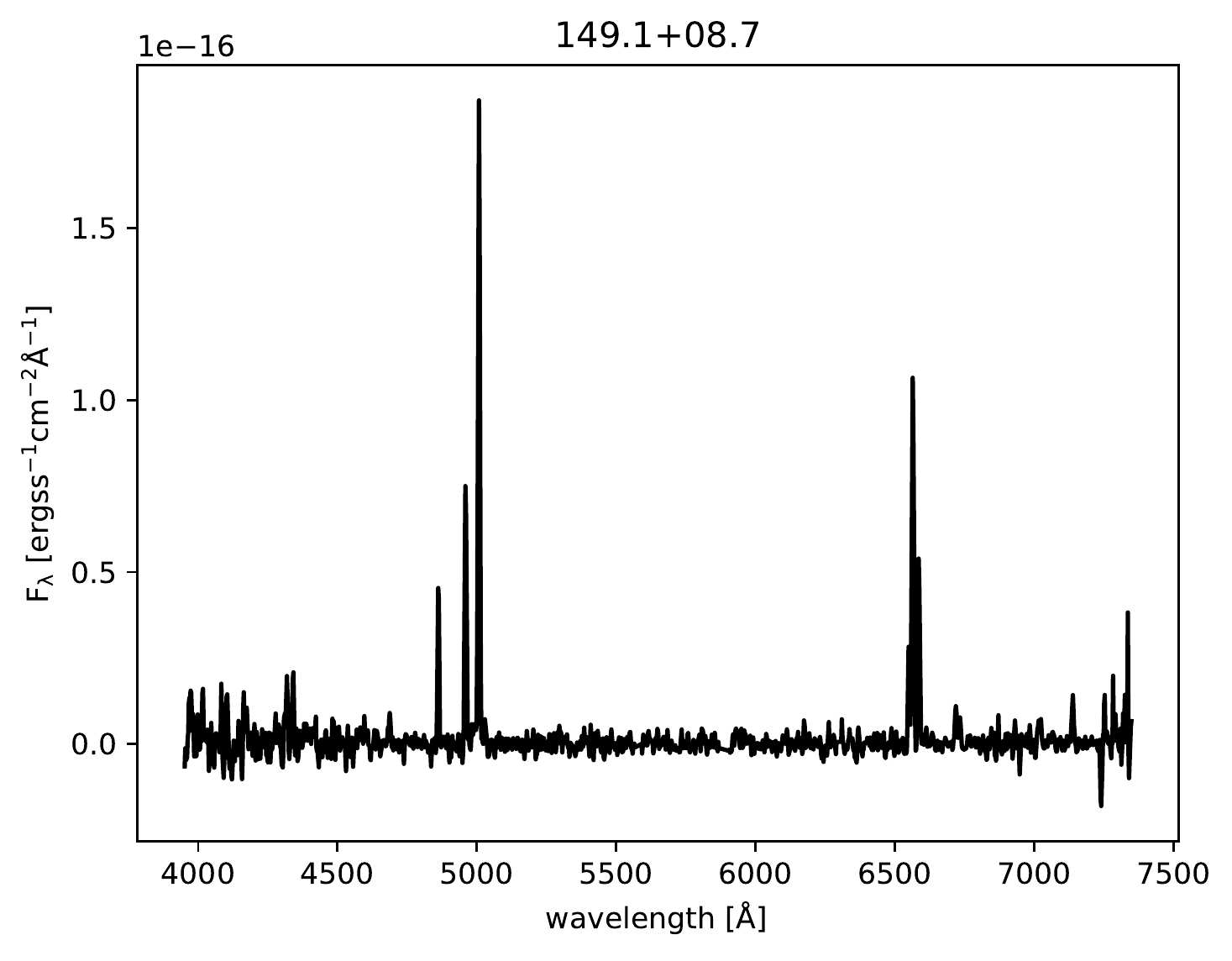}&\includegraphics[width=0.48\textwidth]{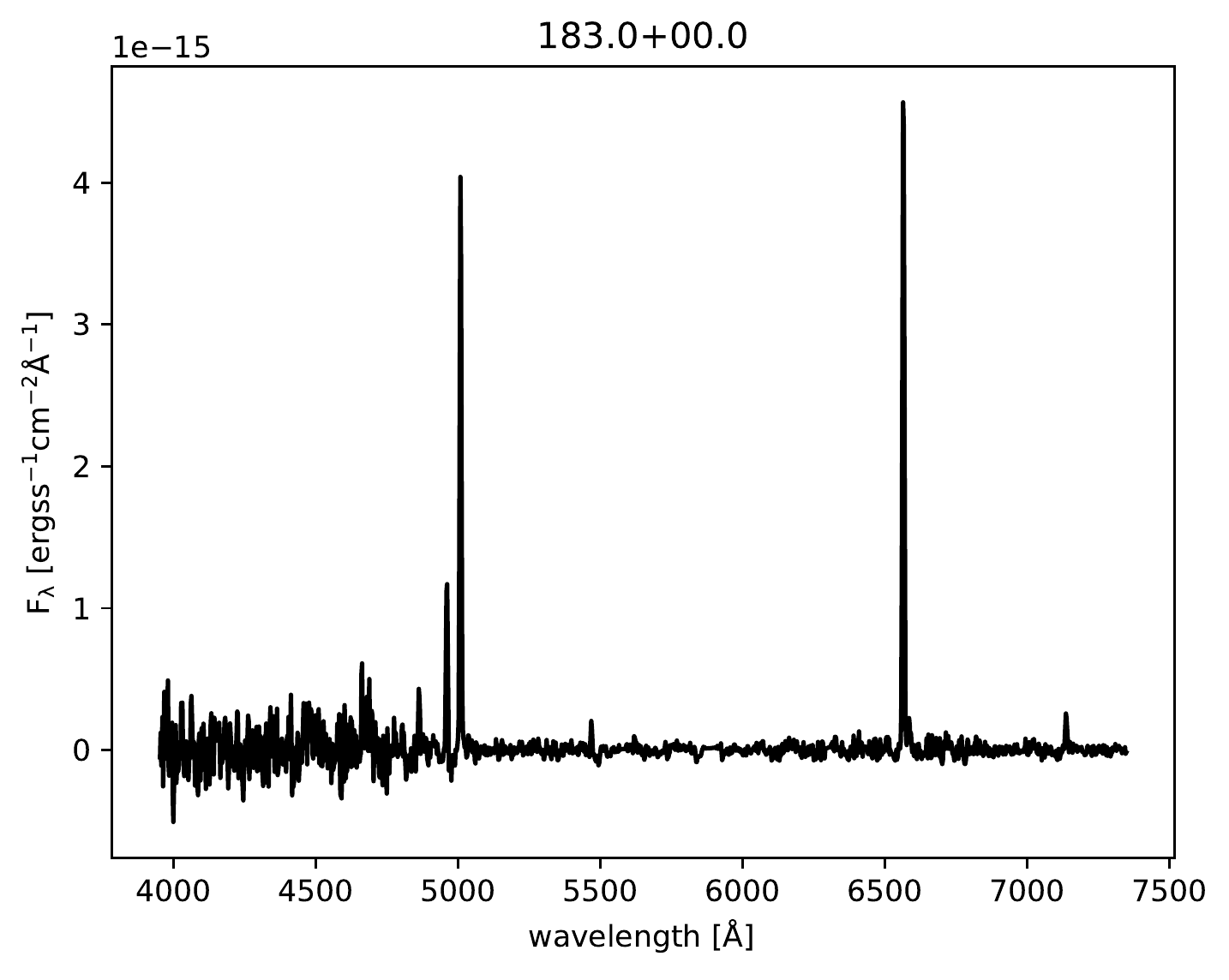}\\
\end{longtable}

\clearpage
\onecolumn
    \begin{longtable}{ | *{7}{l|} }
        \caption{\edited{Images of the 55 newly confirmed PNe. North is up and East is left. The circles mark the major angular diameters given in the HASH database.}}
        \label{tab:images1}\\
        \hline
        \edited{IAU PNG} & HASH ID & optical & $\mathrm{H_\alpha/Sr}$ & WISE321 & NVSS & GALEX\\
        \endhead  
        \hline
\edited{037.6-04.7}& 2502 & \includegraphics[width=19.5mm,height=19.5mm]{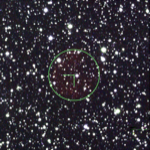} &\includegraphics[width=19.5mm,height=19.5mm]{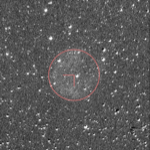} &\includegraphics[width=19.5mm,height=19.5mm]{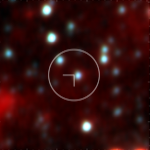} &\includegraphics[width=19.5mm,height=19.5mm]{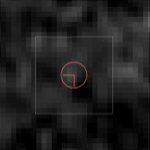} &\\
\hline
\edited{037.9-03.4}&390& \includegraphics[width=19.5mm,height=19.5mm]{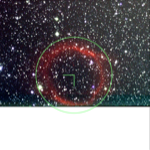} &\includegraphics[width=19.5mm,height=19.5mm]{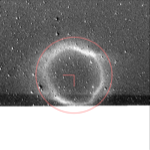} &\includegraphics[width=19.5mm,height=19.5mm]{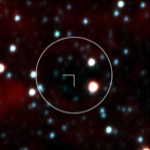} &\includegraphics[width=19.5mm,height=19.5mm]{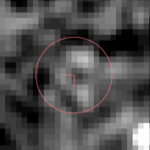} &\\
\hline
\edited{038.7-02.4}& 8193 & \includegraphics[width=19.5mm,height=19.5mm]{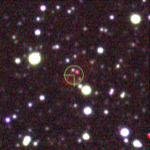} &\includegraphics[width=19.5mm,height=19.5mm]{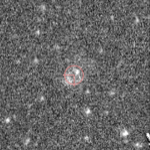} &\includegraphics[width=19.5mm,height=19.5mm]{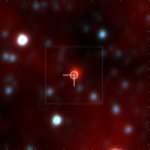} &\includegraphics[width=19.5mm,height=19.5mm]{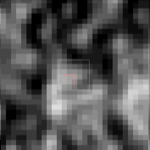} &\includegraphics[width=19.5mm,height=19.5mm]{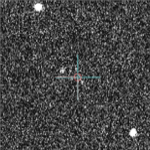} \\
\hline
\edited{040.5-00.0}& 8190 & \includegraphics[width=19.5mm,height=19.5mm]{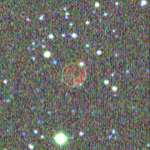} &\includegraphics[width=19.5mm,height=19.5mm]{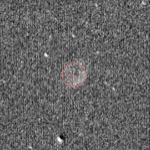} &\includegraphics[width=19.5mm,height=19.5mm]{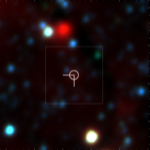} &\includegraphics[width=19.5mm,height=19.5mm]{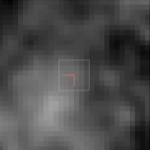} &\includegraphics[width=19.5mm,height=19.5mm]{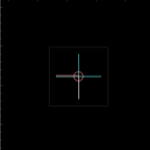} \\
\hline
\edited{040.6-01.5}& 8528 & \includegraphics[width=19.5mm,height=19.5mm]{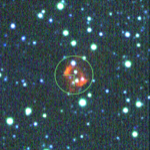} &\includegraphics[width=19.5mm,height=19.5mm]{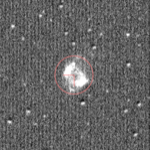} &\includegraphics[width=19.5mm,height=19.5mm]{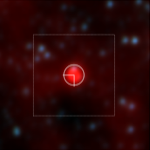} & &\includegraphics[width=19.5mm,height=19.5mm]{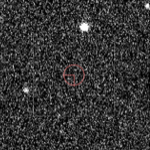} \\
\hline
\edited{043.8+02.1}& 8506 & \includegraphics[width=19.5mm,height=19.5mm]{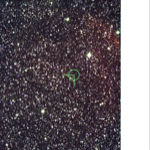} &\includegraphics[width=19.5mm,height=19.5mm]{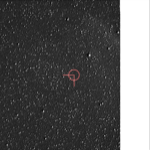} &\includegraphics[width=19.5mm,height=19.5mm]{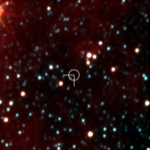} &\includegraphics[width=19.5mm,height=19.5mm]{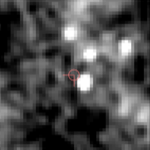} &\\
\hline
\edited{051.3+01.8}& 452 & \includegraphics[width=19.5mm,height=19.5mm]{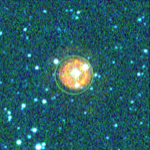} &\includegraphics[width=19.5mm,height=19.5mm]{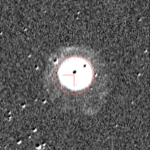} &\includegraphics[width=19.5mm,height=19.5mm]{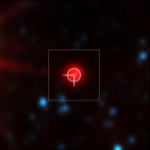} &\includegraphics[width=19.5mm,height=19.5mm]{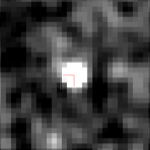} &\\
\hline
\edited{057.6+01.8}& 4819 & \includegraphics[width=19.5mm,height=19.5mm]{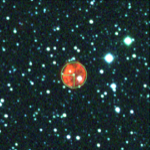} &\includegraphics[width=19.5mm,height=19.5mm]{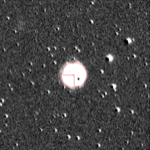} &\includegraphics[width=19.5mm,height=19.5mm]{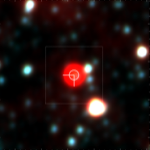} &\includegraphics[width=19.5mm,height=19.5mm]{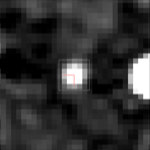} &\\
\hline
\edited{058.1-00.8}& 8566 & \includegraphics[width=19.5mm,height=19.5mm]{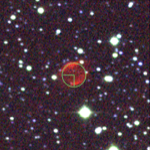} &\includegraphics[width=19.5mm,height=19.5mm]{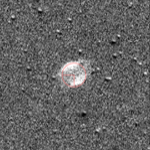} &\includegraphics[width=19.5mm,height=19.5mm]{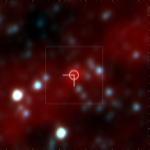} &\includegraphics[width=19.5mm,height=19.5mm]{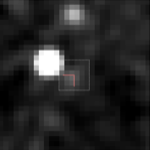} &\\
\hline
\edited{058.9+09.0}& 486 & \includegraphics[width=19.5mm,height=19.5mm]{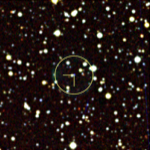} && \includegraphics[width=19.5mm,height=19.5mm]{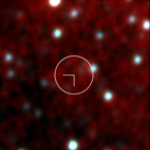} &\includegraphics[width=19.5mm,height=19.5mm]{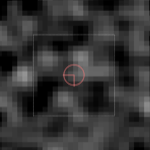} &\includegraphics[width=19.5mm,height=19.5mm]{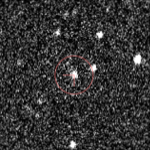} \\
\hline
\edited{059.2+01.0}& 10957 & \includegraphics[width=19.5mm,height=19.5mm]{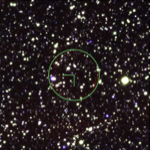} &\includegraphics[width=19.5mm,height=19.5mm]{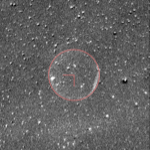} &\includegraphics[width=19.5mm,height=19.5mm]{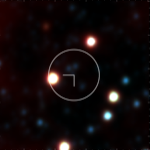} &\includegraphics[width=19.5mm,height=19.5mm]{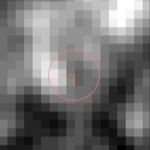} &\\
\hline
\edited{060.0-04.3}&495& \includegraphics[width=19.5mm,height=19.5mm]{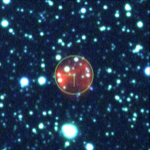} &\includegraphics[width=19.5mm,height=19.5mm]{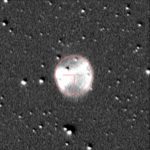} &\includegraphics[width=19.5mm,height=19.5mm]{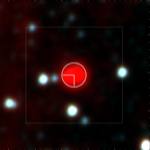} &\includegraphics[width=19.5mm,height=19.5mm]{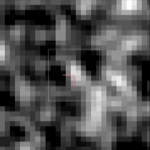} &\\
\hline
\edited{060.5+05.6}& 15561 & \includegraphics[width=19.5mm,height=19.5mm]{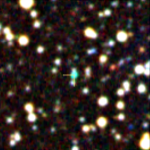} && \includegraphics[width=19.5mm,height=19.5mm]{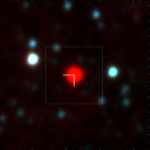} &\includegraphics[width=19.5mm,height=19.5mm]{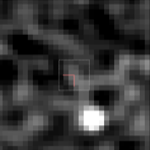} &\\
\hline
\edited{062.1+03.1}& 8206 & \includegraphics[width=19.5mm,height=19.5mm]{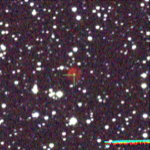} &\includegraphics[width=19.5mm,height=19.5mm]{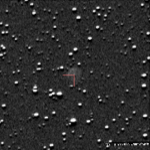} &\includegraphics[width=19.5mm,height=19.5mm]{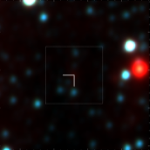} &\includegraphics[width=19.5mm,height=19.5mm]{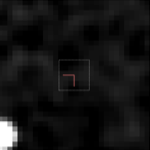} &\\
\hline
\edited{062.4+00.6}& 8214 & \includegraphics[width=19.5mm,height=19.5mm]{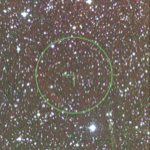} &\includegraphics[width=19.5mm,height=19.5mm]{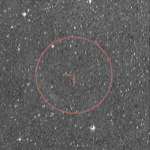} &\includegraphics[width=19.5mm,height=19.5mm]{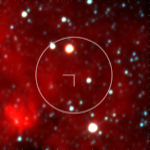} &\includegraphics[width=19.5mm,height=19.5mm]{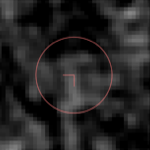} &\\
\hline
\edited{062.5-01.8}& 8219 & \includegraphics[width=19.5mm,height=19.5mm]{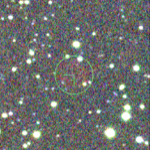} &\includegraphics[width=19.5mm,height=19.5mm]{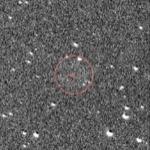} &\includegraphics[width=19.5mm,height=19.5mm]{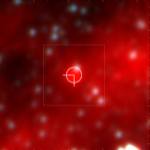} &\includegraphics[width=19.5mm,height=19.5mm]{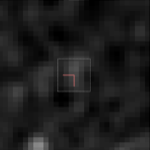} &\\
\hline
\edited{063.1+00.8}& 8217 & \includegraphics[width=19.5mm,height=19.5mm]{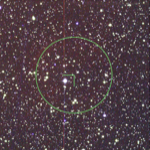} &\includegraphics[width=19.5mm,height=19.5mm]{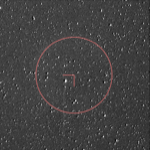} &\includegraphics[width=19.5mm,height=19.5mm]{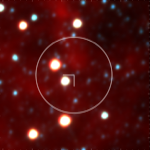} &\includegraphics[width=19.5mm,height=19.5mm]{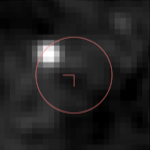} &\\
\hline
\edited{064.9-09.1a}& 15551 & \includegraphics[width=19.5mm,height=19.5mm]{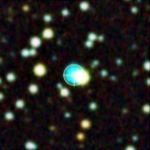} && \includegraphics[width=19.5mm,height=19.5mm]{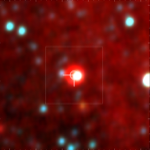} &\includegraphics[width=19.5mm,height=19.5mm]{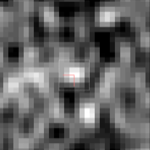} &\includegraphics[width=19.5mm,height=19.5mm]{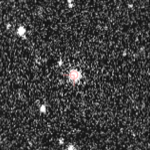} \\
\hline
\edited{066.1+04.7}& 8210 & \includegraphics[width=19.5mm,height=19.5mm]{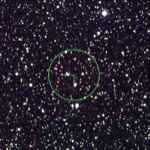} &\includegraphics[width=19.5mm,height=19.5mm]{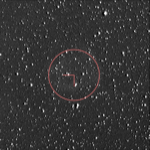} &\includegraphics[width=19.5mm,height=19.5mm]{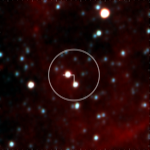} &\includegraphics[width=19.5mm,height=19.5mm]{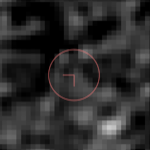} &\\
\hline
\edited{066.5-14.8}& 4359 & \includegraphics[width=19.5mm,height=19.5mm]{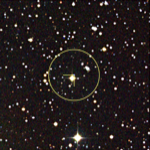} && \includegraphics[width=19.5mm,height=19.5mm]{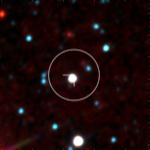} &\includegraphics[width=19.5mm,height=19.5mm]{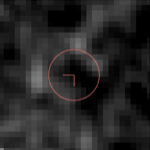} &\includegraphics[width=19.5mm,height=19.5mm]{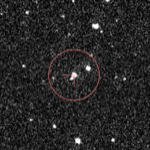} \\
\hline
\edited{066.9-07.8}& 4356 & \includegraphics[width=19.5mm,height=19.5mm]{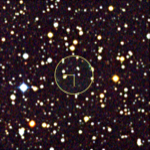} && \includegraphics[width=19.5mm,height=19.5mm]{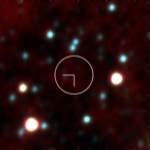} &\includegraphics[width=19.5mm,height=19.5mm]{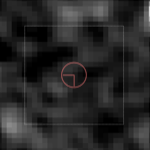} &\includegraphics[width=19.5mm,height=19.5mm]{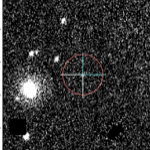} \\
\hline
\edited{067.3-02.6}& 8232 & \includegraphics[width=19.5mm,height=19.5mm]{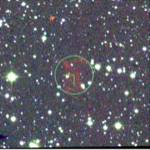} &\includegraphics[width=19.5mm,height=19.5mm]{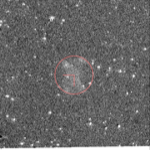} &\includegraphics[width=19.5mm,height=19.5mm]{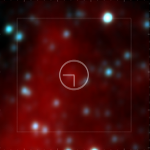} &\includegraphics[width=19.5mm,height=19.5mm]{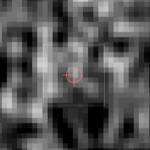} &\\
\hline
\edited{069.6-03.7}& 10881 & \includegraphics[width=19.5mm,height=19.5mm]{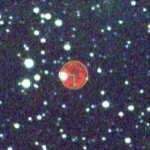} &\includegraphics[width=19.5mm,height=19.5mm]{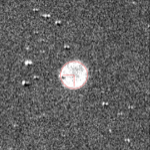} &\includegraphics[width=19.5mm,height=19.5mm]{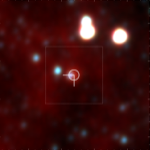} &\includegraphics[width=19.5mm,height=19.5mm]{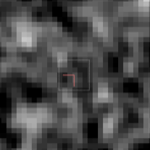} &\includegraphics[width=19.5mm,height=19.5mm]{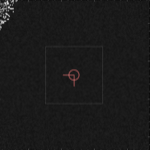} \\
\hline
\edited{070.5+11.0}& 10899 & \includegraphics[width=19.5mm,height=19.5mm]{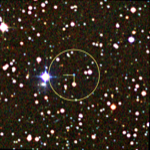} && \includegraphics[width=19.5mm,height=19.5mm]{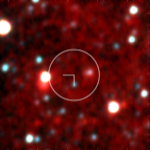} &\includegraphics[width=19.5mm,height=19.5mm]{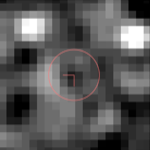} &\\
\hline
\edited{073.4+01.5}& 8230 & \includegraphics[width=19.5mm,height=19.5mm]{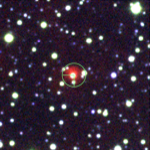} &\includegraphics[width=19.5mm,height=19.5mm]{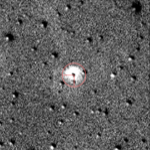} &\includegraphics[width=19.5mm,height=19.5mm]{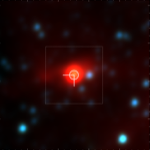} &\includegraphics[width=19.5mm,height=19.5mm]{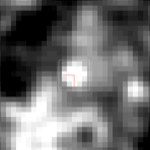} &\\
\hline
\edited{075.0-07.2}& 15564 & \includegraphics[width=19.5mm,height=19.5mm]{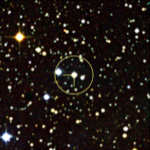} && \includegraphics[width=19.5mm,height=19.5mm]{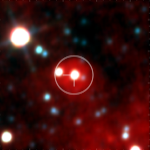} &\includegraphics[width=19.5mm,height=19.5mm]{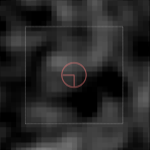} &\includegraphics[width=19.5mm,height=19.5mm]{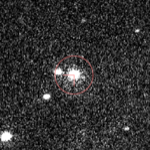} \\
\hline
\edited{075.3+05.5}& 15565 & \includegraphics[width=19.5mm,height=19.5mm]{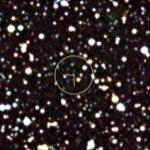} && \includegraphics[width=19.5mm,height=19.5mm]{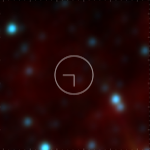} &\includegraphics[width=19.5mm,height=19.5mm]{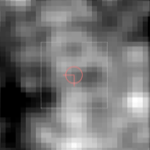} &\\
\hline
\edited{075.5+01.7}& 4408 & \includegraphics[width=19.5mm,height=19.5mm]{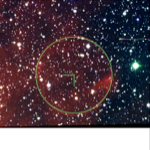} &\includegraphics[width=19.5mm,height=19.5mm]{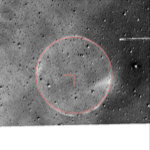} &\includegraphics[width=19.5mm,height=19.5mm]{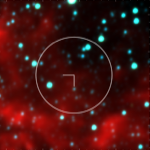} &\includegraphics[width=19.5mm,height=19.5mm]{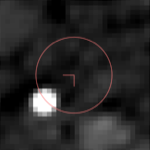} &\\
\hline
\edited{076.8-08.1}& 15566 & \includegraphics[width=19.5mm,height=19.5mm]{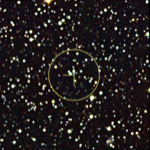} && \includegraphics[width=19.5mm,height=19.5mm]{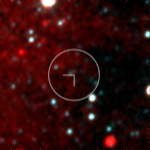} &\includegraphics[width=19.5mm,height=19.5mm]{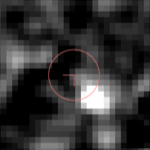} &\includegraphics[width=19.5mm,height=19.5mm]{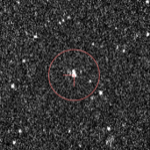} \\
\hline
\edited{078.4-07.2}& 15567 & \includegraphics[width=19.5mm,height=19.5mm]{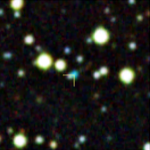} && \includegraphics[width=19.5mm,height=19.5mm]{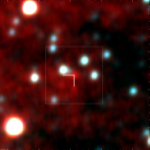} &\includegraphics[width=19.5mm,height=19.5mm]{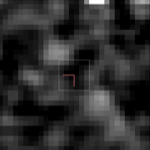} &\\
\hline
\edited{079.8-10.2}&10960& \includegraphics[width=19.5mm,height=19.5mm]{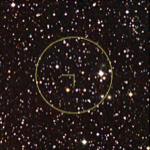} && \includegraphics[width=19.5mm,height=19.5mm]{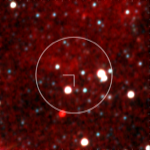} &\includegraphics[width=19.5mm,height=19.5mm]{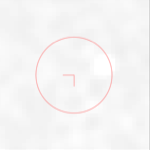} &\includegraphics[width=19.5mm,height=19.5mm]{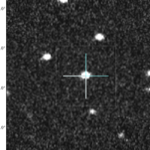} \\
\hline
\edited{082.1-07.8}& 560 & \includegraphics[width=19.5mm,height=19.5mm]{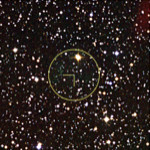} && \includegraphics[width=19.5mm,height=19.5mm]{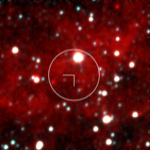} &\includegraphics[width=19.5mm,height=19.5mm]{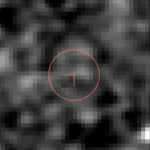} &\\
\hline
\edited{082.5-06.2}& 4362 & \includegraphics[width=19.5mm,height=19.5mm]{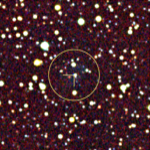} && \includegraphics[width=19.5mm,height=19.5mm]{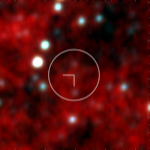} &\includegraphics[width=19.5mm,height=19.5mm]{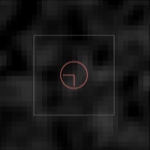} &\\
\hline
\edited{093.8-00.2}& 17066 & \includegraphics[width=19.5mm,height=19.5mm]{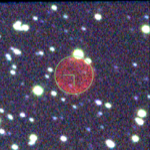} &\includegraphics[width=19.5mm,height=19.5mm]{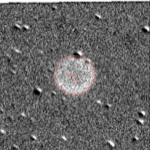} &\includegraphics[width=19.5mm,height=19.5mm]{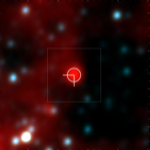} &\includegraphics[width=19.5mm,height=19.5mm]{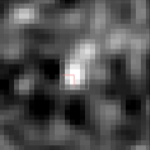} &\\
\hline
\edited{094.5-00.8a}& 10959 & \includegraphics[width=19.5mm,height=19.5mm]{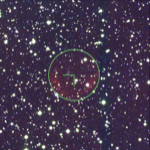} &\includegraphics[width=19.5mm,height=19.5mm]{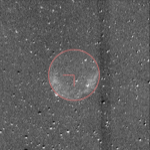} &\includegraphics[width=19.5mm,height=19.5mm]{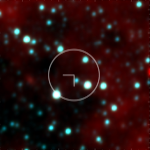} &\includegraphics[width=19.5mm,height=19.5mm]{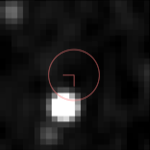} &\\
\hline
\edited{098.3-04.9}& 15568 & \includegraphics[width=19.5mm,height=19.5mm]{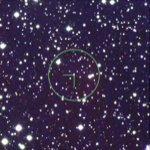} &\includegraphics[width=19.5mm,height=19.5mm]{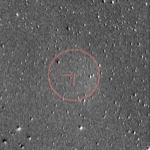} &\includegraphics[width=19.5mm,height=19.5mm]{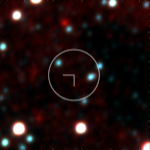} &\includegraphics[width=19.5mm,height=19.5mm]{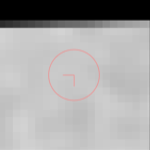} &\includegraphics[width=19.5mm,height=19.5mm]{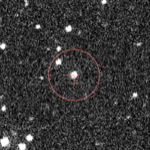} \\
\hline
\edited{099.1+05.7}& 4367 & \includegraphics[width=19.5mm,height=19.5mm]{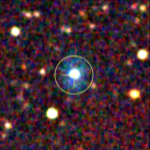} && \includegraphics[width=19.5mm,height=19.5mm]{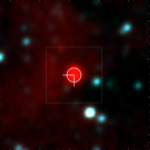} &\includegraphics[width=19.5mm,height=19.5mm]{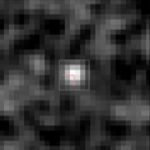} &\includegraphics[width=19.5mm,height=19.5mm]{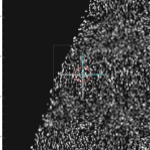} \\
\hline
\edited{099.7-08.8}& 602 & \includegraphics[width=19.5mm,height=19.5mm]{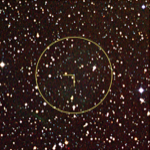} && \includegraphics[width=19.5mm,height=19.5mm]{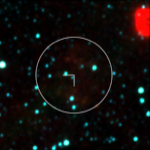} &\includegraphics[width=19.5mm,height=19.5mm]{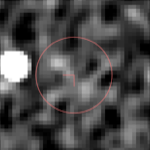} &\includegraphics[width=19.5mm,height=19.5mm]{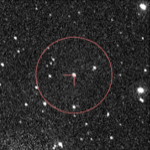} \\
\hline
\edited{103.7+07.2}& 4368 & \includegraphics[width=19.5mm,height=19.5mm]{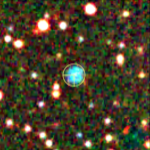} && \includegraphics[width=19.5mm,height=19.5mm]{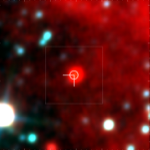} &\includegraphics[width=19.5mm,height=19.5mm]{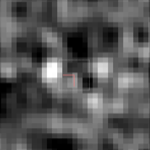} &\includegraphics[width=19.5mm,height=19.5mm]{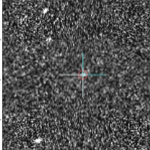} \\
\hline
\edited{107.0+21.3}& 617 & \includegraphics[width=19.5mm,height=19.5mm]{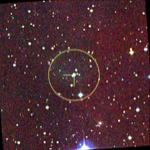} && \includegraphics[width=19.5mm,height=19.5mm]{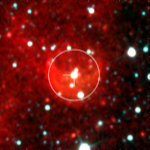} &\includegraphics[width=19.5mm,height=19.5mm]{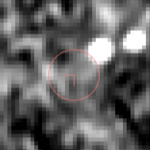} &\includegraphics[width=19.5mm,height=19.5mm]{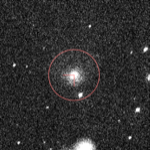} \\
\hline
\edited{111.2-03.0}& 8266 & \includegraphics[width=19.5mm,height=19.5mm]{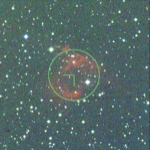} &\includegraphics[width=19.5mm,height=19.5mm]{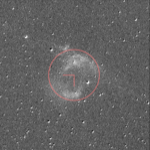} &\includegraphics[width=19.5mm,height=19.5mm]{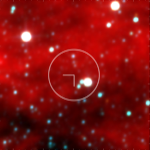} &\includegraphics[width=19.5mm,height=19.5mm]{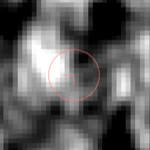} &\includegraphics[width=19.5mm,height=19.5mm]{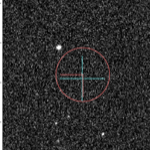} \\
\hline
\edited{120.4-01.3}& 10956 & \includegraphics[width=19.5mm,height=19.5mm]{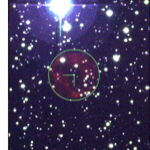} &\includegraphics[width=19.5mm,height=19.5mm]{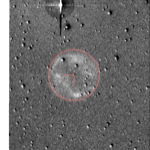} &\includegraphics[width=19.5mm,height=19.5mm]{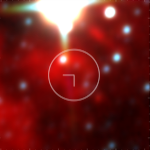} &\includegraphics[width=19.5mm,height=19.5mm]{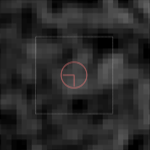} &\includegraphics[width=19.5mm,height=19.5mm]{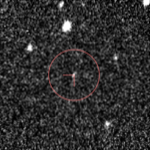} \\
\hline
\edited{129.2-02.0}& 655 & \includegraphics[width=19.5mm,height=19.5mm]{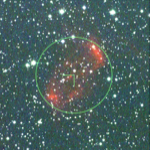} &\includegraphics[width=19.5mm,height=19.5mm]{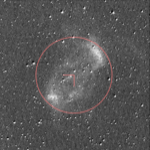} &\includegraphics[width=19.5mm,height=19.5mm]{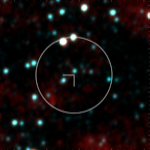} &\includegraphics[width=19.5mm,height=19.5mm]{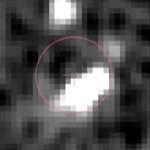} &\includegraphics[width=19.5mm,height=19.5mm]{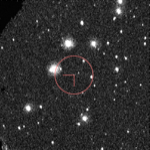} \\
\hline
\edited{129.6-05.6}& 658 & \includegraphics[width=19.5mm,height=19.5mm]{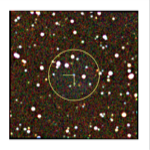} && \includegraphics[width=19.5mm,height=19.5mm]{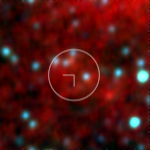} &\includegraphics[width=19.5mm,height=19.5mm]{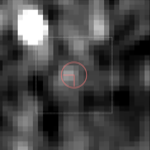} &\includegraphics[width=19.5mm,height=19.5mm]{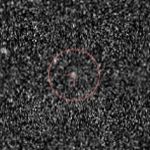} \\
\hline
\edited{136.8-13.2}& 10896 & \includegraphics[width=19.5mm,height=19.5mm]{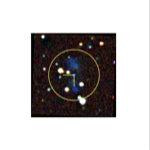} && \includegraphics[width=19.5mm,height=19.5mm]{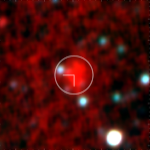} &\includegraphics[width=19.5mm,height=19.5mm]{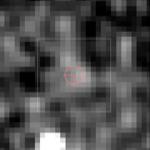} &\includegraphics[width=19.5mm,height=19.5mm]{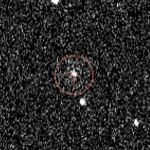} \\
\hline
\edited{138.1+04.1}& 670 & \includegraphics[width=19.5mm,height=19.5mm]{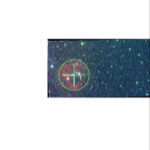} &\includegraphics[width=19.5mm,height=19.5mm]{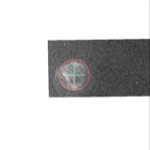} &\includegraphics[width=19.5mm,height=19.5mm]{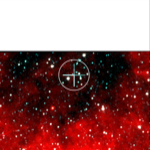} &\includegraphics[width=19.5mm,height=19.5mm]{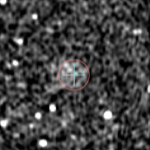} &\includegraphics[width=19.5mm,height=19.5mm]{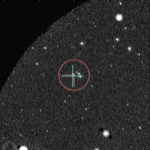} \\
\hline
\edited{139.3+04.8}& 4393 & \includegraphics[width=19.5mm,height=19.5mm]{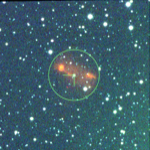} &\includegraphics[width=19.5mm,height=19.5mm]{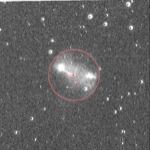} &\includegraphics[width=19.5mm,height=19.5mm]{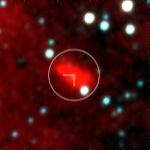} &\includegraphics[width=19.5mm,height=19.5mm]{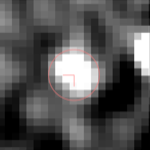} &\includegraphics[width=19.5mm,height=19.5mm]{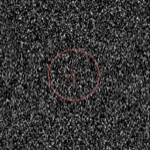} \\
\hline
\edited{147.1-09.0}& 4495 & \includegraphics[width=19.5mm,height=19.5mm]{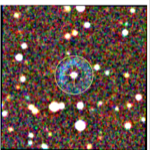} && \includegraphics[width=19.5mm,height=19.5mm]{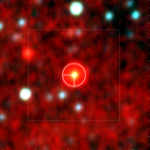} &\includegraphics[width=19.5mm,height=19.5mm]{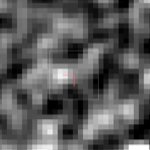} &\includegraphics[width=19.5mm,height=19.5mm]{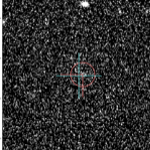} \\
\hline
\edited{147.2+08.3}& 4330 & \includegraphics[width=19.5mm,height=19.5mm]{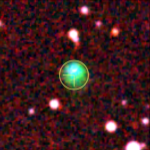} && \includegraphics[width=19.5mm,height=19.5mm]{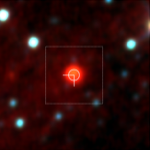} &\includegraphics[width=19.5mm,height=19.5mm]{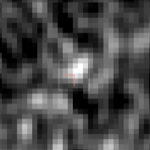} &\includegraphics[width=19.5mm,height=19.5mm]{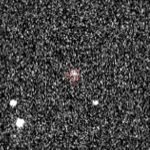} \\
\hline
\edited{151.0-00.4}& 8458 & \includegraphics[width=19.5mm,height=19.5mm]{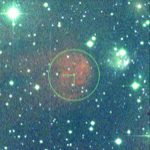} &\includegraphics[width=19.5mm,height=19.5mm]{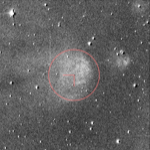} &\includegraphics[width=19.5mm,height=19.5mm]{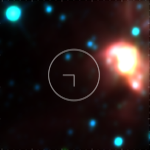} &\includegraphics[width=19.5mm,height=19.5mm]{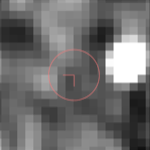} &\includegraphics[width=19.5mm,height=19.5mm]{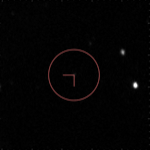} \\
\hline
\edited{154.8+05.9}& 4333 & \includegraphics[width=19.5mm,height=19.5mm]{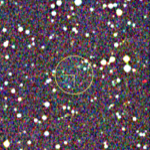} && \includegraphics[width=19.5mm,height=19.5mm]{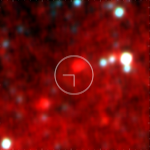} &\includegraphics[width=19.5mm,height=19.5mm]{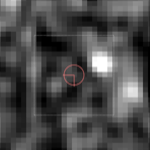} &\includegraphics[width=19.5mm,height=19.5mm]{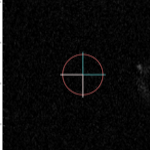} \\
\hline
\edited{164.8-09.8}& 10890 & \includegraphics[width=19.5mm,height=19.5mm]{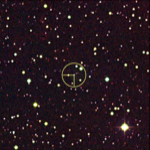} && \includegraphics[width=19.5mm,height=19.5mm]{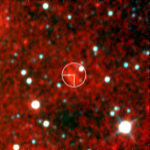} &\includegraphics[width=19.5mm,height=19.5mm]{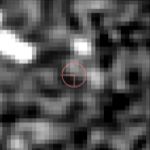} &\includegraphics[width=19.5mm,height=19.5mm]{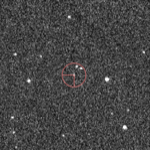} \\
\hline
\edited{174.6-05.2}& 8313 & \includegraphics[width=19.5mm,height=19.5mm]{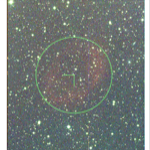} &\includegraphics[width=19.5mm,height=19.5mm]{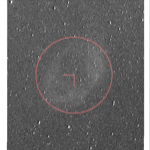} &\includegraphics[width=19.5mm,height=19.5mm]{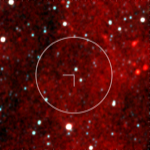} &\includegraphics[width=19.5mm,height=19.5mm]{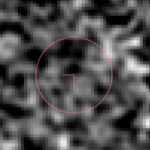} &\includegraphics[width=19.5mm,height=19.5mm]{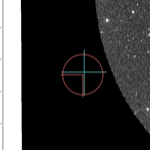} \\
\hline
\edited{175.6+11.4}& 15571 & \includegraphics[width=19.5mm,height=19.5mm]{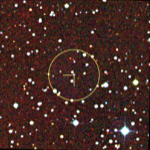} && \includegraphics[width=19.5mm,height=19.5mm]{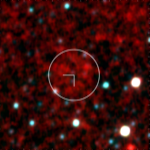} &\includegraphics[width=19.5mm,height=19.5mm]{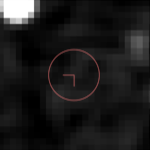} &\includegraphics[width=19.5mm,height=19.5mm]{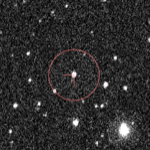} \\
\hline
\edited{182.3-03.7}& 8331 & \includegraphics[width=19.5mm,height=19.5mm]{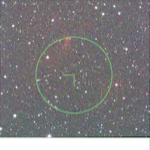} &\includegraphics[width=19.5mm,height=19.5mm]{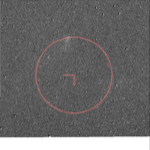} &\includegraphics[width=19.5mm,height=19.5mm]{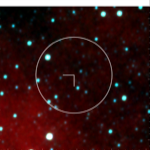} &\includegraphics[width=19.5mm,height=19.5mm]{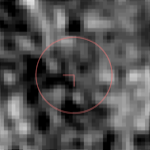} &\includegraphics[width=19.5mm,height=19.5mm]{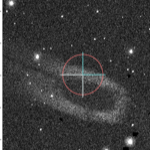}\\
\hline
    \end{longtable}

    \begin{longtable}{ | *{7}{l|} }
        \caption{\edited{Images of the  previously observed candidates that required better spectra for final confirmation. North is up and East is left. The circles mark the major angular diameters given in the HASH database.}}
  \\      \hline
        \edited{IAU PNG} & HASH ID & optical & $\mathrm{H_\alpha/Sr}$ & WISE321 & NVSS & GALEX\\
        \endhead  
        \hline
\edited{040.7+03.4}& 4424 & \includegraphics[width=19.5mm,height=19.5mm]{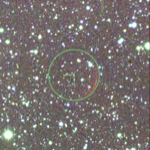} &\includegraphics[width=19.5mm,height=19.5mm]{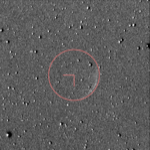} &\includegraphics[width=19.5mm,height=19.5mm]{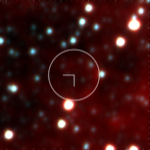} &\includegraphics[width=19.5mm,height=19.5mm]{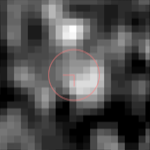} &\\
\hline
\edited{059.1-01.4}& 8215 & \includegraphics[width=19.5mm,height=19.5mm]{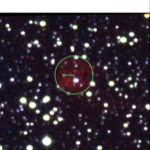} &\includegraphics[width=19.5mm,height=19.5mm]{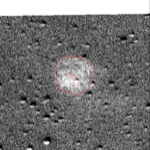} &\includegraphics[width=19.5mm,height=19.5mm]{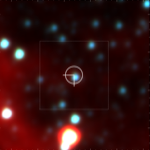} &\includegraphics[width=19.5mm,height=19.5mm]{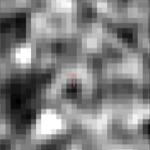} &\\
\hline
\edited{060.2+00.8}& 10878 & \includegraphics[width=19.5mm,height=19.5mm]{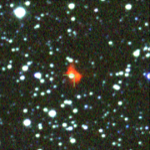} &\includegraphics[width=19.5mm,height=19.5mm]{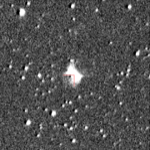} &\includegraphics[width=19.5mm,height=19.5mm]{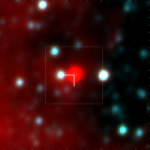} &\includegraphics[width=19.5mm,height=19.5mm]{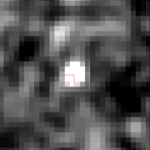} &\\
\hline
\edited{086.1+05.4}& 571 & \includegraphics[width=19.5mm,height=19.5mm]{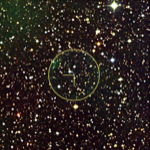} && \includegraphics[width=19.5mm,height=19.5mm]{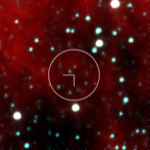} &\includegraphics[width=19.5mm,height=19.5mm]{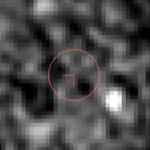} &\\
\hline
\edited{086.9-03.4}& 15806 & \includegraphics[width=19.5mm,height=19.5mm]{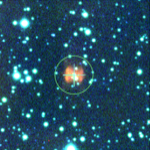} &\includegraphics[width=19.5mm,height=19.5mm]{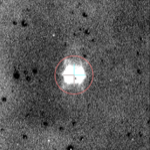} &\includegraphics[width=19.5mm,height=19.5mm]{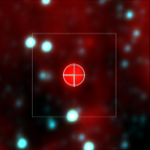} &\includegraphics[width=19.5mm,height=19.5mm]{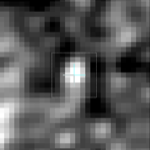} &\\
\hline
\edited{095.1+00.9}& 4431 & \includegraphics[width=19.5mm,height=19.5mm]{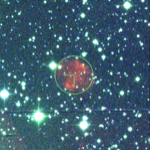} &\includegraphics[width=19.5mm,height=19.5mm]{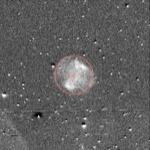} &\includegraphics[width=19.5mm,height=19.5mm]{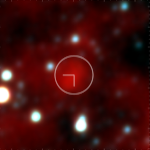} &\includegraphics[width=19.5mm,height=19.5mm]{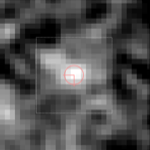} &\includegraphics[width=19.5mm,height=19.5mm]{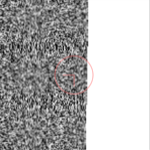} \\
\hline
\edited{095.2+25.4}& 10897 & \includegraphics[width=19.5mm,height=19.5mm]{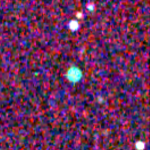} && \includegraphics[width=19.5mm,height=19.5mm]{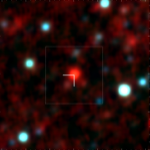} &\includegraphics[width=19.5mm,height=19.5mm]{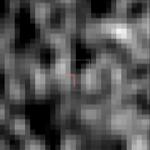} &\includegraphics[width=19.5mm,height=19.5mm]{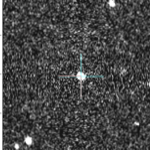} \\
\hline
\edited{097.4+12.3}& 10889 & \includegraphics[width=19.5mm,height=19.5mm]{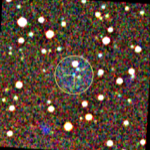} && \includegraphics[width=19.5mm,height=19.5mm]{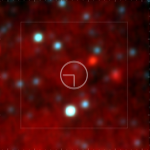} &\includegraphics[width=19.5mm,height=19.5mm]{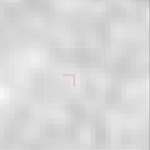} &\includegraphics[width=19.5mm,height=19.5mm]{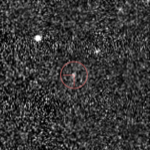} \\
\hline
\edited{098.9+03.0}& 10285 & \includegraphics[width=19.5mm,height=19.5mm]{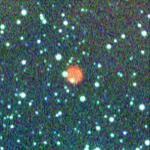} &\includegraphics[width=19.5mm,height=19.5mm]{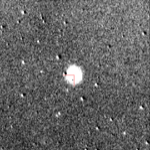} &\includegraphics[width=19.5mm,height=19.5mm]{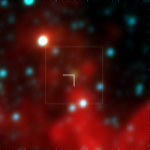} &\includegraphics[width=19.5mm,height=19.5mm]{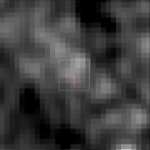} &\\
\hline
\edited{100.3+02.8}& 4386 & \includegraphics[width=19.5mm,height=19.5mm]{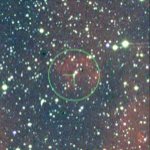} &\includegraphics[width=19.5mm,height=19.5mm]{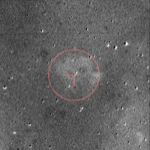} &\includegraphics[width=19.5mm,height=19.5mm]{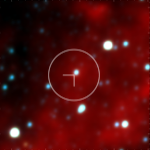} &\includegraphics[width=19.5mm,height=19.5mm]{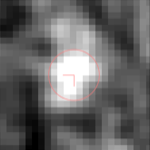} &\includegraphics[width=19.5mm,height=19.5mm]{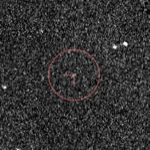} \\
\hline
\edited{105.7+02.2}& 5240 & \includegraphics[width=19.5mm,height=19.5mm]{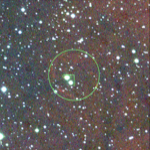} &\includegraphics[width=19.5mm,height=19.5mm]{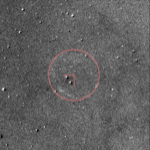} &\includegraphics[width=19.5mm,height=19.5mm]{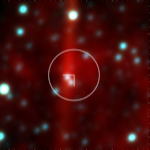} &\includegraphics[width=19.5mm,height=19.5mm]{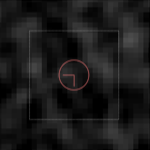} &\includegraphics[width=19.5mm,height=19.5mm]{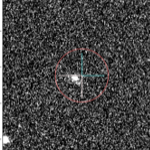} \\
\hline
\edited{109.4+07.7}& 4369 & \includegraphics[width=19.5mm,height=19.5mm]{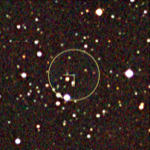} && \includegraphics[width=19.5mm,height=19.5mm]{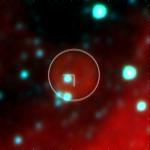} &\includegraphics[width=19.5mm,height=19.5mm]{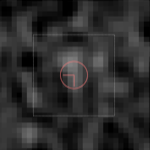} &\includegraphics[width=19.5mm,height=19.5mm]{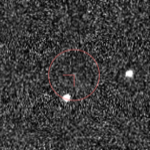} \\
\hline
\edited{123.0+04.6}& 15569 & \includegraphics[width=19.5mm,height=19.5mm]{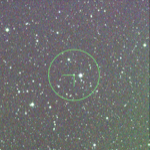} &\includegraphics[width=19.5mm,height=19.5mm]{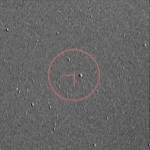} &\includegraphics[width=19.5mm,height=19.5mm]{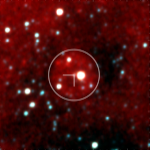} &\includegraphics[width=19.5mm,height=19.5mm]{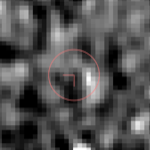} &\includegraphics[width=19.5mm,height=19.5mm]{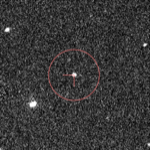} \\
\hline
\edited{134.1+03.0}& 4425 & \includegraphics[width=19.5mm,height=19.5mm]{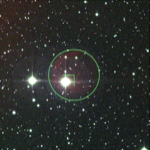} &\includegraphics[width=19.5mm,height=19.5mm]{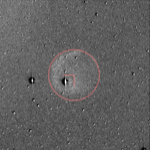} &\includegraphics[width=19.5mm,height=19.5mm]{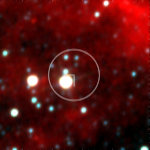} &\includegraphics[width=19.5mm,height=19.5mm]{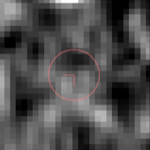} &\includegraphics[width=19.5mm,height=19.5mm]{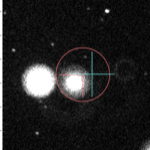} \\
\hline
\edited{149.1+08.7}& 4332 & \includegraphics[width=19.5mm,height=19.5mm]{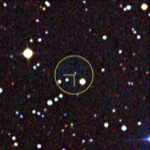} && \includegraphics[width=19.5mm,height=19.5mm]{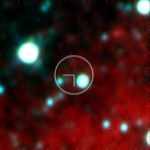} &\includegraphics[width=19.5mm,height=19.5mm]{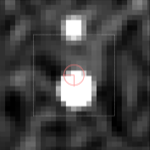} &\includegraphics[width=19.5mm,height=19.5mm]{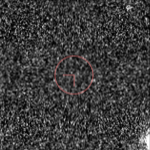} \\
\hline
\edited{183.0+00.0}& 9824 & \includegraphics[width=19.5mm,height=19.5mm]{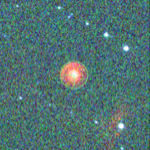} &\includegraphics[width=19.5mm,height=19.5mm]{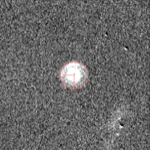} &\includegraphics[width=19.5mm,height=19.5mm]{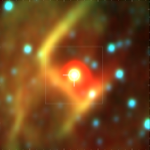} &\includegraphics[width=19.5mm,height=19.5mm]{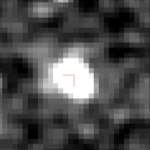} &\includegraphics[width=19.5mm,height=19.5mm]{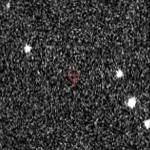}\\
\hline
    \end{longtable}




\bsp	
\label{lastpage}
\end{document}